\documentclass[12pt]{article}
\usepackage{authblk}
\usepackage{amsfonts}
\usepackage{amssymb}
\usepackage{amsmath}
\usepackage{amsthm}
\usepackage{framed}
\usepackage{url}
\usepackage{enumitem}
\usepackage{geometry}
\usepackage{booktabs}

\usepackage{multirow}

\usepackage{mdframed}
\usepackage{mathtools}
\usepackage{algorithm}  
\usepackage{algorithmic}

\newtheorem{theorem}{Theorem}
\newtheorem{lemma}[theorem]{Lemma}

\newcommand{\dmax}{d_\mathrm{max}}

\newcommand{\inn}{\mathrm{in}}
\newcommand{\ex}{\mathrm{ex}}

\newcommand{\typ}{\mathrm{t}}

\newcommand{\ce}{\mathrm{ce}}
\newcommand{\ac}{\mathrm{ac}}

\newcommand{\bd}{\mathrm{bd}}
\newcommand{\bh}{\mathrm{bh}}

\newcommand{\bl}{\mathrm{bl}}
\newcommand{\bc}{\mathrm{bc}}
\newcommand{\Bc}{\mathrm{Bc}}

\newcommand{\UB}{\mathrm{UB}}
\newcommand{\LB}{\mathrm{LB}}
 
\newcommand{\tail}{{\mathrm{tail}}}
\newcommand{\head}{{\mathrm{head}}}
 
\newcommand{\ntreeS}{n_{\mathrm{tree}}^{\mathrm{S}}}
\newcommand{\ntreeT}{n_{\mathrm{tree}}^{\mathrm{T}}}

\newcommand{\G}{\mathcal{G}}

\newcommand{\ta}{{\tt a}}
\newcommand{\tb}{{\tt b}}

\newcommand{\f}{\pmb{f}}
\newcommand{\w}{\pmb{w}}
\newcommand{\x}{\pmb{x}}
\newcommand{\z}{\pmb{z}}

\newcommand{\1}{\pmb{1}}

\newcommand{\val}{\mathrm{val}}
\newcommand{\dia}{\mathrm{dia}}

\newcommand{\inl}{\mathrm{inl}}
\newcommand{\en}{\mathrm{end}}
\newcommand{\pair}{\mathrm{pair}}

\newcommand{\LtoD}{\mathrm{\Lambda\cup \Gamma\cup \Bc\cup \Dg}}
 
\newcommand{\FT}{\mathcal{FT}} 
\newcommand{\T}{\mathcal{T}}  
\newcommand{\W}{\mathrm{W}}

\newcommand{\Dg}{\mathrm{Dg}}
\newcommand{\dg}{\mathrm{dg}}

\begin{document}  
 
\begin{center}
   {\Large\bf A Novel Method for Inference of
Acyclic Chemical Compounds with Bounded Branch-height Based on
Artificial Neural Networks and Integer Programming}
\end{center} 

\begin{center}
Naveed Ahmed Azam$^1$, 
Jianshen Zhu$^1$, 
Yanming Sun$^1$, 
Yu Shi$^1$, \\
Aleksandar Shurbevski$^1$, 
Liang Zhao$^2$,
Hiroshi Nagamochi$^1$, 
Tatsuya Akutsu$^3$, 
\end{center} 
%


\begin{quote}  
{\bf Abstract}\\ 
Analysis of chemical graphs is becoming a major research topic
in computational molecular biology due to its potential applications
to drug design.
One of the major approaches in such a study is
inverse quantitative structure activity/property relationships
(inverse QSAR/QSPR) analysis, which is to infer chemical structures
from given chemical activities/properties.
Recently, a novel framework has been proposed for inverse QSAR/QSPR
using both artificial neural networks (ANN) and
mixed integer linear programming (MILP).
This method consists of a prediction phase and an inverse prediction phase.
In the first phase,  
a feature vector $f(G)$ of a chemical graph $G$ is introduced and 
a prediction function $\psi_{\mathcal{N}}$ on a chemical property $\pi$
is constructed with an ANN $\mathcal{N}$. 
In the second phase, given a target value $y^*$ of the chemical property $\pi$,
a feature vector $x^*$ is inferred  
by solving an MILP formulated from  the trained ANN    $\mathcal{N}$
so that  $\psi_{\mathcal{N}}(x^*)$ is close to  $y^*$ 
and   then a set of chemical structures $G^*$
such that $f(G^*)= x^*$  is enumerated by a graph search algorithm. 
The framework has been applied to the case of 
chemical compounds with cycle index up to 2 so far.
The computational results conducted on instances with
$n$ non-hydrogen atoms show that
a feature vector $x^*$ can be inferred for up to around $n=40$
whereas  graphs $G^*$ can be enumerated  for up to around $n=15$.  
When applied to the case of chemical acyclic graphs, the maximum computable 
diameter of $G^*$ was around up to around 8.  
In this paper, we introduce  a new characterization of graph structure,
called ``branch-height'' based on which a new MILP formulation
and a new graph search algorithm are designed for  chemical acyclic graphs.   
The results of computational experiments using such chemical properties as 
  octanol/water partition coefficient,
  boiling point and
  heat of combustion 
suggest that the proposed method can infer chemical acyclic  graphs $G^*$   
with $n=50$ and diameter 30.   
\\
{\bf Keywords: } QSAR/QSPR,  Molecular Design, 
             Artificial Neural Network, Mixed  Integer Linear Programming, 
             Enumeration of Graphs \\
{\bf  Mathematics Subject Classification: } 
Primary  
05C92,  
92E10, 
Secondary
05C30, 
68T07, 
90C11,  
92-04 
\end{quote}

\section{Introduction}\label{sec:introduction}

In computational molecular biology,
various types of data have been utilized, which include
sequences, gene expression patterns, and protein structures.
Graph structured data have also been extensively utilized,
which include metabolic pathways, protein-protein interaction networks,
gene regulatory networks, and chemical graphs.
Much attention has recently been paid to analysis of chemical graphs
due to its potential applications to computer-aided drug design.
One of the major approaches to computer-aided drug design is
quantitative structure activity/property relationships (QSAR/QSPR) analysis,
the purpose of which is to derive
quantitative relationships between chemical structures and
their activities/properties.
Furthermore, inverse QSAR/QSPR has been extensively studied
\cite{Miyao16,Skvortsova93}, 
the purpose of which is to infer chemical structures from given chemical
activities/properties.
Inverse QSAR/QSPR is often formulated as an optimization problem
to find a chemical structure maximizing (or minimizing) an
objective function under various constraints.

In both QSAR/QSPR and inverse QSAR/QSPR,
chemical compounds are usually represented as vectors of real or integer numbers,
which are often called \emph{descriptors} and
correspond to \emph{feature vectors} in machine learning.
Using these chemical descriptors,
various heuristic and statistical methods have been developed for
finding optimal or nearly optimal graph structures under given
objective functions~\cite{Ikebata17,Miyao16,Rupakheti15}.
Inference or enumeration of graph structures from a given feature vector 
is a crucial subtask in many of such methods.
Various methods have been developed for this enumeration problem
\cite{Fujiwara08,Kerber98,Li18,Reymond15}
and the computational complexity of the inference problem has
been analyzed~\cite{Akutsu12,Nagamochi09}.
On the other hand, enumeration in itself is a challenging task, 
since the number of
molecules (i.e., chemical graphs) with up to 30 atoms (vertices)
{\tt C}, {\tt N}, {\tt O}, and {\tt S},
may exceed~$10^{60}$~\cite{BMG96}.

As a new approach,
artificial neural network (ANN) and deep learning technologies
have recently been applied to inverse QSAR/QSPR.
For example, variational autoencoders~\cite{Gomez18}, 
recurrent neural networks~\cite{Segler18,Yang17}, and
grammar variational autoencoders~\cite{Kusner17} have been applied.
In these approaches, new chemical graphs are generated
by solving a kind of inverse problems on neural networks that are
trained using known chemical compound/activity pairs.
However, the optimality of the solution is not necessarily guaranteed in
these approaches.
In order to guarantee the optimality mathematically,
a novel approach has been proposed~\cite{AN19}
for ANNs, 
 using mixed integer linear programming (MILP).

Recently, a new framework has been proposed
\cite{ACZSNA20,CWZSNA20,ZZCSNA20} 
by combining two previous approaches; efficient enumeration
of tree-like graphs~\cite{Fujiwara08}, and 
MILP-based formulation of the inverse problem on ANNs~\cite{AN19}.
This combined framework for  inverse QSAR/QSPR mainly consists of two phases.
The first phase solves (I) {\sc Prediction Problem}, 
where 
a feature vector $f(G)$ of a chemical graph $G$ is introduced
and 
a prediction function $\psi_{\mathcal{N}}$ on a chemical property $\pi$
is constructed with an ANN $\mathcal{N}$ 
using a data set of  chemical compounds $G$ and their values $a(G)$ of $\pi$.
The second phase solves (II) {\sc Inverse Problem},
  where (II-a) given a target value $y^*$ of the chemical property $\pi$,
   a feature vector $x^*$ is inferred from the trained ANN  $\mathcal{N}$
  so that  $\psi_{\mathcal{N}}(x^*)$ is close to  $y^*$ 
  and (II-b) then a set of chemical structures $G^*$
 such that $f(G^*)= x^*$  is enumerated by a graph search algorithm.  
In (II-a) of the above-mentioned previous methods~\cite{ACZSNA20,CWZSNA20,ZZCSNA20},
an MILP is formulated for acyclic chemical compounds.
 Afterwards, Ito et~al. \cite{IAWSNA20} and Zhu et~al. \cite{ZCSNA20}
 designed a method of  inferring  chemical graphs 
 with  cycle index 1 and 2, respectively 
by formulating a new MILP and using
an efficient algorithm for enumerating  chemical graphs with cycle index 1 
\cite{Suzuki14} and cycle index 2 \cite{2A1B20,2A2B20}. 
The computational results conducted on instances with
$n$ non-hydrogen atoms show that
a feature vector $x^*$ can be inferred for up to around $n=40$
whereas  graphs $G^*$ can be enumerated  for up to around $n=15$. 
 
In this paper, we present a new characterization of graph structure,
called ``branch-height.'' 
Based on this, we can treat a class of acyclic chemical graphs
with a structure that is topologically restricted but frequently appears
in the chemical  database,
formulate a new MILP formulation that can handle
acyclic graphs with a large diameter,
and design a new graph search algorithm
that generates acyclic chemical graphs with up to 50 vertices.  
The results of computational experiments using such chemical properties as  
  octanol/water partition coefficient,
  boiling point and
  heat of combustion 
suggest that the proposed method is much more useful than the previous method.

The paper is organized as follows.  
Section~\ref{sec:preliminary} introduces some notions on graphs,
 a modeling of chemical compounds and a choice of descriptors. 
Section~\ref{sec:inverse_process} reviews the framework for inferring
chemical compounds based on ANNs and MILPs. 
Section~\ref{sec:graph_MILP} introduces a new method of modeling
acyclic  chemical graphs   and  
proposes a new MILP formulation
that  represents an acyclic  chemical graph $G$ with $n$ vertices,
where our MILP requires only 
 $O(n)$  variables and constraints 
 when  the branch-parameter $k$ and the $k$-branch-height  in $G$
 (graph topological parameters newly introduced in this paper)  is constant.   
Section~\ref{sec:graph_search} describes the idea of our new
dynamic programming type of algorithm 
that enumerates a given number of  acyclic  chemical graphs
for a given feature vector. 
Section~\ref{sec:experiment} reports the results on some computational 
experiments conducted for s  chemical properties
such as   octanol/water partition coefficient,
  boiling point and
  heat of combustion.  
Section~\ref{sec:conclude} makes some concluding remarks. 
Appendix~\ref{sec:statistical} provides the statistical feature on
structure of acyclic  chemical graphs in a chemical graph database.  
Appendix~\ref{sec:full_milp} describes the details of all variables and constraints
in our MILP formulation. 
Appendix~\ref{sec:graph_search_appendix} presents descriptions of our new graph search algorithm.

\section{Preliminary}\label{sec:preliminary}

This section  introduces some notions and terminology on graphs,
 a modeling of chemical compounds and our choice of descriptors. 
 
Let $\mathbb{R}$, $\mathbb{Z}$  and $\mathbb{Z}_+$ 
denote the sets of reals, integers and non-negative integers, respectively.
For two integers $a$ and $b$, let $[a,b]$ denote the set of 
integers $i$ with $a\leq i\leq b$.

\subsection{Graphs} 
   
A {\em graph} stands for a simple undirected graph, where an edge joining
two vertices $u$ and $v$ is denoted by $uv$ $(= vu)$.
 The sets of vertices and edges of a graph $G$ are denoted by $V(G)$ and $E(G)$, respectively.
 Let $H=(V,E)$ be a  graph with a set $V$ of vertices and a set $E$ of edges.
For a vertex $v\in V$, the set of neighbors of $v$ in $H$ is denoted by $N_H(v)$,
and the {\em degree} $\deg_H(v)$ of $v$ is defined to be $|N_H(v)|$. 
The length of a path is defined to be the number of edges in the path. 
The {\em distance}  $\mathrm{dist}_H(u,v)$ 
between two vertices $u,v\in V$ is defined to be the minimum length 
of a path connecting $u$ and $v$ in $H$.
The {\em diameter} $\dia(H)$ of $H$ is defined to be 
the maximum distance between two vertices  in $H$;
i.e.,   $\dia(H)\triangleq \max_{u,v\in V} \mathrm{dist}_H(u,v)$.
Denote by $\ell(P)$ the length of a path $P$.

\medskip\noindent{\bf Trees}     
For a tree $T$ with an even (resp., odd) diameter $d$, 
the {\em center} is defined 
to be the vertex $v$ (resp., the adjacent vertex pair $\{v,v'\}$) 
 that situates in the middle of
one of the longest paths with length $d$. 
The center of each tree is uniquely determined. 

\begin{figure}[ht!] \begin{center}
\includegraphics[width=.90\columnwidth]{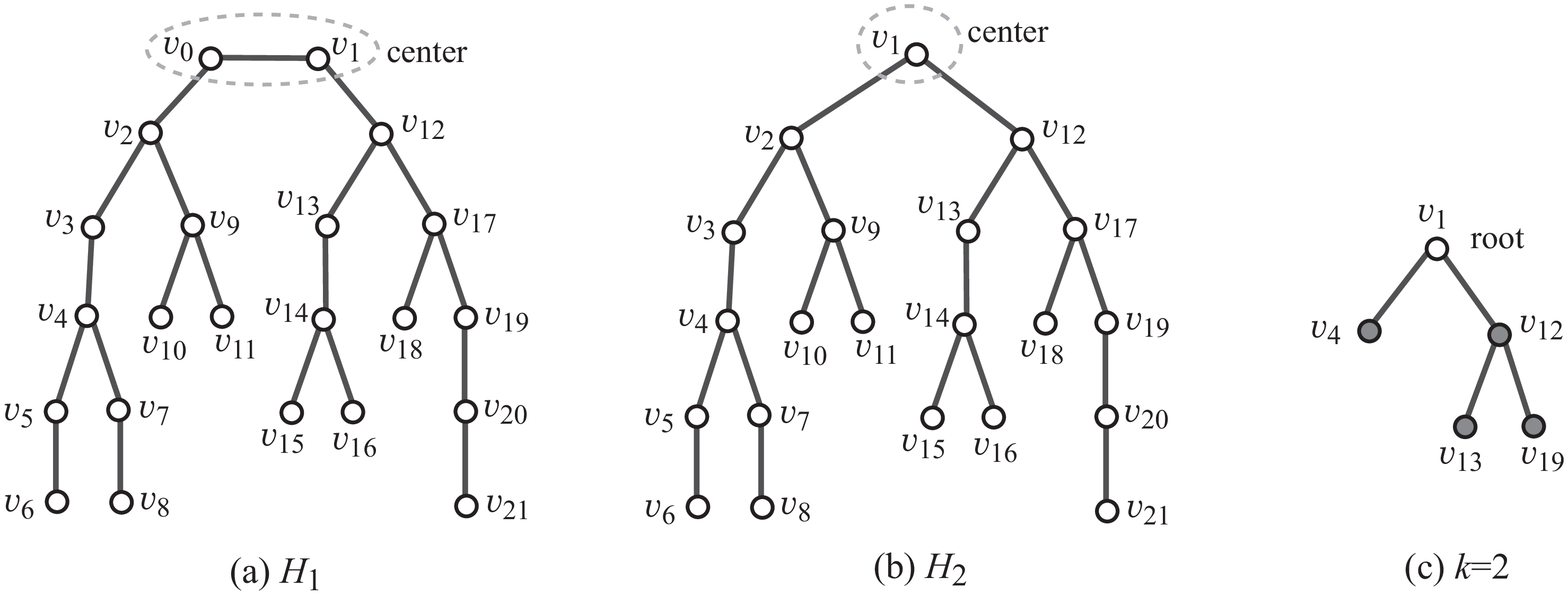}
\end{center}
\caption{An illustration of rooted trees and a 2-branch-tree:  
(a) A tree $H_1$ with odd diameter 11;
(b) A tree $H_2$ with even diameter 10;
(c) The 2-branch-tree of $H_2$.  
}
\label{fig:branch-height_tree} \end{figure}

\medskip\noindent{\bf Rooted Trees}     
A {\em rooted tree} is defined to be
a tree where a   vertex
 (or a  pair of adjacent vertices) is designated as the {\em root}.
Let $T$ be a rooted tree, where for two adjacent vertices
$u$ and $v$, vertex $u$ is called the parent of $v$ if
$u$ is closer to the root than $v$ is. 
The {\em height} $\mathrm{height}(v)$ of a vertex $v$  in $T$
is defined to be the maximum length of a path
from $v$ to a leaf $u$ in the  descendants of $v$,
where  $\mathrm{height}(v)=0$ for each leaf $v$ in $T$. 
Figure~\ref{fig:branch-height_tree}(a) and (b) illustrate
examples of trees rooted at the center.

\medskip\noindent{\bf Degree-bounded Trees}     
For positive integers $a,b$ and $c$ with $b\geq 2$, 
let $T(a,b,c)$ denote the rooted tree
such that the number of children of the root is $a$,
the number of children of each non-root internal vertex is $b$
and the distance from the root to each leaf is $c$.
We see that  the number of vertices in $T(a,b,c)$ is
$a(b^c-1)/(b-1)+1$, and 
the number of non-leaf vertices in $T(a,b,c)$
is $a(b^{c-1}-1)/(b-1)+1$.
%
In the rooted tree  $T(a,b,c)$, we denote the vertices 
by $v_1,v_2,\ldots,v_n$ with a breadth-first-search order,
and denote the edge between a vertex $v_i$ with $i\in [2,n]$ and its parent by $e_i$, 
where $n=a(b^c-1)/(b-1)+1$ and each vertex 
$v_i$ with $i\in [1, a(b^{c-1}-1)/(b-1)+1]$ is a non-leaf vertex.
For each vertex $v_i$ in   $T(a,b,c)$, 
let  $\mathrm{Cld}(i)$ denote the set of indices
$j$ such that $v_j$ is a child of $v_i$, 
 and $\mathrm{prt}(i)$ denote the index $j$
 such that $v_j$ is the parent of $v_i$ when $i\in [2,n]$. 
Let $P_{\mathrm{prc}}(a,b,c)$ be a set of ordered index pairs $(i,j)$
of vertices $v_i$ and $v_j$ in   $T(a,b,c)$. 
We call $P_{\mathrm{prc}}(a,b,c)$ {\em proper}
if the next conditions hold:  
\begin{enumerate}
\item[(a)] For each  subtree $H=(V,E)$ of  $T(a,b,c)$ with $v_1\in V$,
there is at least one subtree $H'=(V',E')$
such that \\
~-~  $H'$ is  isomorphic to $H$ by a graph isomorphism $\psi:V\to V'$ with 
$\psi(v_1)=v_1$; and \\
~-~  for each pair $(i,j)\in P_{\mathrm{prc}}(a,b,c)$,  if $v_j\in V'$ then $v_i\in V'$; and
\item[(b)] For each pair of vertices $v_i$ and $v_j$ in $T(a,b,c)$
such that $v_i$ is the parent of $v_j$,
there is a sequence $(i_1,i_2),(i_2,i_3),\ldots,(i_{k-1},i_k)$
of index pairs in $P_{\mathrm{prc}}(a,b,c)$ 
such that $i_1=i$ and $i_k=j$. 
\end{enumerate} 
Note that a proper set $P_{\mathrm{prc}}(a,b,c)$ is not necessarily unique.   

\begin{figure}[ht!] \begin{center}
\includegraphics[width=.94\columnwidth]{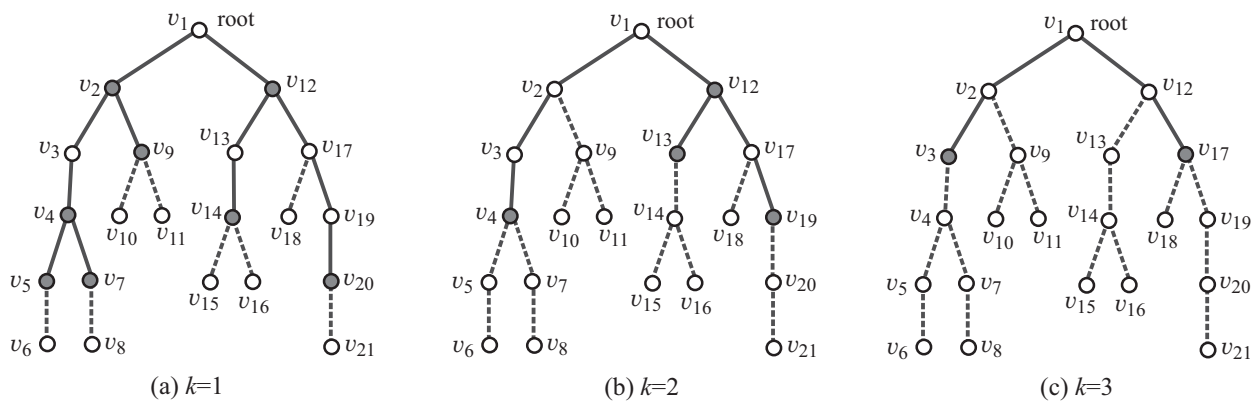}
\end{center}
\caption{An illustration of  the $k$-branches (depicted by gray circles), 
the $k$-branch-subtree (depicted by solid lines)
 and $k$-fringe-trees  (depicted by dashed lines)  of $H_2$: 
(a) $k=1$;
(b) $k=2$;
(c) $k=3$.  
}
\label{fig:branch-height_tree_k123} \end{figure}

\medskip\noindent{\bf Branch-height in Trees}    
In this paper, we introduce ``branch-height'' of a tree 
as a new measure to the ``agglomeration degree'' of trees.
We specify a non-negative integer $k$, called a {\em branch-parameter} 
to define branch-height. 
 First we regard   $T$ as a rooted tree by choosing
 the center of $T$ as the root.
 Figure~\ref{fig:branch-height_tree}(a) and (b) 
illustrate examples of rooted trees.
We introduce the following terminology on a rooted tree $T$.
\begin{itemize}
\item[-] A {\em leaf $k$-branch}:  a non-root vertex $v$ in $T$  such that
  $\mathrm{height}(v)= k$.
  
\item[-] A  {\em non-leaf $k$-branch}:  
 a  vertex  $v$ in $T$ such that $v$ has at least two children $u$  
with  $\mathrm{height}(u)\geq k$.
We call a leaf or non-leaf $k$-branch a {\em  $k$-branch}.
Figure~\ref{fig:branch-height_tree_k123}(a)-(c) illustrate  the $k$-branches of 
the rooted tree 
 $H_2$  in Figure~\ref{fig:branch-height_tree}(b) for $k=1,2$ and $3$,
respectively. 

\item[-] A {\em $k$-branch-path}: a path $P$ in $T$
that joins two vertices $u$ and $u'$ such that 
each of $u$ and $u'$ is the root or a $k$-branch and
$P$ does not contain   the root or a $k$-branch
as an internal vertex. 


\item[-]
The  {\em $k$-branch-subtree} of $T$:
 the subtree of $T$ that consists of 
the edges in all $k$-branch-paths of $T$. 
 We call a vertex (resp., an edge) in $T$ 
 a {\em $k$-internal vertex} (resp., a {\em $k$-internal edge})
  if it is contained in the $k$-branch-subtree  of $T$
 and a {\em $k$-external vertex}   (resp., a {\em $k$-external edge}) otherwise.
 Let $V^\inn$ and $V^\ex$ (resp., $E^\inn$ and $E^\ex$)  
 denote the sets of  $k$-internal and $k$-external vertices (resp., edges) in $T$.
 
\item[-]
The  {\em $k$-branch-tree} of $T$: the rooted tree 
obtained from the $k$-branch-subtree  of $T$
by replacing each $k$-branch-path with a single edge. 
Figure~\ref{fig:branch-height_tree}(c)
illustrates  the $2$-branch-tree of the rooted tree $H_2$ 
 in Figure~\ref{fig:branch-height_tree}(b). 
 
\item[-] 
 A {\em $k$-fringe-tree}: One of the connected components
that consists of the edges not in any  $k$-branch-subtree.
Each $k$-fringe-tree $T'$ contains exactly one vertex $v$ in a  $k$-branch-subtree,
where $T'$ is regarded as a   tree rooted at $v$.
Note that the height of any $k$-fringe-tree is at most $k$.
Figure~\ref{fig:branch-height_tree_k123}(a)-(c) illustrate 
 the $k$-fringe-tree of the rooted tree 
 $H_2$  in Figure~\ref{fig:branch-height_tree}(b) for $k=1,2$ and $3$,
respectively. 
 
\item[-]
The {\em $k$-branch-leaf-number} $\bl_k(T)$: 
the number of leaf $k$-branches  in $T$.
For the trees $H_i$, $i=1,2$ in Figure~\ref{fig:branch-height_tree}(a) and (b), 
it holds that  
$\bl_0(H_1)= \bl_0(H_2)=8$,
$\bl_1(H_1)= \bl_1(H_2)=5$,
$\bl_2(H_1)= \bl_2(H_2)=3$ and 
$\bl_3(H_1)= \bl_3(H_2)=2$.

\item[-]
The {\em $k$-branch-height} $\bh_k(T)$ of $T$: 
the maximum number of non-root $k$-branches along a path
from the root  to a leaf of $T$; i.e., $\bh_k(T)$ is 
the height of the $k$-branch-tree $T^*$ (the maximum length of a path
from the root to a leaf in $T^*$).  
For the example of trees $H_i$, $i=1,2$ in Figure~\ref{fig:branch-height_tree}(a) and (b), 
it holds that 
$\bh_0(H_1)=\bh_0(H_2)=5$,
$\bh_1(H_1)=\bh_1(H_2)=3$,
$\bh_2(H_1)=\bh_2(H_2)=2$ and
$\bh_3(H_1)=\bh_3(H_2)=1$.
\end{itemize}

We observe that most chemical graphs $G$ with at most 50 non-hydrogen atoms
satisfy $\bh_2(G)\leq 2$. 
See Appendix~A for a summary of 
statistical feature of chemical graphs registered   in the chemical  database  PubChem.

\subsection{Modeling of Chemical Compounds}\label{sec:chemical_model}

We represent the graph structure of a chemical compound as a graph
with labels on vertices and multiplicity on edges in a hydrogen-suppressed model.
Let $\Lambda$ be a set of  labels each of which represents a chemical element
 such as
 {\tt C} (carbon), {\tt O} (oxygen), {\tt N} (nitrogen)
 and so on,
 where we assume that $\Lambda$ does not contain {\tt H} (hydrogen).
Let $\mathrm{mass}(\ta)$ and $\mathrm{val}(\ta)$ 
denote the mass and    valence of a  chemical element $\ta\in \Lambda$,
respectively.  
In our model, we   use integers
  $\mathrm{mass}^*(\ta)=\lfloor 10\cdot \mathrm{mass}(\ta)\rfloor$, 
  $\ta\in \Lambda$ and assume that
  each chemical element $\ta\in \Lambda$ has a unique 
  valence  $\val(\ta)\in [1,4]$. 
  
We introduce a total order $<$ over the elements in $\Lambda$
according to their mass values; i.e., we write ${\tt a<b}$
for chemical elements ${\tt a,b}\in \Lambda$ with
 $\mathrm{mass}(\ta)<\mathrm{mass}({\tt b})$.
Choose a set  $\Gamma_{<}$ of tuples 
 $\gamma=({\tt a,b},m)\in\Lambda\times \Lambda\times [1,3]$ 
 such that ${\tt a<b}$.
For a tuple  $\gamma=({\tt a,b},m)\in\Lambda\times \Lambda\times [1,3]$,
let $\overline{\gamma}$ denote the tuple $({\tt b,a},m)$.
 Set $\Gamma_{>}=\{\overline{\gamma}\mid \gamma\in \Gamma_{<}\}$
 and 
  $\Gamma_{=}=\{({\tt a,a},m)\mid \ta\in \Lambda, m\in [1,3]\}$.
 A pair of two atoms $\ta$ and ${\tt b}$ joined with a bond-multiplicity $m$
 is denoted by a tuple $\gamma=({\tt a,b},m)\in \Gamma$, 
 called the {\em adjacency-configuration} of the atom pair. 

 We use  a hydrogen-suppressed model because hydrogen atoms can be
added at the final stage.
A {\em chemical graph} over $\Lambda$ and 
$\Gamma_{<}\cup \Gamma_{=}$ is defined to be 
a  tuple $G=(H,\alpha,\beta)$
of a graph $H=(V,E)$, a function   $\alpha:V\to \Lambda$ 
and a function $\beta: E\to [1,3]$ 
such that 
\begin{enumerate}
\item[(i)] $H$ is connected;  
\item[(ii)]  $\sum_{uv\in E}\beta(uv)\leq  \mathrm{val}(\alpha(u))$ 
   for each vertex $u\in V$; and
\item[(iii)] $(\alpha(u),\alpha(v),\beta(uv))\in \Gamma_{<}\cup \Gamma_{=}$
 for each edge $uv\in E$. 
\end{enumerate}
For a notational convenience, we denote the sum of bond-multiplicities
of edges incident to a vertex as follows:
\[    \beta(u) \triangleq \sum_{uv\in E}\beta(uv) 
\mbox{ for each vertex $u\in V$.}\]
A chemical graph $G=(H,\alpha,\beta)$ is called 
  a ``chemical monocyclic graph'' if
  the graph $H$ is a monocyclic graph.
  Similarly for other types of graphs for $H$.

 We define
the {\em bond-configuration} of an edge $e=uv \in E$ in a chemical graph $G$
to be a tuple $(\deg_H(u),\deg_H(v),\beta(e))$
such that $\deg_H(u)\leq \deg_H(v)$ for
the end-vertices $u$ and $v$ of $e$. 
 Let  $\Bc$ denote 
 the set of  bond-configurations $\mu=(d_1,d_2,m)\in  [1,4]\times[1,4]\times[1,3]$
 such that  $\max\{d_1,d_2\}+m\leq 4$. 
 We regard that $(d_1,d_2,m)=(d_2,d_1,m)$. 
 For two tuples $\mu=(d_1,d_2,m), \mu'=(d'_1,d'_2,m')\in \Bc$,
 we write $\mu\geq  \mu'$ if  
 $\max\{d_1,d_2\}\geq \max\{d'_1,d'_2\}$, 
 $\min\{d_1,d_2\}\geq \min\{d'_1,d'_2\}$ and    $m\geq m'$, 
 and write $\mu> \mu'$ if $\mu\geq \mu'$ and $\mu\neq \mu'$.  
 
\subsection{Descriptors}
In our method, we use only graph-theoretical descriptors for defining a feature vector,
 which facilitates our designing an algorithm for constructing graphs.  
Given a chemical acyclic graph $G=(H,\alpha,\beta)$, we define a {\em feature vector} $f(G)$
that consists of the following 11 kinds of descriptors. 
We choose an integer $k^*\in [1,4]$ as a branch-parameter.

\begin{itemize}
\item[-] $n(G)$: the number $|V|$ of vertices.  
 
\item[-] $\dg_i^\inn(G)$, $i\in [1,4]$: the number of $k^*$-internal vertices of 
degree $i$ in $H$;
i.e., $\dg_i^\inn(G)\triangleq |\{v\in V^\inn \mid \deg_{H}(v)=i\}|$, 
where the multiplicity of edges incident to a vertex $v$ is ignored in the degree of $v$.

\item[-] $\dg_i^\ex(G)$, $i\in [1,4]$: the number of $k^*$-external vertices of 
degree $i$ in $H$;
i.e., $\dg_i^\ex(G)\triangleq |\{v\in V^\ex \mid \deg_{H}(v)=i\}|$. 

\item[-] $\overline{\dia}(G)$: the diameter of $H$ divided by $|V|$;
i.e., $\overline{\dia}(G)\triangleq  \dia(H)/n(G)$. 

\item[-] $\bl_{k^*}(G)$: 
the $k^*$-branch-leaf-number of $G$. 

\item[-] $\bh_{k^*}(G)$: 
the $k^*$-branch-height of $G$. 

\item[-] $\ce_\ta^\inn(G)$,  $\ta\in \Lambda$: 
the number of $k^*$-internal vertices with label $\ta\in \Lambda$; 
i.e., $\ce_\ta^\inn(G)\triangleq |\{ v\in V^\inn \mid \alpha(v)=\ta\}|$.

\item[-] $\ce_\ta^\ex(G)$,  $\ta\in \Lambda$: 
the number of $k^*$-external vertices with label $\ta\in \Lambda$; 
i.e., $\ce_\ta^\ex(G)\triangleq |\{ v\in V^\ex \mid \alpha(v)=\ta\}|$.

\item[-] $\overline{\mathrm{ms}}(G)$: the average mass$^*$ of atoms in $G$; 
i.e., $\overline{\mathrm{ms}}(G)\triangleq 
   \sum_{v\in V}\mathrm{mass}^*(\alpha(v))/n(G)$. 

\item[-] $\bd_m^\inn(G)$, $m=2,3$: the number of 
double and triple bonds of $k^*$-internal edges; 
i.e., $\bd_m^\inn(G)\triangleq \{e\in E^\inn \mid \beta(e)=m\}$, $m=2,3$. 

\item[-] $\bd_m^\ex(G)$, $m=2,3$: the number of double 
and triple bonds of $k^*$-internal edges; 
i.e., $\bd_m^\ex(G)\triangleq \{e\in E^\ex \mid \beta(e)=m\}$, $m=2,3$. 

\item[-] $\ac_{\gamma}^\inn(G)$,  $\gamma=({\tt a,b},m)\in \Gamma$:
 the number of adjacency-configurations 
 $({\tt a,b},m)$ of $k^*$-internal edges in $G$.

\item[-] $\ac_{\gamma}^\ex(G)$,  $\gamma=({\tt a,b},m)\in \Gamma$:
 the number of adjacency-configurations $({\tt a,b},m)$ 
 of $k^*$-external edges   in $G$.
  
\item[-] $\bc_{\mu}^\inn(G)$, $\mu=(d,d',m)\in \Bc$:
the number of bond-configurations $(d,d',m)$  of $k^*$-internal edges in $G$.  

\item[-] $\bc_{\mu}^\ex(G)$, $\mu=(d,d',m)\in \Bc$:
the number of bond-configurations $(d,d',m)$  of $k^*$-external edges in $G$.  

\item[-] $n_{\tt H}(G)$:  the number of hydrogen atoms; i.e., \\
~~~ $\displaystyle{ n_{\tt H}(G)\triangleq 
     \sum_{\ta\in \Lambda, {\tt t}\in\{\inn,\ex\}}
       \mathrm{val}(\ta)\ce_\ta^\typ(G) 
             -   \sum_{\gamma=({\tt a,b},m)\in \Gamma, {\tt t}\in\{\inn,\ex\}}
               2m\cdot  \ac_{\gamma}^\typ(G)) }$\\
~~~~~~~~~~~
$\displaystyle{      = \sum_{\ta\in \Lambda, {\tt t}\in\{\inn,\ex\}}
       \mathrm{val}(\ta)\ce_\ta^\typ(G) 
              -2(n(G)-1 
             + \sum_{m\in [2,3], {\tt t}\in\{\inn,\ex\}}m\cdot \bd_m^\typ(G))  }$.
\end{itemize}

 The number $K$ of descriptors in our feature vector $x=f(G)$ is
 $K=2|\Lambda|+2|\Gamma|+50$. 
 Note that the set of the above $K$ descriptors is not independent in the sense that
 some descriptor depends on the combination of other descriptors in the set.
 For example, descriptor $\bd_i^\inn(G)$ can be determined by 
$\sum_{\gamma=({\tt a,b},m)\in \Gamma: m=i }\ac_{\gamma}^\inn(G)$.
 

\section{A Method for Inferring Chemical Graphs}\label{sec:inverse_process}

\subsection{Framework for the Inverse QSAR/QSPR}
\begin{figure}[!ht]  \begin{center}
\includegraphics[width=.98\columnwidth]{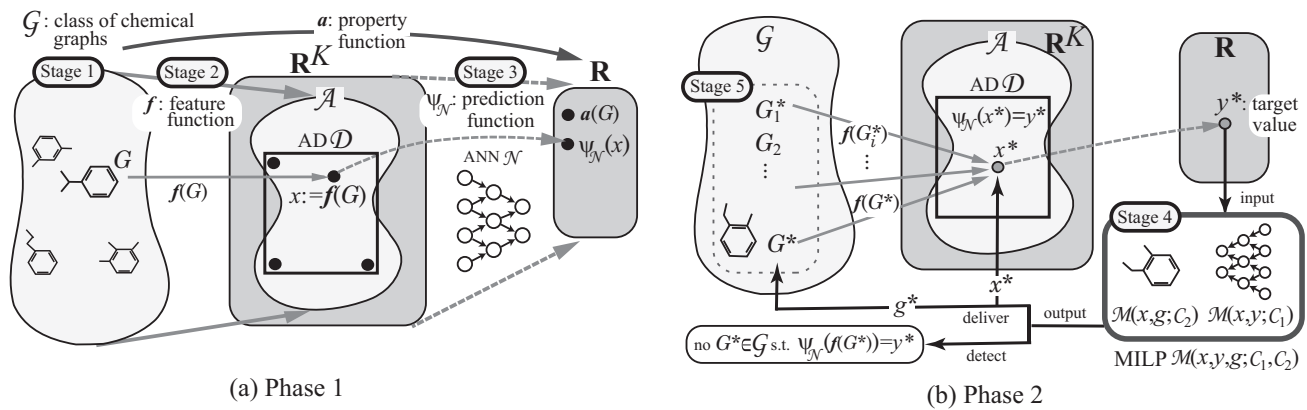}
\end{center} \caption{  (a) An illustration of Phase~1: 
Stage~1 for preparing  a data set $D_{\pi}$ for  a graph class $\G$
and a specified chemical property $\pi$;   
Stage~2 for introducing a feature function $f$ with descriptors;  
Stage~3 for constructing  a prediction function $\psi_\mathcal{N}$ 
with an ANN $\mathcal{N}$;  
(b) An illustration of Phase~2: 
 Stage~4 for formulating 
an  MILP  $\mathcal{M}(x,y,g;\mathcal{C}_1,\mathcal{C}_2)$
and finding   a feasible solution $(x^*,g^*)$ of the MILP
for a target value $y^*$ so that  $\psi_\mathcal{N}(x^*)=y^*$
(possibly detecting that no target graph $G^*$ exists);   
Stage~5 for enumerating graphs $G^*\in \G$ 
such that $f(G^*)=x^*$.  }
\label{fig:framework}  \end{figure}

We review the framework that solves the inverse QSAR/QSPR
by using MILPs ~\cite{IAWSNA20,ZCSNA20}, 
which is illustrated in Figure~\ref{fig:framework}.
%
For a specified chemical property $\pi$  such as boiling point,
we denote by $a(G)$ the observed value of the property $\pi$ for a chemical compound $G$.
As the first phase, we solve (I) {\sc Prediction Problem} 
with the following three steps.
 \\
 
\noindent {\bf Phase~1.} \\
\smallskip\noindent 
{\bf Stage~1:}~ 
Let $\mathrm{DB}$ be a set of chemical graphs.
For a specified chemical property $\pi$, choose a class $\G$ of graphs 
such as acyclic graphs or monocyclic graphs.
Prepare a data set $D_{\pi}=\{G_i\mid i=1,2,\ldots,m\}\subseteq 
\G\cap \mathrm{DB}$ such that 
  the  value $a(G_i)$ of each chemical graph
$G_i$, $i=1,2,\ldots,m$ is available.
Set reals  $\underline{a}, \overline{a}\in \mathbb{R}$
so that $\underline{a}\leq  a(G_i)\leq \overline{a}$, $i=1,2,\ldots,m$.  

\smallskip\noindent 
{\bf Stage~2:}~ 
Introduce a feature function $f: \G\to \mathbb{R}^K$ for a positive integer $K$.
We call $f(G)$ the {\em feature vector} of $G\in \G$, and
 call each entry of  a vector $f(G)$  a {\em descriptor} of $G$.   

\smallskip\noindent 
{\bf Stage~3:}~ 
Construct a prediction function $\psi_\mathcal{N}$ 
with an ANN $\mathcal{N}$ that,  
given a   vector  in $\mathbb{R}^K$, 
returns a real in the range $[\underline{a},\overline{a}]$  
so that $\psi_\mathcal{N}(f(G))$ takes a value nearly equal to $a(G)$ 
for many chemical graphs  in  $D$. 
 See Figure~\ref{fig:framework}(a)  for an illustration 
 of  Stages~1 ,2 and 3 in Phase~1.

In this paper, we use the range-based method 
to define an applicability domain (AD)~\cite{Netzeva05}
 to our inverse  QSAR/QSPR.
Set $\underline{x_j}$ and $\overline{x_j}$ to be the minimum and maximum
values of the $j$-th descriptor $x_j$ in $f(G_i)$ over all graphs $G_i$, $i=1,2,\ldots,m$  
(where we possibly normalize some descriptors 
such as $\ce_\ta^\inn(G)$,
which is normalized with $\ce_\ta^\inn(G)/n(G)$). 
 Define  our AD $\mathcal{D}$ to be the set of vectors $x\in \mathbb{R}^K$  such that 
 $\underline{x_j}\leq x_j\leq \overline{x_j}$
 for the variable $x_j$ of each $j$-th descriptor, $j=1,2,\ldots,k$.

In the second phase,  we try to find a vector  $x^*\in \mathbb{R}^K$  
from a target value $y^*$ of  the chemical propery $\pi$ 
 such that $\psi_\mathcal{N}(x^*)=y^*$. 
Based on the  method due to Akutsu and Nagamochi~\cite{AN19},
  Chiewvanichakorn~et~al.~\cite{CWZSNA20} 
  showed that this problem can be formulated as an MILP. 
By including a set of linear constraints such that $x\in  \mathcal{D}$ 
into their MILP, we obtain the next result. 

\begin{theorem} \label{Th1}{\rm (\cite{IAWSNA20,ZCSNA20})}
Let   $\mathcal{N}$ be an ANN with  a piecewise-linear activation function
for an input vector $x\in \mathbb{R}^K$, 
 $n_A$ denote the number of   nodes in the architecture 
  and   $n_B$ denote the total number of break-points
over all  activation functions. 
Then there is an MILP $\mathcal{M}(x,y;\mathcal{C}_1)$  
that consists of variable vectors
$x\in \mathcal{D}~(\subseteq \mathbb{R}^K)$, 
$y\in \mathbb{R}$, 
and an auxiliary variable vector $z\in \mathbb{R}^p$ 
for some integer  $p=O(n_A+n_B)$
and a set $\mathcal{C}_1$ of $O(n_A+n_B)$ constraints on these variables 
such that:  $\psi_{\mathcal{N}}(x^*)=y^*$ if and only if
 there is a vector    $(x^*,y^*)$   feasible to  $\mathcal{M}(x,y;\mathcal{C}_1)$.
\end{theorem}
 
See Appendix~\ref{sec:AD} 
 for the set of constraints to define  our AD $\mathcal{D}$ 
in the MILP $\mathcal{M}(x,y;\mathcal{C}_1)$ in Theorem~\ref{Th1}.

 A vector $x\in \mathbb{R}^K$ is called {\em admissible}
 if   there is a graph $G\in \G$
 such that $f(G)=x$~\cite{ACZSNA20}.  
 Let $\mathcal{A}$ denote the set of admissible vectors $x\in \mathbb{R}^K$. 
 To ensure that a vector $x^*$ inferred from a given target value $y^*$
  becomes admissible, we introduce
    a new vector variable $g\in \mathbb{R}^{q}$ for an integer $q$.
For the class $\G$ of chemical acyclic graphs,
  Azam~et~al.~\cite{ACZSNA20}  introduced 
  a set $\mathcal{C}_2$ of new constraints 
  with    a new vector variable $g\in \mathbb{R}^{q}$ for an integer $q$ 
 so that a feasible solution $(x^*,g^*)$ of a new MILP
 for a target value $y^*$ delivers a vector $x^*$ with 
$\psi_{\mathcal{N}}(x^*)=y^*$ and
 a vector $g^*$ that represents  a  chemical acyclic graph $G^*\in \G$.
Afterwards, for the classes of chemical graphs with cycle index 1 and 2,
Ito~et~al.~\cite{ACZSNA20} and Zhu~et~al.~\cite{ZCSNA20}
presented such a set $\mathcal{C}_2$ of constraints 
so that a vector $g^*$ in a feasible solution $(x^*,g^*)$ of a new MILP
can represent  a chemical graph $G^*$ in the class $\G$, 
 respectively. 
 
As the second phase, we solve  (II) {\sc Inverse Problem} 
 for the inverse QSAR/QSPR
by treating the following inference problems.   

\smallskip
\noindent (II-a)   Inference of Vectors \\
{\bf Input:} A real $y^*$ with $\underline{a}\leq y^*\leq \overline{a}$. \\ 
{\bf Output:} Vectors $x^*\in  \mathcal{A}\cap   \mathcal{D}$ 
and $g^*\in  \mathbb{R}^{q}$ such that $\psi_\mathcal{N}(x^*)=y^*$
and $g^*$ forms a chemical graph $G^*\in \G$ with 
$f(G^*)=x^*$.

\bigskip  \noindent
 (II-b)  Inference of Graphs \\
{\bf Input:} A vector $x^*\in \mathcal{A}\cap \mathcal{D}$.    \\
{\bf Output:} All graphs $G^*\in \G$ such that
$f(G^*)=x^*$.    
\smallskip

The second phase consists of the next two steps.

\medskip \noindent {\bf Phase~2.}  \\
\smallskip\noindent 
{\bf Stage~4:}~  Formulate Problem (II-a)    
as the above MILP  $\mathcal{M}(x,y,g;\mathcal{C}_1,\mathcal{C}_2)$ 
based on $\G$ and $\mathcal{N}$. 
Find a feasible solution $(x^*,g^*)$ of the MILP 
such that \[\mbox{
  $x^*\in \mathcal{A}\cap \mathcal{D}$  and  $\psi_\mathcal{N}(x^*)=y^*$ }\]
(where the second requirement may be replaced with inequalities  
 $(1-\varepsilon)y^* \leq \psi_\mathcal{N}(x^*) \leq(1+\varepsilon)y^*$
 for a tolerance $\varepsilon>0$). 

\smallskip\noindent 
{\bf Stage~5:}~ To solve Problem (II-b),
enumerate all (or a specified number) of graphs $G^*\in \G$ 
such that $f(G^*)=x^*$ for the inferred vector $x^*$. 
%
See Figure~\ref{fig:framework}(b) for an illustration of  Stages~4 and 5 in Phase~2.

\subsection{Our Target Graph Class}
In this paper, we choose  a branch-parameter $k\geq 1$
and define a class  $\G$ of chemical acyclic graphs $G$
such that \\
- the maximum degree in $G$ is at most 4; \\
-  the $k$-branch height $\bh_k(G)$   
is bounded for a specified branch-parameter $k$; and \\
- the size of each $k$-fringe-tree in $G$ is bounded. 

The reason why we restrict ourselves to the graphs in $\G$  
is that this class $\G$ covers a large part of the acyclic chemical compounds
registered in the chemical database PubChem.
See Appendix~\ref{sec:statistical} for a summary of the statical feature of
the chemical graphs in PubChem in terms of $k$-branch height 
and the size of $2$-fringe-trees.
According to this, over 55\% (resp., 99\%) 
of  acyclic chemical compounds with up to
   100 non-hydrogen atoms in  PubChem
   have the maximum degree 3 (resp., 4); and
   nearly 87\% (resp., 99\%) of  acyclic chemical compounds with up to
   50 non-hydrogen atoms in  PubChem
   has the $2$-branch height at most 1 (resp., 2).
This implies that $k=2$ is sufficient to cover the most of
chemical acyclic graphs.
For $k=2$, 
  over 92\% of 2-fringe-trees of  chemical compounds with up to
   100 non-hydrogen atoms in  PubChem obey the following 
   size constraint: 
\begin{equation}\label{eq:fringe-size}
   \mbox{$n \le 2d + 2$ for each 2-fringe-tree  $T$ 
  with $n$ vertices and $d$ children of the root. }\end{equation}

We formulate an MILP in Stage~4 that, given a target value $y^*$,
 infers a vector $x^*\in \mathbb{Z}_+^K$ with 
 $\psi_\mathcal{N}(x^*)=y^*$   
 and  a chemical acyclic
graph $G^*=(H,\alpha,\beta)\in\G$ with  $f(G^*)=x^*$.
We here specify some of the features of a graph $G^*\in\G$ 
such as the number of non-hydrogen atoms in order to 
control the graph structure of target graphs to be inferred 
and to simplify  MILP formulations. 
In this paper, we specify the following features on a graph $G\in\G$:
a set  $\Lambda$ of chemical elements,
a set $\Gamma_{<}$ of adjacency-configuration, 
the maximum degree, the number of non-hydrogen atoms,
the diameter,  the $k$-branch-height and  
  the $k$-branch-leaf-number for a branch-parameter $k$. 

More formally, given specified integers 
 $n^*,\dmax, \dia^*, k^*, \bh^*,\bl^*\in \mathbb{Z}$
 other than $\Lambda$ and $\Gamma$,
let $\mathcal{H}(n^*, \dmax, \dia^*, k^*, \bh^*, \bl^*)$
denote the set of acyclic graphs  $H$ such that  \\
~~~  the maximum degree of a vertex is at most 3 when $\dmax=3$
  (or equal to 4 when  $\dmax=4$), \\
~~~  the number $n(H)$ of vertices in $H$ is $n^*$, \\
~~~ the diameter $\dia(H)$ of $H$ is $\dia^*$, \\
~~~ the $k^*$-branch-height $\bh_{k^*}(H)$ is $\bh^*$,  \\
~~~ the $k^*$-branch-leaf-number $\bl_{k^*}(H)$ is $\bl^*$ and   \\
~~~ (\ref{eq:fringe-size}) holds. 

To design Stage~4 for our class $\G$,
we formulate an MILP    $\mathcal{M}(x,g;  \mathcal{C}_2)$  that infers
a chemical graph $G^*=(H,\alpha,\beta)\in \G$ with  
$H\in \mathcal{H}(n^*, \dmax, \dia^*, k^*, \bh^*, \bl^*)$
for a given specification $(\Lambda,\Gamma,n^*, \dmax, \dia^*, k^*, \bh^*, \bl^*)$ 
 The details will be given  
in Section~\ref{sec:graph_MILP}  and  Appendix~\ref{sec:full_milp}.

\bigskip
Design of Stage~5; i.e.  generating chemical graphs $G^*$
that satisfy $f(G^*)=x^*$ for a given feature vector $x^*\in\mathbb{Z}_+^K$
 is still challenging for a relatively large instance with
  size $n(G^*)\geq 20$.
There have been proposed  algorithms for generating chemical graphs $G^*$
 in Stage~5  for the classes of graphs with cycle index 0 to 2
~\cite{Fujiwara08,Suzuki14,2A1B20,2A2B20}.
All of these are designed based on the branch-and-bound method
and 
 can generate a target chemical graph with size $n(G^*)\leq 20$.
To break this barrier, we newly employ the dynamic programming method
for designing an algorithm in Stage~5
in order to generate a target chemical graph $G^*$ with size $n(G^*)=50$.
For this, we further restrict the structure of acyclic graphs $G$
so that the number $\bl_2(G)$ of leaf $2$-branches is at most 3. 
Among  all acyclic chemical compounds with up to 50 non-hydrogen atoms
in the chemical database PubChem,
the ratio of the number of  acyclic chemical compounds $G$
with $\bl_2(G)\leq 2$  (resp., $\bl_2(G)\leq 3$)   is  78\%  (resp.,  95\%). 
See Section~\ref{sec:graph_search}  for the details on the new algorithm
 in Stage~5.

\section{MILPs for  Chemical Acyclic Graphs with Bounded Branch-height}
 \label{sec:graph_MILP} 

In this section, we formulate   an  MILP  $\mathcal{M}(x,g;\mathcal{C}_2)$ 
  to infer a chemical acyclic graph $G$ in the class $\G$   
for a given
 specification $(\Lambda,\Gamma,n^*, \dmax, \dia^*, k^*, \bh^*, \bl^*)$
  defined in the previous section.

\subsection{Scheme Graphs}
   
We introduce a directed graph  
with size $O(n^*\cdot (\dmax-1)^{\max\{\bh^*,k^*\}} 
                 + (\dmax-1)^{\bh^*+k^*})$, 
called a {\em scheme graph} $\mathrm{SG}$,
 so that an acyclic graph 
 $H\in \mathcal{H}(n^*, \dmax, \dia^*, k^*, \bh^*, \bl^*)$
  can be chosen from the scheme graph $\mathrm{SG}$.  
Let $t^*$, $s^*$ and $c^*$ be integers such that  
\[ t^*= n^* - (\bh^* -1) - (k^*+1)\bl^*,\]  
\[\mbox{
  $s^*=a(b^c-1)/(b-1)+1$ for
$a=\dmax$, $b=\dmax\!-\!1$ and $c=\bh^*$,}\]  
\[c^*=s^*-1. \]
Let a  scheme graph  
$\mathrm{SG}( \dmax, k^*, \bh^*, t^*)$
  consist  of 
a tree $T_B$, a path $P_{t^*}$, a set $\{S_s\mid  s\in[1,s^*]\}$ of trees,
a set $\{T_t\mid  t\in[1,t^*]\}$ of trees, and a set of directed edges between
$T_B$ and $P_{t^*}$ so that   an acyclic graph 
 $H\in \mathcal{H}(n^*, \dmax, \dia^*, k^*, \bh^*, \bl^*)$
  will be constructed in the following way:
\begin{enumerate}
\item[(i)] 
The $k^*$-branch-tree of $H$ will be chosen as a subtree of $T_B=(V_B,E_B)$; 
\item[(ii)] 
Each $k^*$-fringe-tree rooted at a vertex $u_s\in V(T_B)$  of $H$
will be chosen as a subtree of $S_s$; 
\item[(iii)] 
Each $k$-branch-path of $H$ (except for its end-vertices) will be chosen
as a subpath of  $P_{t^*}$ or as an edge in $T_B$; 
\item[(iv)] 
Each $k^*$-fringe-tree rooted at a vertex $v_t\in V(P_{t^*})$  of $H$
will be chosen as a subtree of $T_t$;   and
\item[(v)] 
An edge $(u,v)$ directed from $T_B$ to $P_{t^*}$ will be selected as 
an initial edge of a $k^*$-branch-path of $H$ and
an edge $(v,u)$ directed from $P_{t^*}$ to $T_B$   will be  selected as 
an ending edge of a $k^*$-branch-path of $H$.
\end{enumerate}

More formally  each component of  a  scheme graph  
$\mathrm{SG}( \dmax, k^*, \bh^*, t^*)$ is defined as follows.

\begin{figure}[ht!]
\begin{center}
\includegraphics[width=.80\columnwidth]{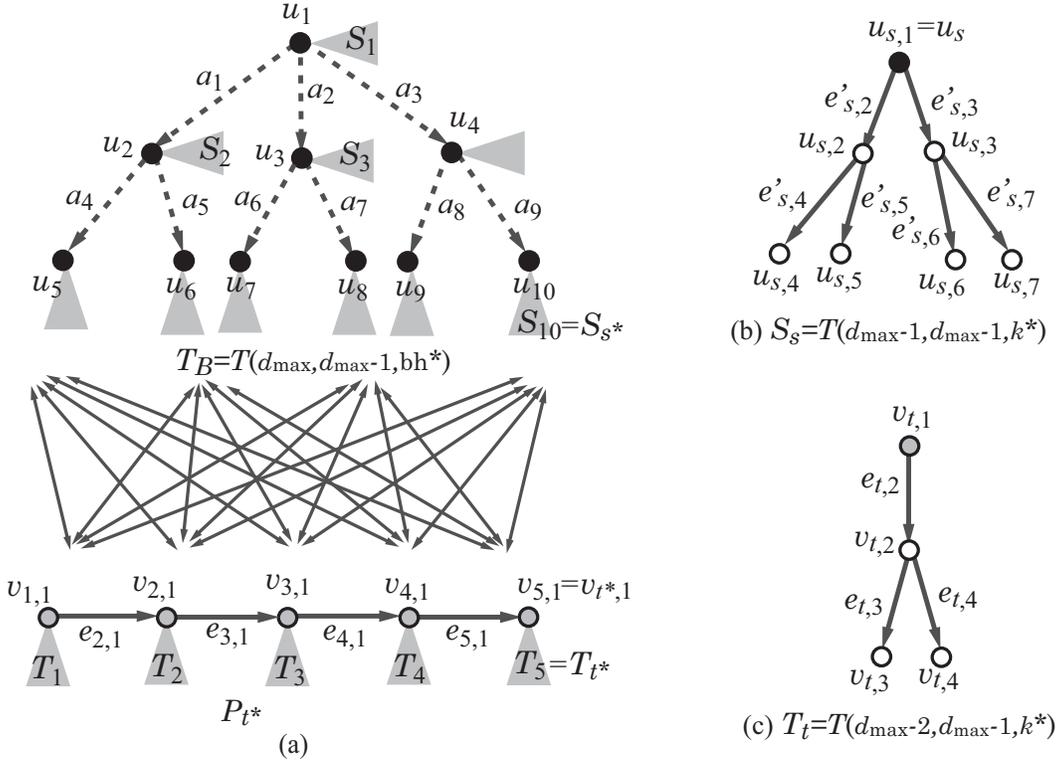}
\end{center}
\caption{An illustration of   scheme graph
$\mathrm{SG}( \dmax, k^*, \bh^*, t^*)$ 
with 
$\dmax=3$, $k^*=2$, $\bh^*=2$, and $t^*=5$,
where the vertices in $T_B$ (resp., in $P_{t^*}$) are depicted
with black (resp., gray) circles: 
(a) A base-tree $T_B$ and a link-path $P_{t^*}$ are joined
with directed edges between them;
(b) A tree $S_s$ rooted at a vertex $u_s=u_{s,1}\in V_B$;
(c) A tree $T_t$ rooted at a vertex $v_t=v_{t,1}\in V_P$. 
}
\label{fig:rank0_BH_scheme}  
\end{figure}

\begin{enumerate}
\item[(i)] 
$T_B =(V_B=\{u_1,u_2,\ldots,u_{s^*}\}, E_B=\{a_1,a_2\ldots,a_{c^*}\})$, called a {\em base-tree}
 is a tree rooted at a vertex $u_1$  that is isomorphic to the rooted tree
$T(\dmax, \dmax\!-\!1, \bh^*)$.
Regard $T_B$ as an ordered tree by introducing a total order for each set of siblings
and call the first (resp., last) child in a set of siblings the leftmost (resp. rightmost) child,
which defines the leftmost (rightmost) path from the root $u_1$ to a leaf in $T_B$, 
as illustrated in  Figure~\ref{fig:rank0_BH_scheme}(a). 

For each vertex $u_s\in V_B$, let 
$E_B(s)$ denote the set of indices $i$ of edges $a(i)\in E_B$ incident to $u_s$
and 
$\mathrm{Cld}_B(s)$ denote the set of
indices $i$ of children $u_i\in V_B$ of $u_s$ in the tree $T_B$. 

For each integer $d\in [0,k^*]$,
let $V_B(d)$ denote the set of indices $s$ of
vertices $u_s\in V_B$ whose depth is $d$ in the tree $T_B$,
where $V_B(\bh^*)$ is the 
set of indices $s$ of leaves $u_s$ of $T_B$. 

Regard each edge $a_i\in E_B$ as a directed edge $(u_s,u_{s'})$
from one end-vertex $u_s$ of $a_i$
to the other end-vertex $u_{s'}$ of $a_i$ such that $s=\mathrm{prt}(s')$
(i.e., $u_s$ is the parent of $u_{s'}$), where
 $\mathrm{head}(i)$ and $\mathrm{tail}(i)$ denote 
the head $u_{s'}$ and tail $u_s$ of  edge $a_i\in E_B$,
respectively.

  For each index $s\in [1,s^*]$, let  
$E_B^+(s)$ (resp., $E_B^-(s)$) denote 
the set of indices  $i$  of edges $a_i\in E_B$ such that 
the tail (resp., head) of $a_i$ is vertex $u_{s}$. 

 Let $L_B$ denote the set of indices of leaves  of $T_B$,
 and $s^\mathrm{left}$ (resp., $s^\mathrm{right}$)
 denote the index $s\in L_B$ of the leaf 
 $u_s$ at which the leftmost (resp., rightmost) path from the root ends. 

 For each leaf $u_s$, $s\in L_B$, 
 let $V_{B,s}$ (resp.,    $E_{B,s}$) denote the set of
indices $s$ of non-root vertices $u_s$ (resp.,   indices $i$ of edges $a(i)\in E_B$) 
 along the path from the root to the leaf $u_s$ in the tree $T_B$. 

For the example of a base-tree $T_B$ with $\bh^*=2$
 in Figure~\ref{fig:rank0_BH_scheme}, it holds that  
$L_B=\{5,6,7,8,9,10\}$, 
$s^\mathrm{left}=5$, $s^\mathrm{right}=10$, 
$E_{B,s^\mathrm{left}}=\{1,4\}$ and   $V_{B,s^\mathrm{left}}=\{2,5\}$.
 
\item[(ii)] 
  $S_s$, $s\in [1,s^*]$ is a tree rooted at  vertex $u_s\in V_B$ in $T_B$
  that is isomorphic to the rooted tree
$T(\dmax\!-\!1, \dmax\!-\!1, k^*)$,
   as illustrated in  Figure~\ref{fig:rank0_BH_scheme}(b).   
Let $u_{s,i}$ and $e'_{s,i}$ denote  the vertex and edge in $S_s$ that correspond to 
the $i$-th vertex and  the $i$-th edge in
$T(\dmax\!-\!1, \dmax\!-\!1, k^*)$, 
  respectively.  
Regard each edge  $e'_{s,i}$ as a directed edge $(u_{s,\mathrm{prt}(i)},u_{s,i})$. 
For this, each vertex $u_s\in V_B$ is also denoted by $u_{s,1}$.

\item[(iii)] 
  $P_{t^*}=(V_P=\{v_1,v_2,\ldots,v_{t^*}\},
    E_P=\{e_2,e_3,\ldots,e_{t^*}\})$, called a {\em link-path}
     with size $t^*$ is a directed path from vertex $v_1$ to vertex $v_{t^*}$,  
 as illustrated in  Figure~\ref{fig:rank0_BH_scheme}(a).  
Each edge $e_t\in E_P$ is directed from vertex $v_{t-1}$ to vertex $v_t$.
     
\item[(iv)]
 $T_t$, $t\in [1,t^*]$ is a tree rooted at  vertex $v_t$ in $P_{t^*}$ 
 that is  isomorphic to the rooted tree 
$T(\dmax\!-\!2, \dmax\!-\!1, k^*)$,
   as illustrated in  Figure~\ref{fig:rank0_BH_scheme}(c).  
Let $v_{t,i}$ and $e_{t,i}$ denote  the vertex and edge in $T_t$ that correspond to 
the $i$-th vertex and  the $i$-th edge in
$T(\dmax\!-\!2, \dmax\!-\!1, k^*)$,
  respectively. 
Regard each edge  $e_{t,i}$ as a directed edge $(v_{t,\mathrm{prt}(i)},u_{t,i})$. 
For this, each vertex $v_t\in V_P$ is also denoted by $v_{t,1}$. 
      
\item[(v)]  
For every pair $(s,t)$ with $s\in [1,s^*]$ and $t\in[1,t^*]$,
join vertices $u_{s}$ and $v_{t}$ with directed edges $(u_{s},v_{t})$
and $(v_{t},u_{s})$, 
   as illustrated in  Figure~\ref{fig:rank0_BH_scheme}(a). 
\end{enumerate} 

\begin{figure}[ht!]
\begin{center}
\includegraphics[width=.99\columnwidth]{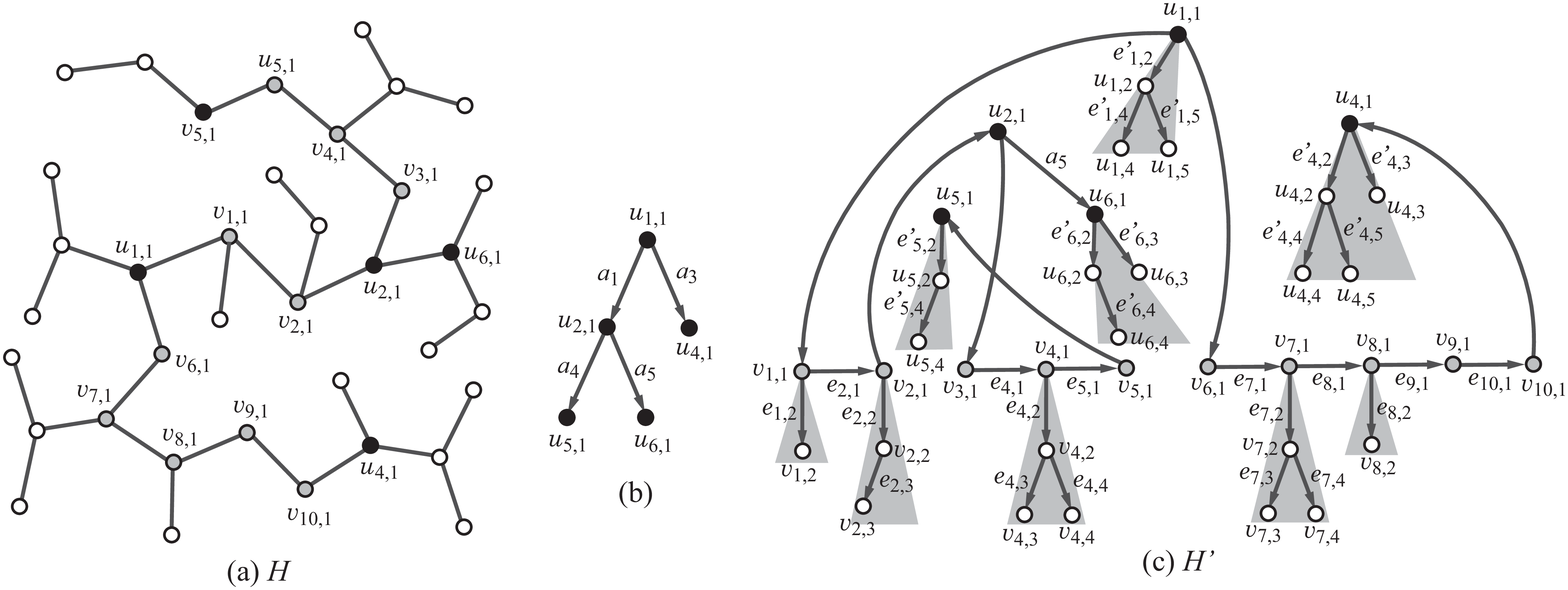}
\end{center}
\caption{An illustration of  selecting a subgraph $H$ from the scheme graph
$\mathrm{SG}( \dmax, k^*, \bh^*, t^*=n^*-\bl^*-1)$: 
(a) An acyclic graph  
 $H\in \mathcal{H}(n^*, \dmax, \dia^*, k^*, \bh^*, \bl^*)$ with
$n^*=37$, $\dmax=3$, $\dia^*(H)=17$,
$k^*=2$,  $\bh^*=2$ and $\bl^*=3$,
where  the labels of some vertices   indicate the corresponding vertices 
in  the scheme graph
$\mathrm{SG}( \dmax, k^*, \bh^*, t^*)$;  
(b) The $k^*$-branch-tree of $H$ for $k^*=2$;
(c) An acyclic graph $H'$ selected from  
$\mathrm{SG}( \dmax, k^*, \bh^*, t^*)$
as a graph that is isomorphic to $H$ in (a).
}
\label{fig:rank0_BH_scheme_example}  
\end{figure}

Figure~\ref{fig:rank0_BH_scheme_example}(a) illustrates 
  an acyclic graph $H$ with $n(H)=37$,  $\dia(H)=17$, 
  $\bh_2(H)=2$ and $\bl_2(H)=3$,
   where the maximum degree of a  vertex is 3. 
Figure~\ref{fig:rank0_BH_scheme_example}(b) illustrates
the $2$-branch-tree of the  acyclic graph $H$ in
Figure~\ref{fig:rank0_BH_scheme_example}(a).
Figure~\ref{fig:rank0_BH_scheme_example}(c)
 illustrates a subgraph $H'$ of the scheme graph
$\mathrm{SG}( \dmax, k^*, \bh^*, t^*=n^*-\bl^*-1)$  such that
$H'$ is isomorphic to the  acyclic graph $H$ in
Figure~\ref{fig:rank0_BH_scheme_example}(a).
    
In this paper, we obtain the following result. 
 
\begin{theorem} \label{Th2} 
Let $\Lambda$ be  a set  of chemical elements,
$\Gamma$  be a set  of adjacency-configurations,
where $|\Lambda|\leq |\Gamma|$,  and   $K=|\Lambda|+|\Gamma|+28$. 
 Given non-negative integers $n^*\geq 3$,  
 $\dmax\in\{3,4\}$, $\dia^*\geq 3$, 
 $k^*\geq 1$,  $\bh^*\geq 1$  and $\bl^*\geq 2$,
  there is an MILP $\mathcal{M}(x,g;\mathcal{C}_2)$  
that consists of variable vectors 
$x\in  \mathbb{R}^K$ and 
$g\in \mathbb{R}^q$ for an integer  
$q=O( |\Gamma|\cdot [
 (\dmax\!-\!1)^{\bh^*+k^*}
 +n^*\cdot(\dmax\!-\!1)^{\max\{\bh^*,k^*\}})])$
and a set $\mathcal{C}_2$ of 
$O(|\Gamma|
 + (\dmax\!-\!1)^{\bh^*+k^*}
 +n^*\cdot(\dmax\!-\!1)^{\max\{\bh^*,k^*\}}) )$
 constraints on $x$ and $g$ 
such that:  
   $(x^*,g^*)$ is feasible to  $\mathcal{M}(x,g;  \mathcal{C}_2)$
  if and only if  $g^*$ forms a chemical acyclic graph 
  $G=(H,\alpha,\beta)\in \mathcal{G}(\Lambda,\Gamma)$ 
  such that  
$H\in  \mathcal{H}(n^*, \dmax, \dia^*, k^*, \bh^*,\bl^*)$
and $f(G)=x^*$. 
\end{theorem}

 Note that  our MILP requires only  $O(n^*)$ variables and constraints 
 when the branch-parameter $k^*$, the  $k^*$-branch height 
  and $|\Gamma|$ are constant.   
 We formulate an MILP in Theorem~\ref{Th2} so that such 
 a graph $H$ is selected as a subgraph of the scheme graph. 

We explain the basic idea of our MILP.  
The MILP mainly consists of the following three types of constraints.
\begin{enumerate}
\item[C1.] 
Constraints for selecting an acyclic graph  $H$
 as a subgraph of the scheme graph
$\mathrm{SG}( \dmax, k^*, \bh^*, t^*)$; 

\item[C2.] 
Constraints for assigning chemical elements to vertices and multiplicity to edges
to determine a chemical graph $G=(H,\alpha,\beta)$;  and

\item[C3.] 
Constraints for computing descriptors
 from the selected acyclic chemical graph $G$.
\end{enumerate}

In the constraints of C1, more formally we prepare the following.

\begin{enumerate}
\item[(i)] 
In the scheme graph $\mathrm{SG}( \dmax, k^*, \bh^*, t^*)$,
we prepare 
a binary variable $u(s,1)$ for each vertex $u_s=u_{s,1}\in V_B$, $s\in [1,s^*]$
so that
 vertex $u_s=u_{s,1}$ becomes a $k^*$-branch of
a selected graph $H$ if and only if $u(s,1)=1$.
 The subgraph of the  base-tree $T_B$ that consists of
 vertices  $u_s=u_{s,1}$   with $u(s,1)=1$  will be the $k^*$-branch-tree
 of the graph $H$. 
 We also prepare 
 a binary variable  $a(i)$, $i\in [1,c^*]$ for each edge $a_i\in E_B$,
where $c^*=s^*-1$.
 For a pair of a vertex $u_{s,1}$ and a child $u_{s',1}$ of $u_{s,1}$
 such that $u(s,1)=u(s',1)=1$, 
 either the edge $a_i=(u_{s,1},u_{s',1})$ is used in
 the selected graph $H$ (when $a(i)=1$) or
 a path $P_i=(u_{s,1},v_{t',1},v_{t'+1,1},\ldots,v_{t'',1},u_{s',1})$
  from vertex $u_{s,1}$ to vertex $u_{s',1}$
 is constructed in $H$ with an edge $(u_{s,1},v_{t',1})$,
 a subpath $(v_{t',1},v_{t'+1,1},\ldots,v_{t'',1})$ of
 the link-path $P_{t^*}$ and an edge $(v_{t'',1},u_{s',1})$
  (when $a(i)=0$).  
  For example,
  vertices $u_{1,1}$ and $u_{2,1}$ are connected
  by a path  $P_1=(u_{1,1},v_{1,1},v_{2,1},u_{2,1})$
  in the selected graph $H'$ in  
Figure~\ref{fig:rank0_BH_scheme_example}(c).

\item[(ii)]
Let 
   \[ n_\mathrm{tree}^{\mathrm{S}} =1+(\dmax\!-\!1)((\dmax\!-\!1)^{k^*}-1)/(\dmax-2),\]
   \[ n_\mathrm{tree}^{\mathrm{T}} =1+(\dmax\!-\!2)((\dmax\!-\!1)^{k^*}-1)/(\dmax-2),\]  
where $n_\mathrm{tree}^{\mathrm{S}} $  
(resp.,  $n_\mathrm{tree}^{\mathrm{T}}$)  
is the numbers of vertices 
 in the rooted tree 
 $T(\dmax\!-\!1, \dmax\!-\!1, k^*)$
 (resp.,  $T(\dmax\!-\!2, \dmax\!-\!1, k^*)$).
 In each tree $S_s$, $s\in [1,s^*]$ (resp., $T_t$, $t\in [1,t^*]$)  in  the scheme graph,
we prepare 
 a binary variable $u(s,i)$ (resp., $v(t,i)$) 
 for each vertex $u_{s,i}$, $i\in [2,  n_\mathrm{tree}^{\mathrm{S}}]$
(resp., $v_{t,i}$, $i\in [2,  n_\mathrm{tree}^{\mathrm{T}}]$) 
so that $u(s,i)=1$ (resp., $v(t,i)=1$) means that
the corresponding vertex $u_{s,i}$ (resp., $v_{t,i}$) is
used as a vertex in a selected graph $H$. 
 The (non-empty) subgraph of a tree $S_s$ (resp.,  $T_t$) that consists of
 vertices  $u_{s,i}$  with $u(s,i)=1$ (resp., $v_{t,i}$ with $v(t,i)=1$) 
 will be a $k^*$-fringe-tree  of a selected graph $H$. 
 
 \item[(iii)] 
 In the  link-path $P_{t^*}$,
 we prepare a binary variable $e(t)$, $t\in [2,t^*]$ for each edge
$e_{t,1}=(v_{t-1,1}, v_{t,1})\in E_P$ 
so that $e(t)=1$ if and only if
edge $e_{t,1}$ is used in some path $P_i=(u_{s,1},v_{t',1},v_{t'+1,1},\ldots,v_{t'',1},u_{s',1})$ constructed in (i).  
 
 \item[(iv)] 
 For each pair $(s,t)$ of $s\in [1,s^*]$ and $t\in [1,t^*]$,
 we prepare a binary variable  $e(s,t)$ (resp., $e(t,s)$)  so that 
 $e(s,t')=1$ (resp., $e(t'',s)=1$) if and only if
 directed edge $(u_{s,1},v_{t',1})$ (resp., $(v_{t'',1},u_{s,1})$)
  is used as the first edge (resp., last edge)
  of  some path $P_i=(u_{s,1},v_{t',1},v_{t'+1,1},\ldots,$ $v_{t'',1},u_{s',1})$
   constructed in (i).  
\end{enumerate}

Based on these, we include constraints with some more additional variables
so that a selected subgraph $H$ is a connected acyclic graph. 
See constraints (\ref{eq:SG_first}) to (\ref{eq:SS_last})
 in Appendix~\ref{sec:full_milp}
 for the details.

In the constraints of C2,  we prepare an integer variable 
$\widetilde{\alpha}(u)$ for each vertex $u$ in the scheme graph
that represents the chemical element $\alpha(u)\in \Lambda$
if $u$ is in a selected graph $H$ (or $\widetilde{\alpha}(u)=0$ otherwise)
and 
an integer variable 
 $\widetilde{\beta}(e)\in [0,3]$ 
 (resp., $\widehat{\beta}(e)\in [0,3]$)  for each edge $e$
 (resp., $e=e(s,t)$ or $e(t,s)$, $s\in [1,s^*]$, $t\in [1,t^*]$)
  in the scheme graph
that represents the multiplicity $\beta(e)\in [1,3]$
if $e$ is in a selected graph $H$ (or $\widetilde{\beta}(e)$ or 
$\widehat{\beta}(e)$ takes $0$ otherwise). 
This determines a chemical graph $G=(H,\alpha,\beta)$. 
Also we include constraints for a selected chemical graph
$G$ to satisfy the valence condition
 $(\alpha(u),\alpha(v),\beta(uv))\in \Gamma$ for each edge $uv\in E$.
See constraints (\ref{eq:AM_first}) to (\ref{eq:AE_last})
 in Appendix~\ref{sec:full_milp} for the details.

In the constraints of C3,  we introduce a variable for each descriptor
and constraints with some more variables to compute the value of 
each descriptor in $f(G)$ for a selected chemical graph $G$.
See constraints (\ref{eq:NE_first}) to (\ref{eq:NBC_last}) 
in Appendix~\ref{sec:full_milp} for the details.

\section{A New Graph Search Algorithm}
\label{sec:graph_search}

  The algorithm used in Stage~5 in the previous methods of inferring
chemical acyclic graphs \cite{ACZSNA20,CWZSNA20,ZZCSNA20} 
are all based on the branch-and-bound algorithm
proposed by Fujiwara et~al.~\cite{Fujiwara08}
where an enormous number of chemical graphs are
constructed by repeatedly appending and removing a vertex one by one
until a target chemical graph is constructed.
Their algorithm  cannot generate even one acyclic chemical graph when $n(G)$
is larger than around 20.

 This section designs a new  dynamic programming method
for designing an algorithm in Stage~5. 
We consider  the following aspects: 
\begin{enumerate}
\item[(a)] Treat acyclic graphs with a certain limited structure  
that frequently appears among chemical compounds registered 
in the chemical data base; and  
\item[(b)]  Instead of manipulating acyclic graphs directly,
first compute the frequency vectors $\f(G')$
(some types of feature vectors)   of subtrees $G'$ of all target acyclic graphs
and then construct  
a limited number of target graphs $G$ from the process of computing the vectors.
\end{enumerate}

\begin{figure}[!ht] \begin{center}
\includegraphics[width=.98\columnwidth]{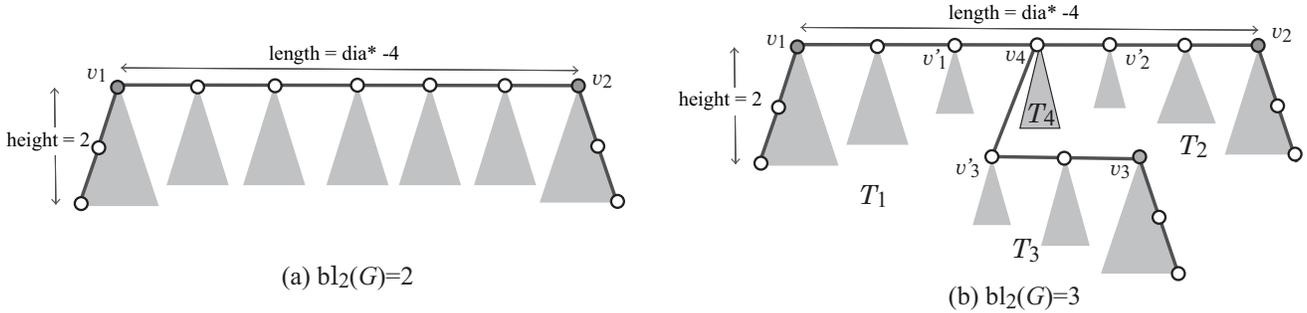}
\end{center}
\caption{An illustration of  chemical acyclic  graphs $G$ with diameter $\dia^*$
and $\bl_2(G)=2,3$: 
(a)~A chemical acyclic graph $G$ with two leaf 2-branches $v_1$ and $v_2$;
(b)~A chemical acyclic graph $G$ with  
 three leaf 2-branches $v_1, v_2$ and $v_3$. 
}
\label{fig:few_leaf_2-branches} 
\end{figure}

In (a), we choose a branch-parameter $k^*=2$
and treat acyclic graphs $G$ that have a small $2$-branch number such as 
  $\bl_2(G)\in [2,3]$. 
  and satisfy the  size constraint (\ref{eq:fringe-size}) on  2-fringe-trees.   
   Figure~\ref{fig:few_leaf_2-branches}(a) and (b) 
  illustrate  chemical acyclic graphs $G$ with 
 $\bl_2(G)=2$ and $\bl_2(G)= 3$, respectively.   

We design a method in (b)  based on the mechanism of dynamic programming 
wherein 
the first phase computes some compressed forms of all substructures of target 
objects   before
the second phase realizes a final object based on the computation process of
the first phase.

Section~\ref{sec:frequency_vector} defines a frequency vector $\f(G)$ 
that represents a feature vector $f(G)$ of a chemical graph $G$.
Section~\ref{sec:search_idea} presents
the idea and a sketch of our new algorithms
for generating  acyclic graphs $G$ with  $\bl_2(G)\in [2,3]$. 
Detailed descriptions of the algorithms are 
presented in Appendix~\ref{sec:graph_search_appendix}.
  
\subsection{Multi-rooted Trees and Frequency Vectors}\label{sec:frequency_vector}

For a finite set $A$ of elements, let
$\mathbb{Z}_+^A$ denote the set of functions
$\w:A\to \mathbb{Z}_+$.
A function $\w\in \mathbb{Z}_+^A$ is called
a {\em non-negative integer vector} (or a vector) on $A$
and the value $\x(a)$ for an element $a\in A$ is called
the {\em entry} of $\x$ for $a\in A$. 
For a vector $\w\in \mathbb{Z}_+^A$ and  an element $a\in A$,
let $\w+\1_{a}$ (resp., $\w-\1_{a}$)
denote the vector  $\w'$ 
such that  $\w'(a)=\w(a)+1$ (resp., $\w'(a)=\w(a)-1$)
 and  $\w'(b)=\w(b)$
 for the other elements   $b\in A\setminus\{a\}$.  
For a vector $\w\in \mathbb{Z}_+^A$ and a subset $B\subseteq A$,
let $\w_{[B]}$ denote the {\em projection} of $\w$ to $B$;
i.e., $\w_{[B]}\in \mathbb{Z}_+^B$ such that 
$\w_{[B]}(b)=\w(b)$, $b\in B$.
 
Let  $\Bc$ denote 
the set of  tuples $\mu=(d_1,d_2,k)\in  [1,4]\times[1,4]\times[1,3]$
(bond-configuration) such that  $\max\{d_1,d_2\}+k\leq 4$. 
We regard that $(d_1,d_2,k)=(d_2,d_1,k)$. 
For two tuples $\mu=(d_1,d_2,k), \mu'=(d'_1,d'_2,k')\in \Bc$,
we write $\mu\geq  \mu'$ if  
$\max\{d_1,d_2\}\geq \max\{d'_1,d'_2\}$, 
$\min\{d_1,d_2\}\geq \min\{d'_1,d'_2\}$ and    $k\geq k'$, 
and write $\mu> \mu'$ if $\mu\geq \mu'$ and $\mu\neq \mu'$. 
 Let $\Dg=\{\dg1, \dg2, \dg3, \dg4\}$, where $\dg i$ denotes
the number of vertices with degree~$i$.

Henceforth  we deal with vectors $\w$  
that have their $\w_\inn$ and $\w_\ex$ components,
both $\w_\inn, \w_\ex \in \mathbb{Z}_+^{\LtoD}$,
and for convenience we write $\w = (\w_\inn, \w_\ex)$
in the sense of concatenation.

For a vector $\x = (\x_\inn, \x_\ex)$ with 
 $\x_\inn, \x_\ex\in \mathbb{Z}_+^{\LtoD}$, let  $\G (\x)$ 
denote the set of  chemical acyclic graphs $G$
 that satisfy the following: \\
~~~~ $\ce_\ta^\inn(G)=\x_\inn(\ta)$ and  $\ce_\ta^\ex(G)=\x_\ex(\ta)$ 
for each chemical element  $\ta\in \Lambda$, \\ 
~~~~ $\ac_{\gamma}^\inn(G)= \x_\inn(\gamma)$ and   
$\ac_{\gamma}^\ex(G)=\x_\ex(\gamma)$  
for each adjacency-configuration $\gamma \in \Gamma$, \\
~~~~ $\bc_{\mu}^\inn(G)=\x_\inn(\mu)$ and   
 $\bc_{\mu}^\ex(G)=\x_\ex(\mu)$
for each bond-configuration   $\mu \in \Bc$, \\
~~~~ $\dg_i^\inn(G)=\x_\inn(\dg i)$ and   
$\dg_i^\ex(G)=\x_\ex(\dg i)$ for each degree $\dg i\in \Dg$. \\ 
   
Throughout the section, let $k^*=2$ be a branch-parameter, 
 $\x^*= (\x_\inn^*, \x_\ex^*)$ be a given feature vector with
 $\x_\inn^*, \x_\ex^*\in \mathbb{Z}_+^\LtoD$, 
and $\dia^*$ be an integer. 
We infer a chemical acyclic graph $G\in \G(\x^*)$
such that $\bl_2(G)\in [2,3]$ and the diameter of $G$ is $\dia^*$, 
 where $n^*=\sum_{\ta\in \Lambda}(\x^*_\inn(\ta)+\x^*_\ex(\ta) )$. 
Note that any other descriptors of $G\in \G(\x^*)$ can be determined
by the entries of vector $\x^*$. 

To infer a chemical acyclic graph $G\in \G(\x^*)$,
we consider a connected subgraph $T$ of $G$   that consists of 
\begin{equation}\begin{array}{l}\label{eq:subtree} 
\mbox{~~ - a subtree of the 2-branch-subtree $G'$ of $G$   and} \\
\mbox{~~ - the 2-fringe-trees rooted at vertices in $G'$. }
\end{array} \end{equation}
Our method first generates a set $\FT$ of all possible rooted trees $T$
that can be a 2-fringe-tree of a chemical graph $G\in \G(\x^*)$,
and then extends the trees $T$  by repeatedly appending a tree in $\FT$
until a chemical graph $G\in \G(\x^*)$ is formed.
In the  extension, we actually manipulate the ``frequency vectors'' of
trees defined below.  

To specify which part of a given tree $T$ plays a role of
2-internal vertices/edges or 2-external vertices/edges in
 a chemical graph $G\in \G(\x^*)$ to be inferred,
we designate at most three vertices $r_1(T)$, $r_2(T)$ and $r_3(T)$ in $T$
as  {\em terminals}, 
and call $T$ {\em rooted}
(resp.,  {\em bi-rooted} and {\em tri-rooted})
 if the number of terminals is one (resp., two and three).
For a rooted tree (resp., bi- or tri-rooted tree) $T$,
let $\widetilde{V}_{\inn}$ denote the set of 
vertices contained in a path between two terminals of $T$,
 $\widetilde{E}_{\inn}$ denote the set of edges in $T$
between two vertices in  $\widetilde{V}_{\inn}$,
and define 
$\widetilde{V}_{\ex}\triangleq V(T)\setminus \widetilde{V}_{\inn}$
and
$\widetilde{E}_{\ex}\triangleq E(T)\setminus \widetilde{E}_{\inn}$.
For a  bi- or tri-rooted tree  $T$,
define the {\em backbone path} $P_T$ of $T$
to be   the path of $T$ between vertices $r_1(T)$ and $r_2(T)$. 
 
Given a chemical acyclic graph $T$,
define $\f_{\tt t}(T)$, ${\tt t}\in\{\inn,\ex\}$ 
to be  the vector $\w\in \mathbb{Z}_+^\LtoD$
 that consists of the following entries:  
\begin{itemize}
\item[-] 
$\w(\ta)=|\{v\in \widetilde{V}_{\tt t} \mid \alpha(v)=\ta\}|$,
$\ta\in\Lambda$, 
\item[-] 
$\w(\gamma)=|\{uv\in \widetilde{E}_{\tt t} \mid 
 \{\alpha(u),\alpha(v)\}=\{{\tt a,b}\}, \beta(uv)=q\}|$,
$\gamma=({\tt a,b}, q)\in\Gamma$,
\item[-] 
$\w(\mu)=|\{uv\in \widetilde{E}_{\tt t}\mid 
 \{\deg_T(u),\deg_T(v)\}=\{ d,d'\}, \beta(uv)=m\}|$,
$\mu=(d,d', m)\in \Bc$,
\item[-] 
$\w(\dg i)=|\{v\in \widetilde{V}_{\tt t}\mid   \deg_T(v)=i\}|$,
$\dg i\in \Dg$.
\end{itemize}
Define $\f(T)\triangleq (\f_\inn(T),\f_\ex(T))$. 
The entry for an element ${\tt e}\in\LtoD$ in
 $\f_{\tt t}(T)$, ${\tt t}\in\{\inn,\ex\}$ is denoted by  $\f_{\tt t}({\tt e}; T)$.  
 For a subset $B$ of $\LtoD$,
 let $\f_{{\tt t}[B]}(T)$  denote the projection of $\f_{\tt t}(T)$ to $B$. 
 
Our aim is to generate all chemical bi-rooted (resp., tri-rooted) trees
$T$ with diameter $\dia^*$ such that $\f(T)=\x^*$.

\subsection{The Idea of New Algorithms}\label{sec:search_idea}
 
This section describes the idea and a sketch of our new graph search algorithms. 
 
\subsubsection{Case of $\bl_2(G)=2$}

We call a chemical graph $G\in \G(x^*)$ 
with diameter $\dia^*$ and $\bl_2(G)=2$ a {\em target graph}.

\begin{figure}[!ht] \begin{center}
\includegraphics[width=.95\columnwidth]{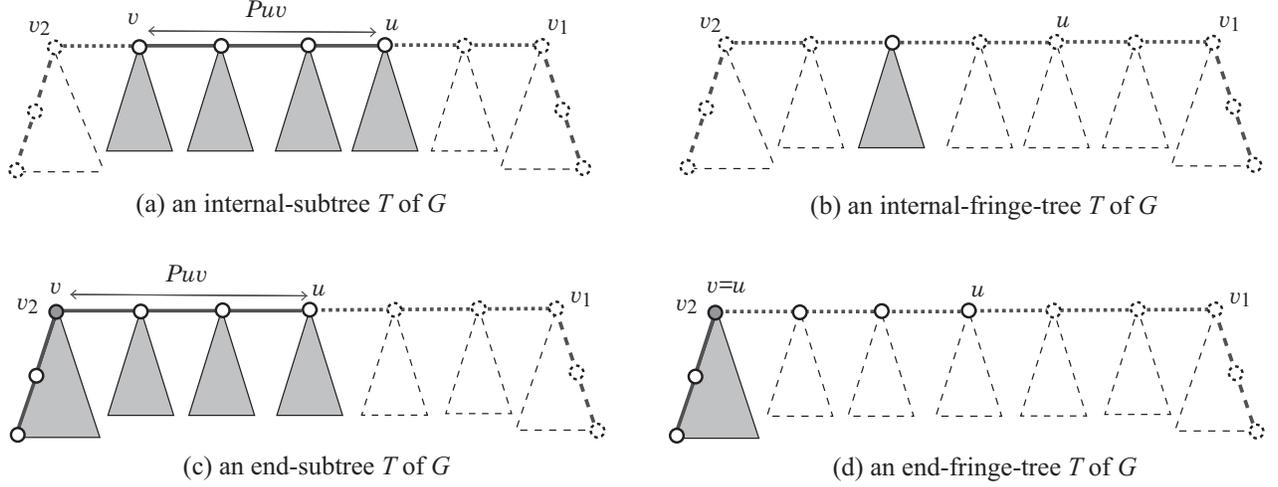}
\end{center}
\caption{An illustration of subtrees $T$ of a chemical acyclic graph $G$
  in Figure~\ref{fig:few_leaf_2-branches}(a), where the vertices/edges in $T$
   are depicted by solid lines:
(a)~An internal-subtree $T$ of $G$;  
(b)~An internal-fringe-tree $T$ of $G$;
(c)~An  end-subtree $T$ of $G$;  
(d)~An end-fringe-tree $T$ of $G$.
}
\label{fig:subtrees_two_2-branches} 
\end{figure}

A chemical acyclic graph $G$ with $\bl_2(G)=2$
has exactly two leaf 2-branches $v_i$, $i=1,2$,
where the length of the path between
the two leaf 2-branch $v_1$ and $v_2$  of a target graph $G$
is $\dia^*-2k^*=\dia^*-4$.
We observe that a connected subgraph $T$ of a target graph $G$ 
that satisfies (\ref{eq:subtree}) for   $\bl_2(G)=2$
is a chemical rooted or bi-rooted tree. 
We call such a subgraph $T$
an {\em internal-subtree} (resp., {\em end-subtree}) of $G$ 
 if  neither (resp., one) of $u$ and $v$  is  a 2-branch in $G$.
When $u=v$, 
we call an  internal-subtree  (resp.,  end-subtree) $T$ of $G$  
an  {\em internal-fringe-tree} (resp., {\em end-fringe-tree}) of  $G$.
 Figure~\ref{fig:subtrees_two_2-branches}(a)-(d) illustrate 
  an internal-subtree,  an internal-fringe-tree, 
  an end-subtree and an end-fringe-tree of $G$. 
 
\begin{figure}[!ht] \begin{center}
\includegraphics[width=.69\columnwidth]{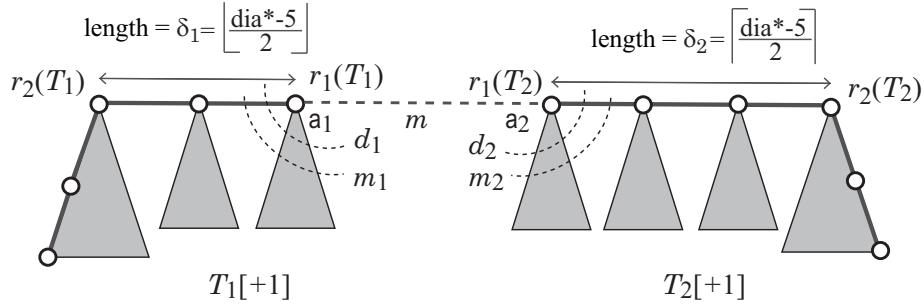}
\end{center}
\caption{An illustration of combining 
two bi-rooted trees $T_1=T_{\w^1}$ and $T_2=T_{\w^2}$
with a new edge with multiplicity $m$ joining vertices $r_1(T_1)$ and $r_1(T_2)$
to construct  a target graph $G$,
where  $\ta_i\in \Lambda$, $d_i\in [1,\dmax-1]$, $m_i\in[d_i,\val(\ta_i)-1]$, 
$i = 1,2$ and $m \in [1, \min\{3, \val(\ta_1) - m_1, \val(\ta_2) - m_2\} ]$.
}
\label{fig:combine_two_2-branches} 
\end{figure}

Let $\delta_1=\lfloor\frac{\dia^* - 5}{2} \rfloor$
  and $\delta_2 =\dia^*-5-\delta_1=\lceil \frac{\dia^* - 5}{2} \rceil$.
We regard a target graph $G\in \G(x^*)$ 
with $\bl_2(G)=2$ and diameter $\dia^*$
as a combination of two   chemical bi-rooted 
trees $T_1$ and $T_2$ with $\ell(P_{T_i})=\delta_i$, 
$i=1,2$ joined by an edge $e=r_1(T_1)r_1(T_2)$, 
as illustrated in Figure~\ref{fig:combine_two_2-branches}. 

We start with generating chemical rooted trees and then
iteratively extend chemical  bi-rooted trees $T$ with $\ell(P_T)=1,2,\dots,\delta_1$
before we finally combine two chemical  bi-rooted trees $T_1$ and $T_2$
with $\ell(P_{T_i})=\delta_i$.
To describe our algorithm, we introduce some notations. 
\begin{itemize}
\item[-]
Let $\T(x^*)$ denote the set of all bi-rooted trees $T$
(where possibly $r_1(T)=r_2(T)$)   such that 
    $\f_{\inn}(T)\leq \x_{\inn}^*$ and 
$\f_{\ex}(T)\leq \x_{\ex}^*$,
which is a necessary condition for $T$ to be an internal-subtree or end-subtree of 
a target graph $G\in \G(x^*)$.
\item[-]
Let $\mathcal{FT}$ denote the set of all rooted trees $T\in \T(x^*)$
that can be a 2-fringe-tree of  a target  graph $G$,
where $T$ satisfies the size constraint (\ref{eq:fringe-size}) of 2-fringe-trees. 
\item[-]
For each integer $h\in  [1,\dia^*-4]$,
  $\T_{\en}^{(h)}$ denote the set of all bi-rooted trees 
$T\in \T(x^*)$ that can be an end-subtree of a  target  graph $G$
such that $\ell(P_T)=h$,
and each 2-fringe-tree $T_v$ rooted at a vertex $v$ in $P_T$ 
belongs to $\mathcal{FT}$. 
\end{itemize}
 
We remark that the size $|\T_{\en}^{(h)}|$ of trees 
will be enormously large for $n^*\geq 25$ and $\dia^*\geq 10$.
This  suggests that construction of  a target graph
$G$ by enumerating trees in $\T_{\en}^{(h)}$ 
directly never works for such a large size of instances.
The idea of our new algorithm is to 
compute only the set $\W_{\en}^{(h)}$ of frequency vectors $\w$
of these trees, whose size $|\W_{\en}^{(h)}|$ is much more restricted than that of 
$\T_{\en}^{(h)}$.
We compute the set $\W^{(h)}$ of frequency vectors $\w$ of trees in 
$\T_{\en}^{(h)}$ iteratively for each integer $h\geq 0$.
During the computation,
we keep  a sample of a tree $T_{\w}$ for each of such
frequency vectors $\w$ so that a final step can construct some number of
 target graphs $G$ by assembling these sample trees.
Based on this, we generate target graphs $G\in\G(x^*)$ 
by the following steps:
\begin{enumerate}
\item[1.] 
(i) Compute $\mathcal{FT}$ by a branch-and-bound procedure 
that generates all possible rooted trees $T\in \T(\x^*)$
 (where $r_1(T)=r_2(T)$)
that can be a 2-fringe-tree of  a target graph $G\in \G(x^*)$; \\
(ii) Compute the set $\W^{(0)}$
 of all vectors $\w=(\w_\inn,\w_\ex)$
such that $\w_\inn=\f_{\inn}(T)$ and
$\w_\ex =\f_{\ex}(T)$
 for some tree $T\in \mathcal{FT}$; \\
 (iii) 
For each vector $\w=(\w_\inn,\w_\ex)\in \W^{(0)}$,
choose a sample tree $T_{\w}\in  \mathcal{FT}$
such that $\w_\inn=\f_{\inn}(T)$ and
$\w_\ex =\f_{\ex}(T)$,
and store these sample trees; 

\item[2.] For each integer $h=1,2,\ldots,  \delta_2$,
iteratively execute the next: \\
(i) 
Compute the set $\W_{\en}^{(h)}$
 of all vectors $\w=(\w_\inn,\w_\ex)$
such that $\w_\inn=\f_{\inn}(T)$ and
$\w_\ex =\f_{\ex}(T)$ 
for some bi-rooted tree $T\in \T_{\en}^{(h)}$,
where such a vector $\w$ is obtained from  a combination of
  vectors $\w'\in \W^{(0)}$ 
and   $\w''\in  \W_{\en}^{(h-1)}$; \\
(ii) For each vector $\w \in \W_{\en}^{(h)}$,  store 
a sample tree $T_{\w}$, which  is obtained from  a combination of
sample trees $T_{\w'}$ with $\w'\in \W^{(0)}$  
and $T_{\w''}$ with $\w''\in  \W_{\en}^{(h-1)}$; 

\item[3.]  
We call a  pair  of vectors $\w^1\in  \W_{\en}^{(\delta_1)}$
and $\w^2\in  \W_{\en}^{(\delta_2)}$ 
 {\em feasible} if  it admits a target graph $G\in \G(\x^*)$ such that
$\w_{\inn}^1+\w_{\inn}^2\leq \x^*_{\inn}$  and
$\w_{\ex}^1+\w_{\ex}^2\leq \x^*_{\ex}$.
Find the set $\W_\pair$ 
of all feasible pairs of vectors  $\w^1$ and $\w^2$;

\item[4.] 
For each feasible vector pair  $(\w^1,\w^2)\in \W_\pair$,
construct a corresponding target graph $G$ by combining 
the corresponding samples trees
$T_{\w^1}$ and  $T_{\w^2}$,
as illustrated in Figure~\ref{fig:combine_two_2-branches}.
\end{enumerate}
 
For a relatively large instance with $n^*\geq 40$ and $\dia^*\geq 20$, 
the number $|\W_\pair|$ 
of feasible vector pairs in Step~4 is still very large.
 In fact, the size $|\W_{\en}^{(h)}|$ of a vector set $\W_{\en}^{(h)}$
  to be computed in Step~2 
 can also be considerably large during an execution of the algorithm.
 For such a case, 
   we impose a time limitation on the running time 
   for computing $\W_{\en}^{(h)}$ 
 and a memory limitation on the number of vectors 
 stored in a vector set $\W_{\en}^{(h)}$.
With these limitations,  
 we can compute only a limited subset $\widehat{\W}_{\en}^{(h)}$ 
 of each vector set $\W_{\en}^{(h)}$  in Step~2.
 Even with such a subset $\widehat{\W}_{\en}^{(h)}$,
 we still can find a large size of a subset $\widehat{\W}_\pair$
  of $\W_\pair$ in Step~3.  
 
Our algorithm also delivers a lower bound on the number of
all target graphs $G\in\G(x^*)$ in the following way.
In Step~1, we also compute the number $t(\w)$ of trees $T\in \FT$
such that $\w=\f(T)$ for each $\w\in \W^{(0)}$.
In Step~2, when a vector $\w$ is constructed from
two vectors $\w'$ and $\w''$, we iteratively compute the number $t(\w)$
of trees $T$ such that $\w=\f(T)$  by 
$t(\w):=t(\w')\times t(\w'')$. 
In Step~3,  when a feasible vector pair $(\w^1,\w^2)\in \W_\pair$ is obtained,
we know that the number of the corresponding target graphs $G$ 
is $t(\w^1)\times t(\w^2)$. 
Possibly we compute a subset $\widehat{\W}_\pair$
  of $\W_\pair$ in Step~3.
Then $(1/2)\sum_{(\w^1,\w^2)\in\widehat{\W}_\pair}t(\w^1)\times t(\w^2)$
gives a lower bound on the number of target graphs $G\in\G(x^*)$,
where we divided by 2 since 
an axially symmetric target graph $G$ can correspond
to   two vector pairs in $\W_\pair$.
 
Detailed descriptions  of the five steps
in the above algorithm can be found in Appendix~\ref{sec:graph_search_appendix}.

\subsubsection{Case of $\bl_2(G)=3$}

We call a chemical graph $G\in \G(x^*)$ 
with diameter $\dia^*$ and $\bl_2(G)=3$ a {\em target graph}.
Let $n^*_{\inl}\triangleq  \sum_{\ta \in \Lambda} \x^*_{\inn}(\ta)$,
which is the number of 2-internal vertices in a target graph $G\in \G(\x^*)$.

A chemical acyclic graph $G$ with $\bl_2(G)=3$
has exactly three leaf 2-branches $v_i$, $i=1,2$
and exactly one 2-internal vertex $v_4$ adjacent to
three  2-internal vertices $v_i$, $i=1,2,3$, 
as illustrated in Figure~\ref{fig:few_leaf_2-branches}(b).
We call vertex $v_4$ the {\em joint-vertex} of $G$.
Without loss of generality assume that
the length of the path $P_{v_1,v_2}$  between $v_1$ and $v_2$
is $\dia^*-4$ and that
 the length of the path $P_{v_1,v'_1}$ is not smaller than that of 
$P_{v_2,v'_2}$. 

Analogously with the case of $\bl_2(G)=2$,
we define {\em internal-subtree} (resp., {\em end-subtree}, 
 {\em internal-fringe-tree} 
and   {\em end-fringe-tree}) of $G$ 
to be a connected subgraph $G'$ that satisfies (\ref{eq:subtree}). 
Observe that $G$ can be partitioned into three end-subtrees
$T_i$, $i=1,2,3$, the 2-fringe-tree $T_4$ rooted at the joint-vertex $v_4$
and three edges $v_iv_4$, $i=1,2,3$, 
where the backbone path $P_{T_i}$ connects leaf 2-branch $v_i$ and vertex $v'_i$.
In particular, we call the end-subtree of $G$ that
consists of $T_1$, $T_2$, $T_4$ and edges $v_iv_4$, $i=1,2$ 
the {\em main-subtree} of $G$, which 
consists of the  path $P_{v_1,v_2}$ and all the 2-fringe-trees
rooted at vertices in $P_{v_1,v_2}$.
We call $T_3$ the {\em co-subtree} of $G$.

Let $\delta_i$, $i=1,2,3$ denote
 the length of the backbone path  of $T_i$.
Note that  
\[  
 \delta_1+\delta_2+2=\dia^*-4 \mbox{ and }  
 \delta_1\geq \delta_2\geq \delta_3= n^*_{\inl} - \dia^* + 2,\]
 from which 
\[
\delta_2 \in [\delta_3, \lfloor \dia^*/2 \rfloor-3 ]  
\mbox{ and }  
\delta_1 \in [\lceil \dia^*/2\rceil -3,  \dia^* - 6 -\delta_3] .\]

We regard a target graph $G\in \G(x^*)$ 
with $\bl_2(G)=3$ and diameter $\dia^*$
as a combination of the main-subtree    and
the co-subtree  joined with an edge.
We represent  the co-subtree  as    a chemical bi-rooted tree $T$ 
with   $\ell(P_T)=\delta_3$. 
We represent  the main-subtree of a target graph $G$ as 
a  tri-rooted tree $T$ with $\ell(P_T)=\dia-4$ so that 
terminals $r_1(T)$, $r_2(T)$ and $r_3(T)$ correspond
to the two leaf 2-branches and the joint-vertex of $G$, respectively.

\begin{figure}[!ht] \begin{center}
\includegraphics[width=.60\columnwidth]{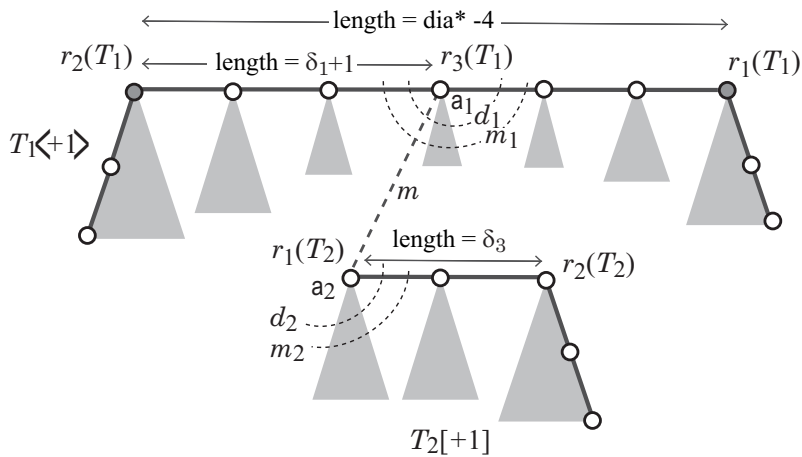}
\end{center}
\caption{An illustration of combining   
a tri-rooted $T_1=T_{\w^1}$ and a bi-rooted tree $T_2=T_{\w^2}$
with a new edge joining vertices $r_1(T_1)$ and $r_1(T_2)$
to construct  a target graph $G$.
}
\label{fig:combine_three_2-branches} 
\end{figure} 

We start with generating chemical rooted trees and then
iteratively extend chemical  bi-rooted trees $T$ 
with $\ell(P_T)=1,2,\dots,\dia^*-6-\delta_3$
before we  combine two chemical  bi-rooted trees $T'_1$ and $T'_2$
to obtain a  chemical  tri-rooted tree $T_1$ 
with $\ell(P_{T_1})=\delta_i$
and finally combine a  chemical  tri-rooted tree $T_1$ 
and a  chemical  bi-rooted trees $T_2$ with  $\ell(P_{T_2})=\delta_3$,
to obtain a target graph $G\in \G(x^*)$. 

Analogously with the case of $\bl_2(G)=2$, we define
  the set $\T(x^*)$ of all bi-rooted trees $T$,
 the set  $\mathcal{FT}$ of all rooted trees $T\in \T(x^*)$
that can be a 2-fringe-tree of  a target  graph $G$ 
and the set  $\T_{\en}^{(h)}$, $h\in [1, \dia^* -6 -\delta_3]$) 
 of all bi-rooted trees $T\in \T(x^*)$ that can be an end-subtree
 of a  target  graph $G$
such that $\ell(P_T)=h$.

We generate target graphs $G\in\G(x^*)$ 
by the following steps:
\begin{enumerate}
\item[1.] Analogously with Step~1 for the case of  $\bl_2(G)=2$,
compute the set $\mathcal{FT}$ and the set $\W^{(0)}$
 of all vectors $\w=(\w_\inn,\w_\ex)$
such that $\w_\inn=\f_{\inn}(T)$ and
$\w_\ex =\f_{\ex}(T)$
 for some tree $T\in \mathcal{FT}$.
 For each vector  $\w \in \W^{(0)}$,
 store a sample  tree  $T_{\w}\in  \mathcal{FT}$;
 
\item[2.]  For each integer $h=1,2,\ldots,  \dia^* -6 -\delta_3$, 
compute the set $\W_{\en}^{(h)}$  
 of all vectors $\w=(\w_\inn,\w_\ex)$
such that $\w_\inn=\f_{\inn}(T)$ and
$\w_\ex =\f_{\ex}(T)$ 
for some bi-rooted tree $T\in \T_{\en}^{(h)}$;
For each vector $\w \in \W_{\en}^{(h)}$, 
store a sample tree $T_{\w}$;

\item[3.]  
For each integer 
$h\in  [\lceil \dia^*/2\rceil -2,  \dia^* - 5 -\delta_3] $,
compute the set  $\W_{\en+2}^{(h)}$ 
of all vectors $\w=(\w_\inn,\w_\ex)$
such that $\w_\inn=\f_{\inn}(T)$ and
$\w_\ex =\f_{\ex}(T)$  of some bi-rooted tree 
$T$ with $\ell(P_T)=h$ 
that represents an end-subtree rooted at the joint-vertex; 
For each vector $\w \in \W_{\en+2}^{(h)}$, 
store a sample tree $T_{\w}$;
 
\item[4.]  
For each integer $\delta_1 \in[\lceil \dia^*/2\rceil -3,  \dia^* - 6 -\delta_3]$,
compute the set $\W_{\mathrm{main}}^{(\delta_1+1)}$  
 of all vectors $\w=(\w_\inn,\w_\ex)$
such that $\w_\inn=\f_{\inn}(T)$ and
$\w_\ex =\f_{\ex}(T)$ 
for some tri-rooted tree $T$ that represents 
the main-subtree 
such that the length of the path $P_{r_2(T),r_3(T)}$ between
terminals $r_2(T)$ and $r_3(T)$ is $\delta_1+1$.
For each vector $\w\in \W_{\mathrm{main}}^{(\delta_1+1)}$, 
store a sample tree $T_{\w}$;

\item[5.] 
We call a  pair  of vectors $\w^1\in  \W_{\mathrm{main}}^{(\delta_1+1)}$
and $\w^2\in  \W_{\en}^{(\delta_3)}$ 
 {\em feasible} if  it admits a target graph $G\in \G(\x^*)$ such that
$\w_{\inn}^1+\w_{\inn}^2\leq \x^*_{\inn}$  and
$\w_{\ex}^1+\w_{\ex}^2\leq \x^*_{\ex}$.
Find the set $\W_\pair$ 
of all feasible pairs of vectors  $\w^1$ and $\w^2$;

\item[6.] 
For each feasible vector pair  $(\w^1,\w^2)\in \W_\pair$,
construct a corresponding target graph $G$ by combining the samples trees
$T_{\w^1}$  and $T_{\w^2}$, which correspond to the main-subtree
and the co-subtree of a target graph $G$, respectively, 
as illustrated in Figure~~\ref{fig:combine_three_2-branches}.
\end{enumerate} 

Detailed descriptions   of the six steps
in the above algorithm can be found in Appendix~\ref{sec:graph_search_appendix}.

\section{Experimental Results}\label{sec:experiment}

We implemented our method of Stages~1 to 5 
for inferring chemical acyclic graphs and
conducted experiments  to evaluate the computational efficiency 
for three chemical properties $\pi$: 
octanol/water partition coefficient  ({\sc K\tiny{ow}}),
boiling point ({\sc Bp})
 and heat of combustion ({\sc Hc}). 
We executed the experiments on a PC with 
     Two Intel Xeon CPUs E5-1660 v3 @3.00GHz, 
     32 GB of RAM 
     running under   OS: Ubuntu 14.04.6 LTS. 
We show 2D drawings of some of the inferred chemical graphs, where  
  ChemDoodle  version 10.2.0 is used for constructing the drawings.

\begin{table}[ht!]\caption{Results of Stage 1 in Phase 1.}
 \begin{center}
 \begin{tabular}{@{} c r c c c c c c @{}}\toprule
  $\pi$ & $\Lambda$ &  $|D_{\pi}|$  & $|\Gamma|$ & 
  $[\underline{n},\overline{n}]$ & 
  $[\underline{\rm bl},\overline{\rm bl}]$ & 
  $[\underline{\rm bh},\overline{\rm bh}]$ & 
  $[\underline{a},\overline{a}]$ \\ \midrule
 {\sc K\tiny{ow}} & {\tt C,O,N} & 216  & 10 & [4, 28] & [0, 2] & [0, 4] & [-4.2, 8.23] \\
 {\sc Bp} & {\tt C,O,N}  & 172 &  10 & [4, 26] & [0, 1] & [0, 3] & [-11.7, 404.84] \\
 {\sc Hc} & {\tt C,O,N}  & 128 &  6 & [4, 26] & [0, 1] & [0, 2] & [1346.4, 13304.5] \\ \bottomrule
 \end{tabular}\end{center}\label{table:stage1}
\end{table}
\bigskip \noindent
{\bf Results on Phase~1.  }
We implemented Stages~1, 2 and 3 in Phase~1 as follows.

\bigskip \noindent
{\bf Stage~1.  }
We set  a graph class   $ \mathcal{G}$   to be
the set of all chemical acyclic graphs,
and set a branch-parameter $k^*$ to be 2. 
For each property 
$\pi\in \{${\sc K\tiny{ow}}, {\sc Bp, Hc}$\}$, 
 we first select a set $\Lambda$ of chemical elements 
 and then collected a  data set  $D_{\pi}$ on chemical acyclic graphs
 over the set $\Lambda$ of chemical elements  provided by HSDB from PubChem. 
To construct the data set,
we eliminated chemical compounds that have  at  most three carbon atoms
  or contain  a  charged element such as ${\tt N}^+$
 or an element ${\tt a}\in \Lambda$ whose valence is different from
 our setting of valence function $\val$.
 
 Table~\ref{table:stage1} shows the size and range of data sets that 
 we prepared for each chemical property in Stage~1,
 where  we denote the following:
 \begin{itemize}
\item[-] $\pi$: one of the chemical properties  
 {\sc K\tiny{ow}},  {\sc Bp} and   {\sc Hc};
 \item[-] $\Lambda$: the set of selected chemical elements 
   (hydrogen atoms are added at the final stage);
 \item[-] $|D_{\pi}|$:  the size of data set $D_{\pi}$ over $\Lambda$
  for  property $\pi$;
\item[-] $|\Gamma|$: the number of different adjacency-configurations
over the   compounds in $D_{\pi}$; 
\item[-] $[\underline{n},\overline{n}]$:  the minimum and maximum number $n(G)$
  of non-hydrogen  atoms over the  compounds $G$ in $D_{\pi}$; 
\item[-]  $[\underline{\bl},\overline{\bl}]$:
  the minimum and maximum numbers $\bl_2(G)$ of leaf 2-branches 
 over the  compounds $G$ in $D_{\pi}$; 
\item[-] $[\underline{\bh},\overline{\bh}]$: 
 the minimum and maximum values of the 2-branch height $\bh_2(G)$ 
 over the  compounds $G$ in $D_{\pi}$;  and 
 \item[-]
$[\underline{a},\overline{a}]$:  the minimum and maximum values
of $a(G)$ in $\pi$ over  compounds $G$  in $D_{\pi}$.    
 \end{itemize}

\bigskip \noindent
{\bf Stage~2.  }
We used a feature function $f$ that consists of the descriptors
 defined in Section~\ref{sec:preliminary}.  
 
\begin{table}[ht!]\caption{Results of Stages~2 and 3 in Phase~1.}
 \begin{center}
 \begin{tabular}{@{} c r c c c c c @{}}\toprule
  $\pi$ & $K$ & Activation & Architecture & L-Time & test R$^2$ (ave.) & test R$^2$ (best) \\ \midrule
  {\sc K\tiny{ow}} & 76 &  ReLU & (76,10,1) & ~~2.12 & 0.901 & 0.951 \\
  {\sc Bp} & 76 & ReLU & (76,10,1) & ~26.07 & 0.935 & 0.965 \\
  {\sc Hc} & 68 & ReLU & (68,10,1) & 234.06 & 0.924 & 0.988 \\ \bottomrule
 \end{tabular}\end{center}\label{table:stages2-3}
\end{table}

\bigskip \noindent
{\bf Stage~3.  }
We used {\tt scikit-learn} version 0.21.6  with Python 3.7.4
to construct ANNs $\mathcal{N}$
where the tool and activation function are set to
be MLPRegressor and ReLU, respectively.
We tested several different architectures of ANNs for each chemical property.
To evaluate the performance of the resulting prediction function
 $\psi_\mathcal{N}$ with cross-validation,
we partition a given data set $D_{\pi}$ into five subsets $D_{\pi}^{(i)}$, $i\in[1,5]$ randomly,
where $D_{\pi}\setminus D_{\pi}^{(i)}$ is used for a training set and
 $D_{\pi}^{(i)}$ is used for a test set in five trials $i\in[1,5]$.  
For  a set $\{y_1,y_2,\ldots,y_N\}$ of observed values and
a set $\{\psi_1,\psi_2,\ldots,\psi_N\}$ of predicted values, 
we define the 
coefficient of determination to be 
$\mathrm{R}^2\triangleq 
1- \frac{\sum_{j\in [1,N]}(y_j-\psi_j)^2} {\sum_{j\in [1,N]}(y_j-\overline{y})^2}$,
where  
$\overline{y}= \frac{1}{N}\sum_{j\in [1,N]}y_j$.  
 Table~\ref{table:stages2-3} shows the results on Stages~2 and 3, where
 \begin{itemize}
\item[-] $K$:  the number of descriptors for the chemical compounds
 in  data set $D_{\pi}$ for  property $\pi$;
\item[-]
 Activation: the choice of activation function;
\item[-]
 Architecture: $(a,b,1)$ consists of an input layer with $a$ nodes, 
 a hidden layer with $b$ nodes
and an output layer with a single node, where $a$ is
 equal to the number $K$ of descriptors;
\item[-]
  L-time: the average time (sec) to construct ANNs  for each trial; 
\item[-] test $\mathrm{R}^2$ (ave): 
the average of coefficient of determination over the five tests; and
\item[-]  test $\mathrm{R}^2$ (best): the largest value 
of coefficient of determination over the five test sets.
 \end{itemize}
 
 From Table~\ref{table:stages2-3}, we see that 
 the execution of Stage~3 was successful, where  
   the average of test $\mathrm{R}^2$ is over  0.9 for all three 
   chemical properties.  
   
For each chemical property $\pi$,
we selected the ANN $\mathcal{N}$ that attained the best  test $\mathrm{R}^2$
score among the five ANNs to formulate 
an MILP $\mathcal{M}(x,y,z;\mathcal{C}_1)$ which will be used in Phase~2.


\bigskip \noindent
{\bf Results on Phase~2.  }
We implemented Stages~4 and 5 in Phase~2 as follows.

\bigskip \noindent
{\bf Stage~4.  }
In this step, we solve
the  MILP  $\mathcal{M}(x,y,g;\mathcal{C}_1,\mathcal{C}_2)$
formulated based on the ANN $\mathcal{N}$ obtained in Phase~1. 
To solve an MILP   in Stage~4, we use
{\tt  CPLEX} version 12.8.
In our experiment,  we choose a target value $y^* \in [\underline{a}, \overline{a}]$.
and fix or bound some descriptors in our feature vector as follows:
\begin{itemize} 
\item[-] Set the 2-leaf-branch number $\bl^*$ to be each of $2$ and $3$; 
\item[-] Fix the instance size $n^*=n(G)$ 
to be each integer in $\{26,32,38,44,50\}$; 
\item[-] 
Set the diameter $\dia^*=\dia(G)$ be one of the integers
in $\{ \lceil (2/5)n^*\rceil,    \lceil (3/5)n^*\rceil \}$.  
\item[-] Set the maximum degree $\dmax:=3$ for $\dia^*=\lceil (2/5)n^*\rceil$ and
       $\dmax:=4$ for $\dia^*= \lceil (3/5)n^*\rceil $;
\item[-] For each instance size $n^*$, 
 test a target value $y^*_{\pi}$ for each chemical property  
$\pi\in \{${\sc K\tiny{ow}}, {\sc Bp, Hc}$\}$. 
\end{itemize}
Based on the above setting, we generated six instances for each instance size $n^*$.
We set $\varepsilon=0.02$ in Stage~4.

 Tables~\ref{table_2_3_2-5} to  \ref{table_2_4_3-5}
 (resp.,  Tables~\ref{table_3_3_2-5} to  \ref{table_3_4_3-5})
  show  the results on Stage~4  for $\bl^*=2$ (resp., $\bl^*=3$), 
 where we denote the following:
 \begin{itemize}
 \item[-]
  $y^*_{\pi}$: a target value in $[\underline{a},\overline{a}]$ for a property $\pi$;
  \item[-]
  $n^*$: a specified number of vertices  in $[\underline{n},\overline{n}]$;
  \item[-]
  $\dia^*$: a specified diameter in
   $\{ \lceil (2/5)n^*\rceil,   \lceil (3/5)n^*\rceil \}$;
 \item[-]  
 IP-time: the   time (sec.) to an  MILP  instance 
  to find   vectors $x^*$ and $g^*$.
 \end{itemize}
 
Observe that most of the MILP instances with
 $\bl^*=2$, $n^*\leq 50$ and $\dia^*\leq 30$
(resp.,  $\bl^*=3$, $n^*\leq 50$ and $\dia^*\leq 30$) 
in one minute (resp., in a few minutes).
The previously most efficient MILP formulation
for inferring chemical acyclic graphs due to 
Zhang~et~al.~\cite{ZZCSNA20} could solve
an instance with  only up to $n^*=20$ 
for the case of $\dmax=4$ and $\dia^*=9$.
Our new MILP formulation on chemical acyclic graphs 
with bounded 2-branch height
considerably improved the tractable size of chemical acyclic graphs
in Stage~4 for the inference problem (II-a). 

\begin{figure}[!htb]
\begin{center} 
\includegraphics[width=.88\columnwidth]{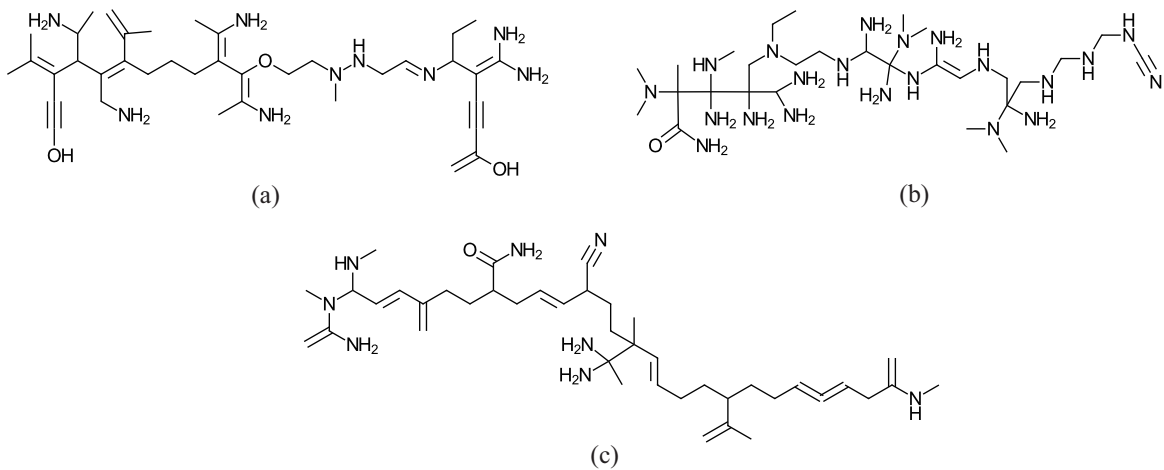}
\end{center}
\caption{An illustration of   chemical acyclic graphs $G$
with $n(G)=50$,  $\bl_2(G)=2$ and $\dmax=4$
 obtained in Stage~4 by solving an MILP: 
(a)  $y^*_{{\rm Kow}}=9$,   $\dia(G)=\lceil (2/5)n^*\rceil =20$; 
(b) $y^*_{{\rm Bp}}=880$,  $\dia(G)= n^*/2 =25$; 
(c)  $y^*_{{\rm Hc}}=25000$,  $\dia(G)=\lceil (3/5)n^*\rceil =30$.
}
\label{fig:inferred-2}  
\end{figure}

 Figure~\ref{fig:inferred-2}(a)-(c) illustrate some chemical acyclic graphs  $G$
 with $\bl_2(G)=2$ obtained in Stage~4 by solving an MILP.  
 Remember that these chemical graphs obey the AD $\mathcal{D}$
 defined in Appendix~A. 
 
\begin{figure}[!htb]
\begin{center} 
\includegraphics[width=.88\columnwidth]{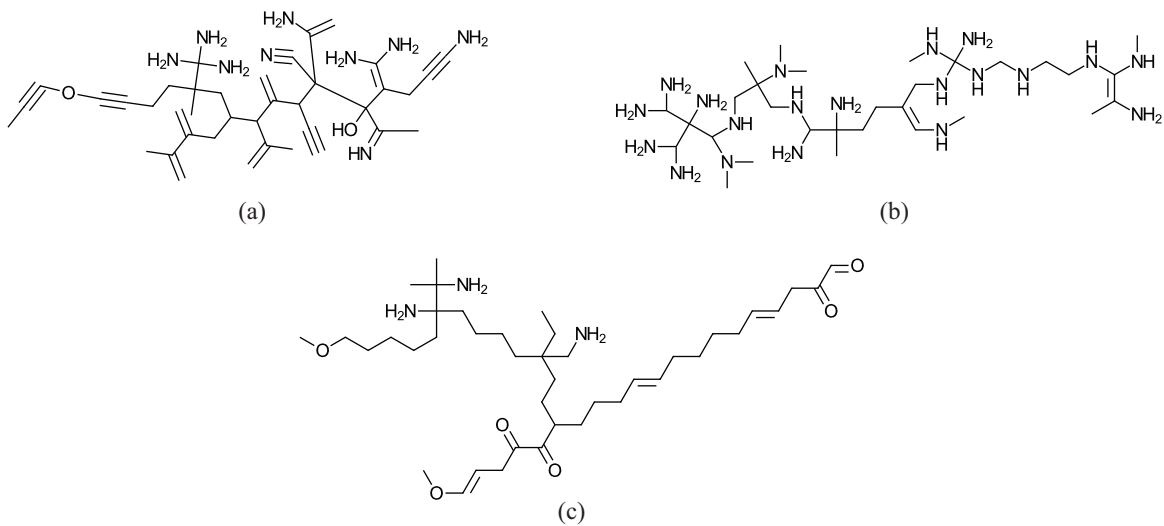}
\end{center}
\caption{An illustration of  chemical acyclic graphs $G$
with $n(G)=50$,  $\bl_2(G)=3$ and $\dmax=4$
 obtained in Stage~4 by solving an MILP: 
(a)  $y^*_{{\rm Kow}}=9$,   $\dia(G)=\lceil (2/5)n^*\rceil =20$; 
(b) $y^*_{{\rm Bp}}=880$,   $\dia(G)= n^*/2 =25$; 
(c)  $y^*_{{\rm Hc}}=25000$,   $\dia(G)=\lceil (3/5)n^*\rceil =30$.
}
\label{fig:inferred-3}  
\end{figure}

 Figure~\ref{fig:inferred-3}(a)-(c) illustrate some  chemical acyclic graphs  $G$
 with $\bl_2(G)=3$ obtained in Stage~4 by solving an MILP.  
 
\bigskip \noindent
{\bf Stage~5.  } 
In this stage, we execute our new graph search algorithms
for generating target graphs $G\in \G(\x^*)$ with $\bl_2(G)\in \{2,3\}$
for a given feature vector $\x^*$ obtained in Stage~4.

We introduce a time limit of 10 minute for 
each iteration $h$ in Step~2
and an execution of Steps~1 and 3    for $\bl^*=2$
(resp., each iteration $h$ in Steps~2 and 3  and $\delta_1$ in Step~4
and an execution of Steps~1  and 5  for $\bl^*=3$).
In the last step, we choose at most 100 feasible vector pairs
and generate a target graph from each of these feasible vector pairs.  
We also impose  an upper bound $\mathrm{UB}$ on the size $|\W|$ of a vector set $\W$ 
that we maintain during an execution of the algorithm.
We executed the algorithm for each of the three bounds $\mathrm{UB}=10^6, 10^7, 10^8$
until a feasible vector pair is found or the running time exceeds a global time limitation of
two hours.  

When no feasible vector pair is found by the graph search algorithms,
we output the target graph $G^*$ constructed from the vector $g^*$
in Stage~4.

 Tables~\ref{table_2_3_2-5} to  \ref{table_2_4_3-5}
 (resp.,  Tables~\ref{table_3_3_2-5} to  \ref{table_3_4_3-5})
  show  the results on Stage~5  for $\bl^*=2$ (resp., $\bl^*=3$), 
 where we denote the following:
  \begin{itemize}
 \item[-]  
 $\#$FP: the number of feasible vector pairs obtained
 by an execution of graph search algorithm 
 for a given  feature vector $\x^*$;
 \item[-]  
 G-LB: a lower bound on the number of all target graphs $G\in \G(\x^*)$
 for a given  feature vector $\x^*$;  
 \item[-]   
 $\#$G: the number of all (or up to 100) chemical acyclic graphs $G$
such that $f(G)=x^*$   (where at least one such graph  $G$ 
has been found from the vector $g^*$ in Stage~4);
 \item[-]  
 G-time: the running time (sec.) to execute  Stage~5 
 for a given  feature vector $\x^*$. 
 ``$>$ 2 hours'' means that the running time exceeds two hours.
 \end{itemize}
 
 Previously 
 an instance of chemical acyclic graphs with size $n^*$ up to 16 
 was solved in Stage~5 by Azam~et~al.~\cite{ACZSNA20}.
 For the classes of chemical graphs with cycle index 1 and 2,
 the maximum size of instances solved in Stage~5
 by Ito~et~al.~\cite{ACZSNA20} and Zhu~et~al.~\cite{ZCSNA20}
 was around 18 and 15,  respectively. 
 Our new algorithm based on dynamic programming 
 solve instances with $n^*=50$.
 In our experiments, we also computed a lower bound G-LP
 on the number of target graphs.
 Observe that there are over $10^{10}$ or $10^{14}$ target graphs
 in some cases.
 Remember that these lower bounds are computed without
 actually generating each target graph one by one.
 So when a lower bound is enormously  large,
 this would suggest that we may need to impose some more
 constraints on the structure of graphs or the range of descriptors
 to narrower a family of target graphs to be inferred.

\begin{table}[ht!] \caption{Results of Stages~4 and 5  for  $\bl^*= 2$,
 $\dmax=3$ and $\dia^* = \lceil \frac{2}{5}n^* \rceil$. }

    \begin{center}
    \begin{tabular}{ @{} c r r r r r r r r @{}}\toprule
    $\pi$  &   $y^*$  & $n^*$  &  $\dia^*$ &~{\small IP-time}\hspace{-1mm} 
      & {\small $\#$FP~} & {\small G-LB~~~} &  {\small $\#$G}~ & {\small G-time} \\  \midrule

 \multirow{5}{*}  {\sc K\tiny{ow}}
&	4	&	26	&	11	&	3.95	&	11,780	 &	 $2.4 \times 10^{6}$	 &	100	 &	0.91	\\
&	5	&	32	&	13	&	4.81	&	216	 &	 $2.7 \times 10^{4}$	 &	100	 &	10.64	\\
&	7	&	38	&	16	&	7.27	&	19,931	 &	 $4.2 \times 10^{7}$	 &	100	 &	48.29	\\
&	8	&	44	&	18	&	9.33	&	241,956	 &	 $1.2 \times 10^{13}$	 &	100	 &	119.01	\\
&	9	&	50	&	20	&	21.57	&	58,365	 &	 $1.7 \times 10^{10}$	 &	100	 &	110.38	\\

\midrule
  \multirow{5}{*}  {\sc Bp}
&	440	&	26	&	11	&	2.09	&	22,342	 &	 $3.6 \times 10^{7}$	 &	100	 &	2.9	\\
&	550	&	32	&	13	&	3.94	&	748	 &	 $5.9 \times 10^{6}$	 &	100	 &	3.77	\\
&	660	&	38	&	16	&	6.4	&	39,228	 &	 $7.3 \times 10^{8}$	 &	100	 &	151.25	\\
&	770	&	44	&	18	&	7.21	&	138,076	 &	 $3.0 \times 10^{12}$	 &	100	 &	182.66	\\
&	880	&	50	&	20	&	9.49	&	106,394	 &	 $3.0 \times 10^{10}$	 &	100	 &	217.18	\\
 \midrule
   \multirow{5}{*} {\sc Hc}
&	13000	&	26	&	11	&	2.94	&	12	 &	 $2.0 \times 10^{1}$	 &	12	 &	0.04	\\
&	16500	&	32	&	13	&	7.67	&	2,722	 &	 $1.2 \times 10^{7}$	 &	100	 &	0.31	\\
&	20000	&	38	&	16	&	10.5	&	1,830	 &	 $9.7 \times 10^{5}$	 &	100	 &	1.06	\\
&	23000	&	44	&	18	&	13.62	&	 12,336	 &	 $4.7 \times 10^{8}$	 &	100	 &	142.02
						\\
&	25000	&	50	&	20	&	15.1	&	136,702	 &	 $5.3 \times 10^{14}$	 &	100	 &	22.26	\\
 \bottomrule
    \end{tabular} \end{center}\label{table_2_3_2-5}
    \end{table}

\begin{table}[ht!] \caption{Results of Stages~4 and 5  for $\bl^*= 2$, 
 $\dmax=4$ and $\dia^*  =  \lceil \frac{3}{5}n^* \rceil$. }

    \begin{center}
    \begin{tabular}{ @{} c r r r r r r r r @{}}\toprule
    $\pi$  &   $y^*$  & $n^*$  &  $\dia^*$ &~{\small IP-time}\hspace{-1mm}  
     & {\small $\#$FP~} & {\small G-LB~~~} &  {\small $\#$G}~ & {\small G-time} \\  \midrule

 \multirow{5}{*}  {\sc K\tiny{ow}}
&	4	&	26	&	16	&	16.21	&	4,198	 &	 $3.5 \times 10^{5}$	 &	100	 &	1.18	\\
&	5	&	32	&	20	&	24.74	&	1,650	 &	 $5.3 \times 10^{6}$	 &	100	 &	0.69	\\
&	7	&	38	&	23	&	38.88	&	154,408	 &	 $9.5 \times 10^{9}$	 &	100	 &	67.31	\\
&	8	&	44	&	27	&	38.73	&	1,122,126	 &	 $8.5 \times 10^{13}$	 &	100	 &	660.37	\\
&	9	&	50	&	30	&	31.59	&	690,814	 &	 $1.1 \times 10^{15}$	 &	100	 &	238.02	\\
\midrule
   \multirow{5}{*}  {\sc Bp}
&	440	&	26	&	16	&	12.44	&	8,156	 &	 $2.6 \times 10^{6}$	 &	100	 &	2.74	\\
&	550	&	32	&	20	&	23.22	&	38,600	 &	 $4.4 \times 10^{8}$	 &	100	 &	12.72	\\
&	660	&	38	&	23	&	20.62	&	52,406	 &	 $1.1 \times 10^{9}$	 &	100	 &	197.89	\\
&	770	&	44	&	27	&	50.55	&	23,638	 &	 $6.8 \times 10^{8}$	 &	100	 &	244.56	\\
&	880	&	50	&	30	&	48.37	&	40,382	 &	 $2.2 \times 10^{11}$	 &	100	 &	884.99	\\
   \midrule
   \multirow{5}{*} {\sc Hc} 
&	13000	&	26	&	16	&	23.26	&	249	 &	 $2.7 \times 10^{3}$	 &	100	 &	0.06	\\
&	16500	&	32	&	20	&	44.2	&	448	 &	 $6.9 \times 10^{4}$	 &	100	 &	0.63	\\
&	20000	&	38	&	23	&	96.02	&	3,330	 &	 $6.1 \times 10^{6}$	 &	100	 &	15.16	\\
&	23000	&	44	&	27	&	82.34	&	43,686	 &	 $1.5 \times 10^{10}$	 &	100	 &	152.96	\\
&	25000	&	50	&	30	&	83.81	&	311,166	 &	 $1.3 \times 10^{13}$	 &	100	 &	287.95	\\
 \bottomrule
    \end{tabular} \end{center}\label{table_2_4_3-5}
    \end{table}   

\begin{table}[ht!] \caption{Results of Stages~4 and 5  for  $\bl^*= 3$,
 $\dmax=3$ and $\dia^*  = \lceil \frac{2}{5}n^* \rceil$.   }
    \begin{center}
    \begin{tabular}{ @{} c r r r r r r r r @{}}\toprule
    $\pi$  &   $y^*$  & $n^*$  &  $\dia^*$ &~{\small IP-time}\hspace{-1mm}
       & {\small $\#$FP~} & {\small G-LB~~~} &  {\small $\#$G}~ & {\small G-time} \\  \midrule
 \multirow{5}{*}  {\sc K\tiny{ow}}
&	4	&	26	&	11	&	3.1	&	511	 &	 $3.6 \times 10^{3}$	 &	100	 &	14.31	\\
&	5	&	32	&	13	&	4.72	&	3,510	 &	 $6.8 \times 10^{6}$	 &	100	 &	851.21	\\
&	7	&	38	&	16	&	5.82	&	11,648	 &	 $1.2 \times 10^{8}$	 &	100	 &	612.86	\\
&	8	&	44	&	18	&	9.69	&	17,239	 &	 $2.2 \times 10^{8}$	 &	100	 &	703.92	\\
&	9	&	50	&	20	&	22.53	&	60,792	 &	 $3.9 \times 10^{12}$	 &	100	 &	762.17	\\
 \midrule
   \multirow{5}{*}  {\sc Bp}
&	440	&	26	&	11	&	3.01	&	66	 &	 $9.0 \times 10^{2}$	 &	66	 &	902.77	\\
&	550	&	32	&	13	&	4.29	&	308	 &	 $1.0 \times 10^{7}$	 &	100	 &	2238.62	\\
&	660	&	38	&	16	&	5.86	&	303	 &	 $1.8 \times 10^{7}$	 &	100	 &	3061.11	\\
&	770	&	44	&	18	&	14.39	&	19,952	 &	 $4.7 \times 10^{10}$	 &	100	 &	678.26	\\
&	880	&	50	&	20	&	10.39	&	17,993	 &	 $7.1 \times 10^{12}$	 &	100	 &	4151.07	\\
  \midrule
   \multirow{5}{*} {\sc Hc} 
&	13000	&	26	&	11	&	3.05	&	340	 &	 $1.5 \times 10^{4}$	 &	100	 &	1.57	\\
&	16500	&	32	&	13	&	5.81	&	600	 &	 $3.1 \times 10^{8}$	 &	100	 &	921.55	\\
&	20000	&	38	&	16	&	15.67&18,502	 &	 $6.2 \times 10^{8}$	 &	100	 &	1212.54	\\
&	23000	&	44	&	18	&	21.15&5,064	 &	 $6.9 \times 10^{9}$	 &	100	 &	1279.95	\\
&	25000	&	50	&	20	&	31.90&41,291	 &	 $2.4 \times 10^{12}$	 &	100	 &	668.5	\\
   \bottomrule
    \end{tabular} \end{center}\label{table_3_3_2-5}
    \end{table}

\begin{table}[ht!] \caption{Results of Stages~4 and 5  for  $\bl^*= 3$,
 $\dmax=4$ and $\dia^*  =  \lceil \frac{3}{5}n^* \rceil$. }
    \begin{center}
    \begin{tabular}{ @{} c r r r r r r r r @{}}\toprule
    $\pi$  &   $y^*$  & $n^*$  &  $\dia^*$ &~{\small IP-time}\hspace{-1mm}
       & {\small $\#$FP~} & {\small G-LB~~~} &  {\small $\#$G}~ & {\small G-time} \\  \midrule

 \multirow{5}{*}  {\sc K\tiny{ow}}
&	4	&	26	&	16	&	9.94	&	100	 &	 $2.5 \times 10^{4}$	 &	100	 &	6.73	\\
&	5	&	32	&	20	&	16.58	&	348	 &	 $1.4 \times 10^{8}$	 &	100	 &	3400.74	\\
&	7	&	38	&	23	&	33.71	&	17,557	 &	 $1.2 \times 10^{11}$	 &	100	 &	2652.38	\\
&	8	&	44	&	27	&	34.28	&	        0 & 	   0                 & 1 & {\rm $>$2 hours}\\
&	9	&	50	&	30	&	68.74	&	80,411	 &	 $6.4 \times 10^{15}$	 &	100	 &	6423.85	\\
\midrule
   \multirow{5}{*}  {\sc Bp}
&	440	&	26	&	16	&	14.16	&	150	 &	 $1.8 \times 10^{5}$	 &	100	 &	29.72	\\
&	550	&	32	&	20	&	18.94	&	305	 &	 $1.4 \times 10^{7}$	 &	100	 &	2641.9	\\
&	660	&	38	&	23	&	21.15	&	1,155	 &	 $2.0 \times 10^{9}$	 &	100	 &	4521.66	\\
&	770	&	44	&	27	&	25.6	&	1,620	 &	 $4.3 \times 10^{8}$	 &	100	 &	175.2	\\
&	880	&	50	&	30	&	63.22	&		0   &               	0 &          1 &  {\rm $>$2 hours}				\\
  \midrule
   \multirow{5}{*} {\sc Hc} 
&	13000	&	26	&	16	&	31.87	&	12	 &	 $2.7 \times 10^{4}$	 &	12	 &	0.66	\\
&	16500	&	32	&	20	&	41.03	&	392	 &	 $3.4 \times 10^{8}$	 &	100	 &	2480.34	\\
&	20000	&	38	&	23	&	48.48	&	630	 &	 $1.4 \times 10^{5}$	 &	100	 &	105.59	\\
&	23000	&	44	&	27	&	143.75	&	 341	 &	 $7.8 \times 10^{8}$	 &	100	 &	5269.1	\\
&	25000	&	50	&	30	&	315.91	&	10,195	 &	 $3.8 \times 10^{9}$	 &	100	 &	5697.08	\\
   \bottomrule
    \end{tabular} \end{center}\label{table_3_4_3-5}
    \end{table}   

\begin{figure}[!htb]
\begin{center} 
 \includegraphics[width=.45\columnwidth]{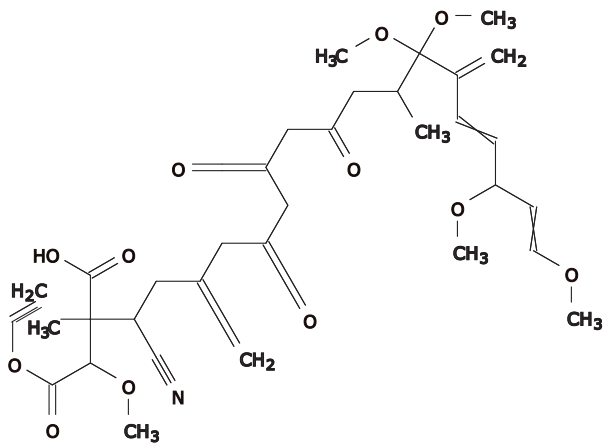}
\end{center}
\caption{An illustration of a chemical acyclic graph $G$
inferred  for three chemical properties  {\sc K\tiny{ow}},
 {\sc Bp}  and  {\sc Hc} simultaneously,
 where     $y^*_{{\rm Kow}} =4.0$,  
 $y^*_{{\rm Bp}} =400.0$ and  
 $y^*_{{\rm Hc}}  =13000.0$, 
$n^* =50$,  $\bl^* =2$,   $\dia^* = 25$ and $\dmax =4$.
}
\label{fig:inferred-triple}  
\end{figure}

\bigskip \noindent
{\bf An Additional Experiment.  } 
 We also conducted some additional experiment to demonstrate that 
 our MILP-based method is flexible to control conditions
  on inference of chemical graphs.
 In Stage~3, we constructed an ANN $\mathcal{N}_{\pi}$ for each of
 the three chemical properties $\pi\in\{${\sc K\tiny{ow}},  {\sc Bp}, {\sc Hc}$\}$,
 and formulated the inverse problem of each ANN $\mathcal{N}_{\pi}$
 as an MILP $\mathcal{M}_{\pi}$.
Since the set of descriptors is common to all three properties
 {\sc K\tiny{ow}},  {\sc Bp} and  {\sc Hc}, it is possible to 
 infer a chemical acyclic graph $G$ that satisfies
 a target value $y^*_{\pi}$ for each of the three properties
 at the same time (if one exists).
 We specify the size of graph so that 
  $n^* =50$,  $\bl^* =2$,   $\dia^* = 25$ and $\dmax =4$, 
  and set target values with
 $y^*_{{\rm Kow}} =4.0$,  
 $y^*_{{\rm Bp}} =400.0$ and  
 $y^*_{{\rm Hc}}  =13000.0$ 
 in   an MILP that consists of the three MILP
  $\mathcal{M}_{{\rm Kow}}$, $\mathcal{M}_{{\rm Hc}}$
   and $\mathcal{M}_{{\rm Bp}}$. 
The MILP was solved    in   18930 (sec) and 
  we obtained a chemical acyclic graph $G$ illustrated 
  in Figure~\ref{fig:inferred-triple}.
We continued to execute Stage~5 for this instance to generate more target graphs
$G^*$.
Table~\ref{table_triple_target} shows that 100 target graphs are generated
by our new dynamic programming algorithm. 

\begin{table}[ht!] \caption{Results of Stages~4 and 5  for  $\bl^*= 2$,
 $\dmax=4$, $n^* = 50$ and $\dia^*  =  25$. }
    \begin{center}
    \begin{tabular}{ @{} c   r | r r r r r r r @{}}\toprule
    $\pi$  &   $y^*$  & $n^*$  &  $\dia^*$ &~{\small IP-time}\hspace{-1mm}
       & {\small $\#$FP~} & {\small G-LB~~~} &  {\small $\#$G}~ & {\small G-time} \\  \midrule
{\sc K\tiny{ow}} & 4  & \multirow{3}{*} {50} 	& 
                                     \multirow{3}{*} {25} &  \multirow{3}{*} {18930.46} &\multirow{3}{*} {117,548} &\multirow{3}{*}{$2.4 \times 10^{11}$}  &\multirow{3}{*}{100} & 
                                     \multirow{3}{*}{423.53}\\
       {\sc Bp} & 400 &\\
       {\sc Hc} &1300 & \\ \bottomrule

\end{tabular} \end{center}\label{table_triple_target}
\end{table}  

\clearpage

\section{Concluding Remarks}\label{sec:conclude}

In this paper, we introduced a new measure, branch-height  of a tree, 
and showed that many of chemical compounds in the chemical database
have a simple structure where the number of 2-branches is small. 
Based on this, we proposed a new method of applying 
the framework for inverse QSAR/QSPR  
\cite{ACZSNA20,CWZSNA20,ZZCSNA20}
to the case of acyclic chemical graphs where 
Azam et al.~\cite{ACZSNA20} inferred chemical graphs 
with around 20  non-hydrogen atoms 
and 
Zhang et al.~\cite{ZZCSNA20} solved 
an MILP of inferring a feature vector for an instance
with   up to around 50  non-hydrogen atoms 
and   diameter 8.
In our method, we formulated a new MILP in Stage~4 specialized
 for acyclic chemical graphs with a small branch number
 and designed a new graph search algorithm in Stage~5
 that computes frequency vectors of graphs 
 in a dynamic programming scheme.
 We implemented our new method and conducted some experiments
 on chemical properties such as 
  octanol/water partition coefficient,
  boiling point and
  heat of combustion. 
The resulting method improved the performance so that  chemical graphs 
with around 50  non-hydrogen atoms 
and around diameter 30 can be inferred.
Since there are many acyclic chemical compounds having large diameters,
this is a significant improvement.

It is left as a future work to design MILPs
and graph search algorithms based on the new idea
of the paper for classes of graphs with a higher rank.

\bigskip
\noindent
{\bf Abbreviations } 
ANN: artificial neural network; MILP: mixed integer linear programming

\bigskip
\noindent
{\bf Acknowledgements}
This research was supported, in part, by 
Japan Society for the Promotion of Science, Japan, under Grant \#18H04113.

\bigskip
\noindent
{\bf Authors' contributions} 
Conceptualization, H.N. and T.A.; 
methodology, H.N.; 
software, N.A.A.,  J.Z., Y.Sun, Y.Shi, A.S. and L.Z.; 
validation, N.A.A., J.Z.,  A.S.  and H.N.; 
formal analysis, H.N.; 
data resources,  A.S., L.Z., H.N. and T.A.; 
writing--original draft preparation, H.N.; 
writing--review and editing, N.A.A., A.S. and T.A.; 
project administration, H.N.; 
funding acquisition, T.A. 
All authors have read and agreed to the published version of the manuscript.

\bigskip
\noindent
{\bf Availability of data and materials} 
Source code of the implementation of our algorithm is freely available from 
{\tt  https://github.com/ku-dml/mol-infer}.

\bigskip
\noindent
{\bf Competing interests} 
The authors declare that they have no competing interests.

\bigskip
\noindent
{\bf Author details}  
$^1$ Department of Applied Mathematics and Physics, Kyoto University, Kyoto 606-8501, Japan. 
$^2$ Graduate School of Advanced Integrated Studies in Human Survavibility,
Kyoto University, Kyoto 606-8306. 
$^3$ Bioinformatics Center,  Institute for Chemical Research, 
Kyoto University, Uji 611-0011, Japan. 

\appendix

\section{Statistical Feature of Molecular Structure}\label{sec:statistical}

We observe the following  features of  the graph-theoretical structure
of chemical graphs registered   in the chemical  database  PubChem. 
%
Let $\mathrm{DB}^{(\leq n)}$ denote the set of
 chemical graphs with at most $n$ non-hydrogen atoms
 that are registered   in  chemical  database  PubChem.  
The {\em cycle index} (or {\em rank}) of
 a chemical graph $G=(H=(V,E),\alpha,\beta)$
 is defined to be $|E|-(|V|-1)$ (i.e., the minimum number of edges
 to be removed to make the graph $H$ acyclic).
 We call a chemical graph  
a {\em rank-$r$} chemical graph if the rank of the graph is $r$.
The {\em core} of a chemical cyclic graph $G$ is defined to be
the induced subgraph $G'$ of $G$ such that
$G'$ consists of vertices in a cycle or vertices in a path
joining two cycles.
A vertex in the core (not in the core) is called a {\em core vertex}
(resp., a non-core vertex).
The edges not in the core of a chemical cyclic graph $G$
form a collection of   trees $T$, which we call a {\em non-core tree}.
Each non-core tree contains
exactly one core vertex and is regarded as a tree rooted at the core vertex.
The {\em $k$-branch height} of a chemical cyclic graph $G$
is defined to be the maximum of $k$-branch heights over
all non-core trees. 

Let $\rho_r$ (\%) denote the ratio  of the number of  chemical graphs 
with  rank at most $r\in [0,4]$  
to the number of all  chemical graphs    in    PubChem.  
See Table~\ref{table1}. 

\begin{table}[!th]
 \caption{The percentage $\rho_r$ of the number of
 chemical compounds with rank at most $r\in [0,4]$ 
 over all chemical compounds in PubChem.}\begin{center}
\begin{tabular}{ c c c  c c } \hline
$\rho_0$ & $\rho_1$  & $\rho_2$ &  $\rho_3$ &   $\rho_4$    \\ \hline 
$2.9\%$ & $16.3\%$  & $44.5\%$ & $68.8\%$  & $84.7\%$\\ \hline 
\end{tabular}  \end{center}\label{table1}
\end{table}

Let
 $\rho_0^{(d)}$ (\%) denote the ratio of the number of  chemical graphs 
in  $\mathrm{DB}^{(\leq 100)}$  
such that the maximum degree is at most $d\in [3,4]$ 
to the number of all  chemical graphs in $\mathrm{DB}^{(\leq 100)}$.
Let 
  $\rho_r^{(d)}$ (\%), $r\in [1,4]$
   denote the ratio of the number of rank-$r$ chemical graphs
    in $\mathrm{DB}^{(\leq 100)}$
      such that the maximum degree of a non-core vertex
 is at most $d\in [3,4]$ 
to the number of all  rank-$r$ chemical graphs  
 in $\mathrm{DB}^{(\leq 100)}$.
See Table~\ref{table2}. 
 
\begin{table}[!th]
 \caption{The percentage $\rho_r^{(d)}$ of the number of
 chemical compounds with rank $r\in [0,4]$ 
 such that the maximum degree of a non-core vertex is at most $d\in [3,4]$
 over all rank-$r$ chemical compounds in $\mathrm{DB}^{(\leq 100)}$.}
 \begin{center}
\begin{tabular}{ c c c c c   c c c c c} \hline 
 $\rho_0^{(3)}$ & $\rho_0^{(4)}$ &  $\rho_1^{(3)}$ & $\rho_1^{(4)}$ & 
 $\rho_2^{(3)}$ & $\rho_2^{(4)}$ &  $\rho_3^{(3)}$ &  $\rho_3^{(4)}$ & 
 $\rho_4^{(3)}$ &  $\rho_4^{(4)}$     \\ \hline 
 $55.55\%$ &  $99.85\%$   &  $68.30\%$  &  $99.97\%$  &  $84.46\%$
  &  $99.99\%$  &  $87.11\%$ &  $99.99\%$  &  $87.75\%$ &  $99.99\%$\\ \hline 
\end{tabular}  \end{center}\label{table2}
\end{table}
    
 
Let 
  $\rho_r(k,h)$ (\%), $r\in [0,4]$, $k=2$, $h\in[1,2]$ 
   denote the ratio of the number of rank-$r$ chemical graphs 
  in $\mathrm{DB}^{(\leq 50)}$
   such that the $k$-branch height is at most $h$ 
   to the number of all  rank-$r$ chemical graphs     
  in $\mathrm{DB}^{(\leq 50)}$. 
See Table~\ref{table3}. 
We see that most chemical graphs $G$ with at most 50 non-hydrogen atoms
satisfy $\mathrm{bh}_2(G)\leq 2$. 
  
\begin{table}[!th]
 \caption{The percentage $\rho_r(k,h)$ (\%)
 of the number of rank-$r$ chemical graphs 
  in $\mathrm{DB}^{(\leq 50)}$
   such that the $k$-branch height is at most $h$ 
   to the number of all  rank-$r$ chemical graphs     
  in $\mathrm{DB}^{(\leq 50)}$.}
 \begin{center}
\begin{tabular}{ c c c c c   c c  } \hline 
$\rho_0(2,1)$  &  $\rho_0(2,2)$  &  $\rho_1(2,1)$  &   $\rho_1(2,2)$  &  
$\rho_2(2,1)$  &  $\rho_3(2,1)$  &  $\rho_4(2,1)$   \\ \hline 
$87.23\%$  &  $99.46\%$  &  $88.13\%$  &  $98.76 \%$  &  $96.39\%$
  &  $99.17\%$  &  $99.43\%$ \\ \hline 
\end{tabular}  \end{center}\label{table3}
\end{table}
 
We show the distribution of 2-branch-height over alkans
 {\tt C}$_n${\tt H}$_{2n+2}$. 
Let $\mathrm{Aln}(n)$ denote the set of all alkans with $n$ carbon atoms,
where $|\mathrm{Aln}(25)|=36,797,588$.
Let   $\rho_{\mathrm{Aln}}(2,h)$ (\%), $h\in[1,4]$ 
   denote the ratio of the number of alkans  in $\mathrm{Aln}(25)$ 
   such that the 2-branch height is at most $h$ 
   to the number of alkans  in $\mathrm{Aln}(25)$. 
See Table~\ref{table5}.

\begin{table}[!th]
 \caption{The percentage $\rho_{\mathrm{Aln}}(2,h)$ (\%)
 of the number of alkans  in $\mathrm{Aln}(25)$ 
   such that the 2-branch height is at most $h$ 
   to the number of alkans  in $\mathrm{Aln}(25)$. }
 \begin{center}
\begin{tabular}{ c c c c } \hline 
$\rho_{\mathrm{Aln}}(2,1)$ & $\rho_{\mathrm{Aln}}(2,2)$ & 
$\rho_{\mathrm{Aln}}(2,3)$ & $\rho_{\mathrm{Aln}}(2,4)$ \\ \hline 
$49.03\%$  &  $97.67\%$  &  $99.99\%$ &  $100.00\%$ \\ \hline 
\end{tabular}  \end{center}\label{table5}
\end{table}

 Let $\rho_{\mathrm{2bt}}(\delta)$ denote the ratio  
 of the number of acyclic chemical graphs 
  in $\mathrm{DB}^{(\leq 50)}$
   such that the degree of the root of 
   the $2$-branch-tree is $\delta\in[1,4]$   
   to the number of all acyclic  chemical graphs     
  in $\mathrm{DB}^{(\leq 50)}$.
See Table~\ref{table4}. 
  
\begin{table}[!th]
 \caption{The percentage $\rho_{\mathrm{2bt}}(\delta)$
  of the number of acyclic chemical graphs 
  in $\mathrm{DB}^{(\leq 50)}$
   such that the degree of the root of 
   the $2$-branch-tree is $\delta\in[1,4]$   
   to the number of all acyclic  chemical graphs     
  in $\mathrm{DB}^{(\leq 50)}$.}
 \begin{center}
\begin{tabular}{ c c c c  } \hline 
$\rho_{\mathrm{2bt}}(1)$  & $\rho_{\mathrm{2bt}}(2)$  & 
$\rho_{\mathrm{2bt}}(3)$  & $\rho_{\mathrm{2bt}}(4)$ \\ \hline 
$6.39\%$  &  $   83.58\%$  &  $  9.30\%$  &  $   0.73\%$ \\ \hline 
\end{tabular}  \end{center}\label{table4}
\end{table}

Among the 2-fringe-trees $T$ of all acyclic chemical graphs
  in $\mathrm{DB}^{(\leq 100)}$, over $90\%$ of them satisfy
 $n\leq 2d+2$ 
for the number $n=|V(T)|$ of non-hydrogen atoms in
a 2-fringe-tree $T$ and
 the number $d$ of non-hydrogen atoms
adjacent to the root  in $T$. 

Let
$\mathcal{FT}_{0,2}$ denote the set of all 2-fringe-trees
that appear in an acyclic chemical graph in $\mathrm{DB}^{(\leq 100)}$,
and 
$\mathcal{FT}_{0,2}^{(\delta)}$, $\delta\in [1,3]$
denote the  set of all 2-fringe-trees $T\in \mathcal{FT}_{0,2}$
that has $\delta$ children (i.e., the degree of the root is $\delta$).
Let  $\rho_{2\delta+2}^{(\delta)}$ (\%) denote the ratio
of the number of 2-fringe-trees in $\mathcal{FT}_{0,2}^{(\delta)}$
 that has at most $2d+2$ vertices  
   to  the number of 2-fringe-trees in $\mathcal{FT}_{0,2}^{(\delta)}$.
See Table~\ref{table6}. 
   
\begin{table}[!th]
 \caption{The percentage $\rho_{2\delta+2}^{(\delta)}$ (\%)
 of the number of 2-fringe-trees in $\mathcal{FT}_{0,2}^{(\delta)}$
 that has at most $2d+2$ vertices  
   to  the number of 2-fringe-trees in $\mathcal{FT}_{0,2}^{(\delta)}$.}
 \begin{center}
\begin{tabular}{ c c c  } \hline 
$\rho_{4}^{(1)}$ & $\rho_{6}^{(2)}$ & $\rho_{8}^{(3)}$ \\ \hline 
$93.77\%$  &  $93.99\%$  &  $92.01\%$ \\ \hline 
\end{tabular}  \end{center}\label{table6}
\end{table}

\newpage

\section{All Constraints in an MILP Formulation for Chemical Acyclic Graphs}
\label{sec:full_milp}

To formulate an MILP that represents a chemical graph, 
 we   distinguish a tuple $({\tt a,b},m)$ from a tuple $({\tt b,a},m)$.
For a tuple  $\gamma=({\tt a,b},m)\in\Lambda\times \Lambda\times \{1,2,3\}$,
let $\overline{\gamma}$ denote the tuple $({\tt b,a},m)$.
Let    $\Gamma_{<}\triangleq \{\overline{\gamma}\mid \gamma\in \Gamma_{>}\}$. 
 We call a tuple $\gamma=({\tt a,b},m)
\in \Lambda\times \Lambda\times \{1,2,3\}$ 
{\em proper} if 
\[\mbox{
 $m\leq \min \{ \val(\ta), \val({\tt b}) \}$
 and  
 $m\leq \max \{ \val(\ta), \val({\tt b}) \}-1$,}\]
 where the latter is assumed because otherwise $G$ must consist  of
 two atoms of ${\tt a=b}$.
 Assume  that each tuple $\gamma\in \Gamma$  is proper.   
  Let $\epsilon$ be a fictitious chemical element
  that represents null,   call a tuple $({\tt a,b},0)$ 
  with  ${\tt a,b}\in    \Lambda\cup\{\epsilon\}$ {\em fictitious}, 
   and define $\Gamma_0$ to be   the set of all fictitious tuples;
   i.e., $\Gamma_0=\{({\tt a,b},0) \mid {\tt a,b}\in \Lambda\cup\{\epsilon\}\}$.
 To represent chemical elements
  ${\tt e}\in \Lambda\cup\{\epsilon\}\cup   \Gamma $ 
 in an MILP, we encode these elements ${\tt e}$ into some integers denoted by $[{\tt e}]$.
 Assume that, for each element $\ta\in \Lambda$,
  $[\ta]$ is a positive integer and that $[\epsilon]=0$.

\subsection{Upper and Lower Bounds on Descriptors} 
\label{sec:AD}

In our formulation of an MILP for inferring a vector $x^*$ in Stage~4,
we fix the following descriptors as specified constants:
the number $n(G)$ of vertices , the diameter $\dia(G)$, 
and the number $\bl_{k^*}(G)$ of leaf $k^*$-leaf branches,
which are set to be given integers $n^*$, $\dia^*$ and $\bl^*$,
respectively.
For each of the other descriptors, 
we specify a  lower bound $\LB$
and an upper bound $\UB$ on the value 
so that the descriptor takes a value from the range
between $\LB$ and $\UB$.

\bigskip   
\noindent  
{\bf constants: } \\
~~ $n^*\geq 5$: the size $n(G)$ of $G$; \\ 
~~ $\LB_\dg^\inn(i), \UB_\dg^\inn(i)\in [0,n^*]$, $i\in[1,4]$: 
lower and upper bounds on
 the number $\dg_i^\inn(G)$ \\
 ~~~~~~ of $k^*$-internal vertices   of degree $i$ in $G$; \\ 
~~ $\LB_\dg^\ex(i), \UB_\dg^\ex(i)\in [0,n^*]$, $i\in[1,4]$:   
lower and upper bounds on
  the number $\dg_i^\ex(G)$ \\
 ~~~~~~ of $k^*$-internal vertices 
  of degree $i$ in $G$; \\ 
~~ $\LB_\ce^\inn(\ta), \UB_\ce^\inn(\ta)\in [0,n^*]$, $\ta\in \Lambda$:   
lower and upper bounds on
  the number $\ce_\ta^\inn(G)$ \\
 ~~~~~~ of $k^*$-internal vertices  $v$
  with $\alpha(v)=\ta$ in $G$; \\ 
~~ $\LB_\ce^\ex(\ta), \UB_\ce^\ex(\ta)\in [0,n^*]$, $\ta\in \Lambda$:   
lower and upper bounds on
  the number $\ce_\ta^\ex(G)$ \\
 ~~~~~~ of $k^*$-external vertices  $v$
  with $\alpha(v)=\ta$ in $G$; \\ 
~~ $\LB_\bd^\inn(m), \UB_\bd^\inn(m)\in [0,n^*-1]$, $m\in [2,3]$:   
lower and upper bounds on
  the number $\bd_m^\inn(G)$ \\
 ~~~~~~ of $k^*$-internal edges $e$
  with $\beta(e)=m$ in $G$; \\ 
~~ $\LB_\bd^\ex(m), \UB_\bd^\ex(m)\in [0,n^*-1]$, $m\in [2,3]$:  
lower and upper bounds on
  the number $\bd_m^\ex(G)$ \\
 ~~~~~~ of $k^*$-external edges $e$
  with $\beta(e)=m$ in $G$; \\ 
~~ $\LB_\ac^\inn(\gamma), \UB_\ac^\inn(\gamma)\in [0,n^*-1]$,
$\gamma \in \Gamma_{<}\cup \Gamma_{=}$:  
lower and upper bounds on
  the number $ \ac_{\gamma}^\inn(G)$ \\
 ~~~~~~  of $k^*$-internal edges $e$
  with   adjacency-configuration $\gamma$ in $G$; \\ 
~~ $\LB_\ac^\ex(\gamma), \UB_\ac^\ex(\gamma)\in [0,n^*-1]$,
$\gamma \in \Gamma_{<}\cup \Gamma_{=}$:  
lower and upper bounds on
  the number $ \ac_{\gamma}^\ex(G)$\\
 ~~~~~~  of $k^*$-external edges $e$
  with adjacency-configuration $\gamma$ in $G$; \\ 
~~ $\LB_\bc^\typ(\mu),  \UB_\bc^\typ(\mu)\in [0,n^*-1]$,
$\mu\in \Bc$:  
lower and upper bounds on
  the number $\bc_{\mu}^\inn(G)$ \\
 ~~~~~~ of $k^*$-internal edges $e$
  with   bond-configuration $\mu$  in $G$; \\ 
~~ $\LB_\bc^\ex(\mu),  \UB_\bc^\ex(\mu)\in [0,n^*-1]$,
$\mu\in \Bc$:  
lower and upper bounds on   the number $\bc_{\mu}^\ex(G)$ \\
 ~~~~~~ of $k^*$-internal edges $e$
  with   bond-configuration $\mu$  in $G$; \\

\smallskip  
\noindent 
{\bf variables $x$ for descriptors: } \\  
~~ $\dg^\inn(i), \dg^\ex(i)\in[0,n^*]$,    $i\in [1,4]$: 
$\dg^\inn(i)$ (resp.,  $\dg^\ex(i)$) represents
 $\dg_i^\inn(G)$   (resp.,  $\dg_i^\ex(G)$); \\ 
~~
  $\ce^\inn(\ta), \ce^\ex(\ta)\in[0,n^*]$, $\ta\in \Lambda$: 
   $\ce^\inn(\ta)$ (resp., $\ce^\ex(\ta)$)    represents 
    $\ce_\ta^\inn(G)$   (resp., $\ce_\ta^\ex(G)$); \\  
~~ $\bd^\inn(m), \bd^\ex(m)\in[0,2n^*]$, $m\in[1,3]$: 
  $\bd^\inn(m)$ (resp., $\bd^\ex(m)$) \\
~~~~~  
   represents $\bd_m^\inn(G)$ (resp., $\bd_m^\ex(G)$); \\ 
~~   $\ac^\inn(\gamma), \ac^\ex(\gamma)\in[0,n^*]$,
   $\gamma\in \Gamma_{<}\cup \Gamma_{=}$:
    $\ac^\inn(\gamma)$ (resp., $\ac^\ex(\gamma)$) represents 
   represents $ \ac_{\gamma}^\inn(G)$  \\
~~~~~(resp., $ \ac_{\gamma}^\ex(G)$); \\ 
~~ $\bc^\inn(\mu), \bc^\ex(\mu)\in[0,n^*-1]$, $\mu\in \Bc$:  
$\bc^\inn(\mu)$ (resp., $\bc^\ex(\mu)$) represents 
   represents $\bc_{\mu}^\inn(G)$ \\
~~~~~  (resp., $\bc_{\mu}^\ex(G)$); \\

\noindent
{\bf constraints: }     
\begin{align}  
 %
%
 \LB_\dg^\typ(i)    \leq    \dg^\typ(i)    \leq \UB_\dg^\typ(i),  
 &&   i\in[1,4],   \typ\in\{\inn,\ex\}, \label{eq:AD_first} \\
 \LB_\ce^\typ(\ta)    \leq     \ce^\typ(\ta)    \leq \UB_\ce^\typ(\ta) ,  
  &&   \ta\in \Lambda, \typ\in\{\inn,\ex\},  \label{eq:AD2}  \\
 %
  %
 \LB_\bd^\typ(m)   \leq     \bd^\typ(m)     \leq \UB_\bd^\typ(m), 
 &&  m\in [2,3], \typ\in\{\inn,\ex\},  \label{eq:AD3} \\
 \LB_\ac^\typ(\gamma)   \leq     \ac^\typ(\gamma)    \leq \UB_\ac^\typ(\gamma), 
  && \gamma\in \Gamma, \typ\in\{\inn,\ex\},   \label{eq:AD4}  \\
 \LB_\bc^\typ(\mu)   \leq     \bc^\typ(\mu)    \leq \UB_\bc^\typ(\mu), 
 && \mu\in \Bc, \typ\in\{\inn,\ex\}.   \label{eq:AD5}  
\end{align}

We use the range-based method to define 
an applicability domain for our method.
For this, we find 
the range (the minimum and maximum)
of each descriptor over all relevant chemical compounds 
and represent each range as a set of linear constraints 
in the constraint set $\mathcal{C}_1$ of our MILP formulation. 
Recall that $D_{\pi}$ stands for a set of chemical graphs used
for constructing a prediction function.
However, the number of examples in $D_{\pi}$ may not be
large enough to capture a general feature on the structure of chemical graphs.
For this, we also use some data set from the whole set $\mathrm{DB}$
of chemical graphs in a data base.
Let $\mathrm{DB}_{\mathcal{G}}^{(i)}$ denote
the set of chemical graphs $G\in \mathrm{DB}\cap \mathcal{G}$
such that $n(G)=i$ for each integer $i\geq 1$.
Based on this, we assume that the given lower and upper bounds on
the above descriptors satisfy the following.
For each $\typ\in\{\inn,\ex\}$, 
\begin{align}  
    n^*  \min_{G\in D_{\pi}\cup \mathrm{DB}_{\mathcal{G}}^{(n^*)}}
       \frac{\dg_i^\typ(G)}{n(G)}  
 \leq  \LB_\dg^\typ(i)    
  \leq \UB_\dg^\typ(i) \leq 
  n^*  \max_{G\in D_{\pi}\cup \mathrm{DB}_{\mathcal{G}}^{(n^*)}}
 \frac{\dg_i^\typ(G)}{n(G)},  
 &&   i\in[1,4],  \label{eq:LU_first} \\
   n^*  \min_{G\in D_{\pi}\cup \mathrm{DB}_{\mathcal{G}}^{(n^*)}}
    \frac{\ce_\ta^\typ(G)}{n(G)} 
    \leq \LB_\ce^\typ(\ta) 
    \leq \UB_\ce^\typ(\ta)  \leq 
   n^*   \max_{G\in D_{\pi}\cup \mathrm{DB}_{\mathcal{G}}^{(n^*)}}
  \frac{\ce_\ta^\typ(G)}{n(G)},  
  &&   \ta\in \Lambda,    \label{eq:LU2}   
\end{align}  

\begin{align}   
 (n^*\!-\!1)\!\!\! \min_{G\in D_{\pi}\cup \mathrm{DB}_{\mathcal{G}}^{(n^*)}}
   \frac{\bd_m^\typ(G)}{n(G)\!-\!1} \leq
     \LB_\bd^\typ(m)    \leq  
     \UB_\bd^\typ(m)   \leq 
 (n^*\!-\!1)\!\!\! \max_{G\in D_{\pi}\cup \mathrm{DB}_{\mathcal{G}}^{(n^*)}} 
   \frac{\bd_m^\typ(G)}{n(G)\!-\!1}, &&  m\in [2,3],  \label{eq:LU3} \\
   (n^*\!-\!1)\!\!\! \min_{G\in D_{\pi}\cup \mathrm{DB}_{\mathcal{G}}^{(n^*)}}
     \frac{\ac_{\gamma}^\typ(G)}{n(G)\!-\!1} \leq
      \LB_\ac^\typ(\gamma)   \leq    
      \UB_\ac^\typ(\gamma)   \leq 
     (n^*\!-\!1)\!\!\! \max_{G\in D_{\pi}\cup \mathrm{DB}_{\mathcal{G}}^{(n^*)}}
 \frac{\ac_{\gamma}^\typ(G)}{n(G)\!-\!1},
 && \gamma\in \Gamma,     \label{eq:LU4}  \\
  (n^*\!-\!1)\!\!\!  \min_{G\in D_{\pi}\cup \mathrm{DB}_{\mathcal{G}}^{(n^*)}}
     \frac{\bc_{\mu}^\typ(G)}{n(G)\!-\!1} \leq
     \LB_\bc^\typ(\mu)   \leq  
     \UB_\bc^\typ(\mu)   \leq 
   (n^*\!-\!1)\!\!\!   \max_{G\in D_{\pi}\cup \mathrm{DB}_{\mathcal{G}}^{(n^*)}}
 \frac{\bc_{\mu}^\typ(G)}{n(G)\!-\!1},
 && \mu\in \Bc.   \label{eq:AD_last}  
\end{align}

%
 
\subsection{Construction of Scheme Graph} 
\label{sec:SG}

We infer a subgraph  $H$  such that
 the maximum degree is $\dmax\in \{3,4\}$,     $n(H)=n^*$, 
     $\bh_{k^*}(H)=\bh^*$ 
     and $\bl_{k^*}(H)=\bl^*$.
For this, we first construct the  scheme graph  
$\mathrm{SG}( \dmax, k^*, \bh^*, t^*)$.
We then prepare a binary variable $u(s,i)$ (resp., $v(t,i)$) 
for each vertex $u_{s,i}$ in tree $S_s$ (resp., $v_{t,i}$ in tree $T_t$).  

Recall that when the two end-vertices of edge $a_i=(u_{s,1},u_{s',1}) \in E_B=\{a_1,a_2\ldots,a_{c^*}\}$ 
is connected in a selected subgraph $H$,
either edge $a_i$ is directly used in $H$ or a path $P_i=(u_{s,1},v_{t',1},v_{t'+1,1},\ldots,$ $v_{t'',1},u_{s',1})$
 from $u_{s,1}$ to $u_{s',1}$ visiting some vertices in $P_{t^*}$ is constructed in $H$.
We regard the index $i$ of each edge $a_i\in E_B=\{a_1,a_2\ldots,a_{c^*}\}$
as the ``color'' of the edge, and
 define the color set of $E_B$ to be $[1,c^*]$.
To introduce necessary linear constraints 
that can construct such a path $P_i$ properly   in our MILP,
we assign the color $i$ to the vertices $v_{t',1},v_{t'+1,1},\ldots,$ $v_{t'',1}$ in $P_{t^*}$
when   a path  $P_i=(u_{s,1},v_{t',1},v_{t'+1,1},\ldots,$ $v_{t'',1},u_{s',1})$ is used in $H$.\\

\noindent
{\bf constants: } \\  
 ~~ Integers $\dmax\in \{3,4\}$,
$n^*\geq 3$, $\dia^*\geq 3$, $k^*\geq 1$,  $\bh^*\geq 1$ and $\bl^*\geq 2$;\\  

\noindent
{\bf variables: } \\
~~ $a(i)\in\{0,1\}$,  $i\in E_B$: 
$a(i)$ represents edge $a_i\in E_B$  ($a(i)=1$, $i\in E_B$)    \\
~~~~~~~  
  ($a(i)=1$ $\Leftrightarrow $   edge $a_i$ is  used in  $H$);   \\
~~ $e(s,t),e(t,s)\in\{0,1\}$, $s\in [1,s^*]$, $t\in [1,t^*]$:  
$e(s,t)$ (resp., $e(t,s)$) represents  \\
~~~~~~~  direction  $(u_{s,1}, v_{t,1})$ (resp.,  $(v_{t,1}, u_{s,1})$), where  
  $e(s,t)=1$ (resp., $e(t,s)=1$)  $\Leftrightarrow $   \\
~~~~~~~   edge $u_{s,1},v_{t,1}$   is  used in  $H$   and 
  direction  $(u_{s,1}, v_{t,1})$ (resp.,  $(v_{t,1}, u_{s,1})$) is assigned\\
~~~~~~~   to edge $u_{s,1} v_{t,1}$;   \\
~~ $\chi(t)\in [0,c^*]$, $t\in [1,t^*]$: $\chi(t)$ represents the color $c\in [0,c^*]$
   assigned to vertex $v_{t,1}$ \\
~~~~~~ 
  ($\chi(t)=c$ $\Leftrightarrow $  vertex $v_{t,1}$ is  assigned color $c$,
  where  $\chi(t)=c=0$ iff $v_{t,1}$ is not in $H$);   \\
~~   $\delta_{\mathrm{clr}}(t,c)\in\{0,1\}$,
 $t\in [1,t^*]$, $c\in [0,c^*]$  
  ($\delta_{\mathrm{clr}}(t,c)=1$    $\Leftrightarrow $ $\chi(t)=c$); \\ 
~~ $\mathrm{clr}(c)\in [0,t^*]$, $c\in [0,c^*]$: 
the number of vertices $v_{t,i}$ with color $c$;\\
~~ $\deg^\mathrm{b+}(s)\in [0,4]$, $s\in [1,s^*]$: 
the out-degree of vertex $u_{s,1}$ in the $k^*$-branch-subtree of $H$; \\
~~ $\deg^\mathrm{b-}(s)\in [0,4]$, $s\in [1,s^*]$: 
the in-degree of vertex $u_{s,1}$ in the $k^*$-branch-subtree of $H$; \\

\noindent
{\bf constraints: }   
\begin{align}
\sum_{c\in [0,c^*]} \delta_\mathrm{clr}(t,c)=1,  ~
\sum_{c\in [0,c^*]}c\cdot \delta_\mathrm{clr}(t,c)=\chi(t),  
&& t\in[1,t^*], \label{eq:SG_first} 
\end{align}   

\begin{align}
\sum_{t\in[1,t^*]} \delta_\mathrm{clr}(t,c)=\mathrm{clr}(c), &&  c\in [0,c^*], \label{eq:A} \\ 
t^*(1-a(i))\geq \mathrm{clr}(i), &&  i\in [1,c^*], \label{eq:SG2} \\ 
 e(s,t)+e(t,s)\leq 1, && s\in [1,s^*],  t\in[1,t^*], \label{eq:SG3} 
\end{align}   

\begin{align}
\sum_{s\in [1,s^*]\setminus\{\mathrm{head}(c)\}}\!\!\! \!\!\! 
   e(t,s)\leq 1-\delta_\mathrm{clr}(t,c),
\sum_{s\in [1,s^*]\setminus\{\mathrm{tail}(c)\}}\!\!\! \!\!\! 
  e(s,t)\leq 1-\delta_\mathrm{clr}(t,c),
&&   c\in [1,c^*], t\in [1,t^*],  \label{eq:SG4} 
\end{align}   

\begin{align}
\sum_{ i\in E_B^-(s)}a(i)+ \sum_{t\in [1,t^*]  }\!\!\! e(t,s) = \deg^\mathrm{b-}(s), ~~
\sum_{ i\in E_B^+(s)}a(i)+ \sum_{t\in [1,t^*]  }\!\!\! e(s,t) = \deg^\mathrm{b+}(s),  &&   
  \notag \\ 
 \deg^\mathrm{b-}(s)+\deg^\mathrm{b+}(s)
  \leq \dmax,  &&   s\in [1,s^*].  \label{eq:SG_last}  
\end{align}

\subsection{Selecting a Subgraph} 
\label{sec:SS}
 
From  the  scheme graph  $\mathrm{SG}( \dmax, k^*, \bh^*, t^*)$, 
we select a subgraph  $H$  such that
     $n(H)=n^*$, $\dia(H)=\dia^*$, 
       $\bh_{k^*}(H)=\bh^*$   and     $\bl_{k^*}(H)=\bl^*$. 
 \bigskip
     
\noindent
{\bf constants: } \\ 
~~ Integers  $\dmax\in \{3,4\}$,
$n^*\geq 3$, $\dia^*\geq 3$, 
$k^*\geq 1$, $\bh^*\geq 1$ and $\bl^*\geq 2$; \\
~~ 
Prepare the following:\\
~~~~
 For each tree $S_s=T(\dmax\!-\!1, \dmax\!-\!1, k^*)$, \\
~~~~~~ the set $\mathrm{Cld}_{\mathrm{S}}(i)$ of the indices of children of a vertex $v_i$;\\
~~~~~~ the index $\mathrm{prt}(i)$ of  the parent of a non-root vertex $v_i$;   \\
~~~~~~ the set $\mathrm{Dsn}_S(d)$ of indices $i$ of a vertex $v_i$ whose
depth is $d$;  \\ 
~~~~~~  a proper set 
$P_{\mathrm{prc}}(\dmax\!-\!1, \dmax\!-\!1, k^*)$ of index pairs, \\
~~~~~~ 
where we denote   
$P_{\mathrm{prc}}(\dmax\!-\!1, \dmax\!-\!1, k^*)$ 
by $P_{S,\mathrm{prc}}$; \\
~~~~
 For each tree $T_t=T(\dmax\!-\!2, \dmax\!-\!1, k^*)$, \\
~~~~~~ the set $\mathrm{Cld}_{\mathrm{T}}(i)$ of the indices of children of a vertex $v_i$;\\
~~~~~~ the index $\mathrm{prt}(i)$ of  the parent of a non-root vertex $v_i$;   \\
~~~~~~  a proper set 
$P_{\mathrm{prc}}(\dmax\!-\!2, \dmax\!-\!1, k^*)$ of index pairs, \\
~~~~~~ 
where we denote   
$P_{\mathrm{prc}}(\dmax\!-\!2, \dmax\!-\!1, k^*)$ 
by $P_{T,\mathrm{prc}}$; \\
\noindent
{\bf variables: } \\
~~ $\sigma(s)\in\{0,1\}$, $s\in [1,s^* ]$: 
  ($\sigma(s)=1$$\Leftrightarrow $ vertex $u_{s,1}$
   is a non-leaf $k^*$-branch or a root); \\
~~ $u(s,i)\in\{0,1\}$, $s\in [1,s^* ]$,     $i\in [1,n_\mathrm{tree}^{\mathrm{S}}]$: 
$u(s,i)$ represents vertex $u_{s,i}$   \\
~~~~~~  
  ($u(s,i)=1$ $\Leftrightarrow $ vertex $u_{s,i}$ is 
   used in  $H$ and edge $e'_{s,i}$ $(i\geq 2)$ is used in $H$), \\
~~~~~~
   ($u(s,1)=1$ and $\sigma(s)=0$ $\Leftrightarrow $ 
    vertex $u_{s,1}$ is a leaf $k^*$-branch);   \\
~~ $v(t,i)\in\{0,1\}$, $t\in [1,t^* ]$,     $i\in [1,n_\mathrm{tree}^{\mathrm{T}}]$: 
$v(t,i)$ represents vertex $v_{t,i}$   \\
~~~~~~  
  ($v(t,i)=1$ $\Leftrightarrow $ vertex $v_{t,i}$ is 
   used in  $H$ and edge $e_{t,i}$ $(i\geq 2)$ is used in $H$);   \\
~~ $e(t)\in\{0,1\}$, $t\in [1,t^*+1]$:  $e(t)$ represents edge $e_{t,1}=v_{t-1,1} v_{t,1}$,  \\
~~~~~~ 
where $e_{1,1}$ and $e_{t^*+1,1}$ are fictitious edges 
  ($e(t)=1$ $\Leftrightarrow $   edge $e_{t,1}$ is
   used in  $H$);   \\
 
\noindent
{\bf constraints: } 
\begin{align}   
%
u(s,i)\geq u(s,j), && s\in [1,s^*], (i,j)\in P_{S,\mathrm{prc}}, \label{eq:SS_first}  \\
%
v(t,i)\geq v(t,j), && t\in [1,t^*],  (i,j)\in P_{T,\mathrm{prc}}, \label{eq:SS1}  \\
\sum_{ s\in [1,s^*], i\in[1, \ntreeS]}u(s,i)
+ \sum_{t\in[1,t^*], i\in[1, \ntreeT]}v(t,i)=n^*, && \label{eq:SS2} \\ 
\sum_{i\in[1, \ntreeS]}u(s,i)
\leq 2+ 2 \sum_{j\in \mathrm{Cld}_{\mathrm{S}}(1)}u(s,j),  &&  s\in [1,s^*], \label{eq:SS3} \\ 
\sum_{i\in[1, \ntreeT]}v(t,i)
\leq 2+ 2 \sum_{j\in \mathrm{Cld}_{\mathrm{T}}(1)}v(t,j),  &&  t\in [1,t^*], \label{eq:SS4}  
\end{align}  

\begin{align}   
 e(t+1)+\sum_{s\in [1,s^*]}e(t,s) =v(t,1) , ~~ 
 e(t) +\sum_{s\in [1,s^*]}e(s,t) =v(t,1) ,  &&  \notag \\ 
 \mbox{ ~(where } e(1)=e(t^*+1)=0),  && t\in [1,t^*],  \label{eq:chi_t} 
\end{align}  

\begin{align}    
 \sum_{c\in [1,c^*]} \delta_\mathrm{clr}(t,c) = v(t,1),    && t\in [1,t^*],  \label{eq:chi_t_0} 
\end{align}  

\begin{align}    
 c^*\cdot ( 1-e(t+1)) \geq \chi(t) - \chi(t+1)  \geq   v(t,1) - e(t+1), && t\in[1,t^*-1], \label{eq:edge+color}  
\end{align}  

\begin{align}    
  a(i)+\sum_{t\in [1,t^*]} e(t,i+1)= u(i+1,1), && i\in[1,c^*],   \label{eq:a+clr} 
\end{align}  

\begin{align}   
\sigma(s)=u(s,1)=1, && \mbox{if $u_s$ is the root}, \label{eq:SS5}  \\
\sigma(s)\leq u(s,1), && s\in [1,s^*], \label{eq:SS6}  
\end{align}  

\begin{align}   
(\dmax\!-\!1)\sigma(s)\geq \!\!\! \sum_{s'\in\mathrm{Cld}_B(s)}\!\!\!  u(s',1)\geq 2\sigma(s), ~ 
  \sum_{i\in\mathrm{Dsn}_S(k^*)}\!\!\! u(s,i)\geq u(s,1)-\sigma(s), ~~~~~ \notag \\
 s\in [1,s^*], u_s\neq \mathrm{root},  \label{eq:SS7} 
\end{align}  

\begin{align}   
 \sum_{ s\in [2,s^*] }(u(s,1) -\sigma(s))  = \bl^* , ~~  
 \sum_{s\in V_B(\bh^*)} u(s,1) \geq 1, &&    \label{eq:SS8} 
 \end{align}  

\begin{align}   
 \sum_{s\in V_{B,s^\mathrm{left}}}\!\!\!  u(s,1) +  \!\!\! \sum_{i\in E_{B,s^\mathrm{left}}} \!\!\! \mathrm{clr}(i)
 = \Big\lceil \frac{\dia^*}{2} \Big\rceil  \! - k^*, ~~~ 
 \sum_{s\in V_{B,s^\mathrm{right}}}\!\!\!  u(s,1) 
 +  \!\!\! \sum_{i\in E_{B,s^\mathrm{right}}}\!\!\!  \mathrm{clr}(i)
 = \Big\lfloor \frac{\dia^*}{2} \Big\rfloor \!  - k^*, &&    \label{eq:SS9} 
 \end{align}  

\begin{align}   
 \sum_{i\in V_{B,s}} u(i,1) +  \sum_{i\in E_{B,s}} \mathrm{clr}(i)
 \leq \Big\lfloor \frac{\dia^*}{2} \Big\rfloor  - k^*, &&
  s\in
 L_B\setminus\{s^\mathrm{left},s^\mathrm{right}\}.   \label{eq:SS_last}  
\end{align}  

Constraints (\ref{eq:SS3}) and 
(\ref{eq:SS4}) represent an extension of the constraint
 (\ref{eq:fringe-size}) on the size of 2-fringe-tree
  to the case of the general branch-parameter $k^*$. 

\subsection{Assigning Multiplicity}  
\label{sec:AM}
  
 We prepare an integer variable $\widetilde{\beta}(e)$ or  $\widehat{\beta}(e)$
 for each edge $e$ in the scheme graph
$\mathrm{SG}( \dmax, k^*, \bh^*, t^*)$   
 to denote the multiplicity of $e$ in a selected graph $H$ and
 include necessary constraints for the variables to satisfy in $H$. 
 \medskip

\noindent
{\bf constants: } \\ 
~~ Prepare functions $\mathrm{tail}$ and $\mathrm{head}$
such that $a_{i}=(u_{\mathrm{tail}(i)}, u_{\mathrm{head}(i)})\in E_B$;\\
~~ Assume that each edge  in a tree $S_s$, $s\in [1,s^*]$
(resp., $T_t$, $t\in [1,t^*]$) is denoted by \\
~~~  $e'_{s,i}$ (resp., $e_{t,i}$) 
with the integer $i\in [2,n_\mathrm{tree}^{\mathrm{S}}]$ of the head $u_{s,i}$
(resp., $v_{t,i}$) of the edge.\\

\noindent
{\bf variables: } \\
~~  $\widetilde{\beta}(i)\in [0,3]$, $i\in [1,c^*]$: 
 $\widetilde{\beta}(i)$ represents the   multiplicity of   edge $a_{i}$, \\
~~~~~~~  where $\widetilde{\beta}(i)=0$   if edge $a_{i}$ is  not 
 in an inferred chemical graph $G$;  \\  
~~  $\widetilde{\beta}(p,i)\in [0,3]$, $p\in [1,s^*\!+\! t^*]$,  $i\in [2,n_\mathrm{tree}^{\mathrm{S}}]$: 
 $\widetilde{\beta}(p,i)$ with $p\leq s^*$ (resp., $p>s^*$)
 represents \\
~~~~~~~  the   multiplicity of   edge $e'_{p,i}$ (resp., $e_{p-s^*,i}$);  \\ 
~~  $\widetilde{\beta}(t,1)\in [0,3]$, $t\in [1, t^*+1]$: 
 $\widetilde{\beta}(t,1)$  
 represents the   multiplicity of   edge $e_{t,1}$;  \\ 
  ~~  $\widehat{\beta}(s,t)\in [0,3]$, $s\in [1,s^*]$,  $t\in [1,t^*]$:
  $\widehat{\beta}(s,t)$  represents
  the   multiplicity of   edge $u_{s,1}v_{t,1}$;  \\

\noindent
{\bf constraints: } 
\begin{align}  
%
  a(i)\leq \widetilde{\beta}(i)\leq 3 a(i), 
  && i\in [1,c^*],  \label{eq:AM_first} \\ 
  u(s,i) \leq \widetilde{\beta}(s,i)\leq 3 u(s,i), 
  && s\in [1,s^*], i\in [2,n_\mathrm{tree}^{\mathrm{S}}],  \label{eq:AM2}  \\
  v(t,i)\leq \widetilde{\beta}(s^*\!+\! t,i)\leq 3 v(t,i), 
  && t\in [1,t^*], i\in [2,n_\mathrm{tree}^{\mathrm{T}}],  \label{eq:AM3}  \\
  e(t)\leq \widetilde{\beta}(t,1)\leq 3 e(t), 
  && t\in [1,t^*+1],  \label{eq:AM4} \\
  e(s,t)+e(t,s)\leq \widehat{\beta}(s,t)\leq 3  e(s,t)+3 e(t,s) , 
  && s\in [1,s^*], t\in [1,t^*]. \label{eq:AM_last}   
\end{align}
 
\subsection{Assigning Chemical Elements and  Valence Condition} 
\label{sec:AE}

We include constraints so that each vertex $v$ in a selected graph $H$
satisfies the valence condition; i.e., $\sum_{uv\in E(H)}\beta(uv)\leq  \val(\alpha(u))$. 
With these constraints, a chemical acyclic graph
 $G=(H,\alpha,\beta)$ on a selected subgraph $H$
will be constructed. 

 \medskip
\noindent
{\bf constants: } \\ 
~~ A set $\Lambda\cup\{\epsilon\}$ of chemical elements, where $\epsilon$ denotes null;\\
~~ A coding $[\ta]$, $\ta\in \Lambda\cup\{\epsilon\}$ such that
    $[\epsilon]=0$; 
    $[\ta]\geq 1$,  $\ta\in \Lambda$; and 
      $[\ta]\neq [{\tt b}]$ if  $\ta\neq {\tt b}$; \\
~~~~~ 
      Let $[\Lambda]$ and $[\Lambda\cup\{\epsilon\}]$ denote 
      $\{[\ta]\mid \ta\in \Lambda\}$ and 
        $\{[\ta]\mid \ta\in \Lambda\cup\{\epsilon\}\}$,
       respectively;  \\
~~ A valence function: $\val: \Lambda\to [1,4]$; \\
~~ Let $E_B(s)$ denote the set of indices $i$ of all edges $a_i\in E_B$
adjacent to vertex $u_{s,1}$ in $T_B$.
      
\noindent
{\bf variables: } \\
~~ $\widetilde{\alpha}(p,i)\in [\Lambda\cup\{\epsilon\}]$, 
 $p\in [1,s^*\!+\! t^*]$, $i\in [1,n_\mathrm{tree}^{\mathrm{S}}]$: \\
 ~~~~~~~~ 
$\widetilde{\alpha}(p,i)$ with $p\leq s^*$ (resp., $p>s^*$)
 represents $\alpha(u_{p,i})$  (resp.,  $\alpha(v_{p-s^*,i})$); \\
 ~~  $\delta_{\alpha}(p,i,\ta)\in\{0,1\}$, $p\in [1,s^*\!+\! t^*]$,
  $i\in [1,n_\mathrm{tree}^{\mathrm{S}}]$, $\ta\in \Lambda\cup \{\epsilon\}$:   \\
~~~~~~ 
     $\delta_{\alpha}(p,i,\ta)=1$  $\Leftrightarrow$ 
      $\alpha(u_{p,i})=\ta$ for $p\leq s^*$ and $\alpha(v_{p-s^*,i})=\ta$ for $p> s^*$;   \\  
~~ 
   $\delta_{\widetilde{\beta}}(i,m)\in\{0,1\}$, $p\in [1,s^*\!+\! t^*]$,  
    $i\in [1,c^*]$, $m\in[0,3]$:   \\
~~~~~~     
      $\delta_{\widetilde{\beta}}(i,m)=1$ $\Leftrightarrow$
       the multiplicity of edge $a_i$ in an inferred chemical graph $G$ is $m$;  \\
~~ 
   $\delta_{\widetilde{\beta}}(p,i,m)\in\{0,1\}$, $p\in [1,s^*\!+\! t^*]$,  
    $i\in [2,n_\mathrm{tree}^{\mathrm{S}}]$, $m\in[0,3]$:   \\
~~~~~~     
      $\delta_{\widetilde{\beta}}(p,i,m)=1$ $\Leftrightarrow$
       the multiplicity of edge $e'_{p,i}$, $p\leq s^*$ (or $e_{p-s^*,i}$, $p>s^*$) in $G$ is $m$;  \\
~~ 
   $\delta_{\widetilde{\beta}}(t,1,m)\in\{0,1\}$, $t\in [1, t^*+1]$,     $m\in[0,3]$:   \\
~~~~~~     
      $\delta_{\widetilde{\beta}}(t,1,m)=1$ $\Leftrightarrow$
       the multiplicity of edge  $e_{t}$  in $G$ is $q$;  \\
~~   $\delta_{\widehat{\beta}}(s,t,m)\in\{0,1\}$, $s\in [1,s^*]$, $t\in [1,t^*]$, 
    $m\in[0,3]$:  \\
~~~~~~ 
      $\delta_{\widehat{\beta}}(s,t,m)=1$ $\Leftrightarrow$
       the multiplicity of edge $u_{s,1}v_{t,1}$  in $G$ is $m$;  \\
      
\noindent
{\bf constraints: } \\
\begin{align} 
 \sum_{\ta\in \Lambda\cup \{\epsilon\}} \delta_{\alpha}(p,i,\ta)=1, 
 &&    p\in [1,s^*\!+\! t^*],    i\in [1,n_\mathrm{tree}^{\mathrm{S}}],   \label{eq:AE_first} \\ 
 \sum_{\ta\in \Lambda\cup \{\epsilon\}}[\ta]\cdot \delta_{\alpha}(p,i,\ta)
      =\widetilde{\alpha}(p,i),      &&  p\in [1,s^*\!+\! t^*],    i\in [1,n_\mathrm{tree}^{\mathrm{S}}],  \label{eq:AE2} 
\end{align}  

\begin{align}
 \sum_{m\in[0,3]} \delta_{\widetilde{\beta}}(i,q)=1,  ~~~ 
 \sum_{m\in[1,3]} m\cdot\delta_{\widetilde{\beta}}(i,m)=\widetilde{\beta}(i),  
&& i\in [1,c^*],  \label{eq:AE3}
\end{align}  

\begin{align}
 \sum_{m\in[0,3]} \delta_{\widetilde{\beta}}(p,i,m)= 1,  ~~
 \sum_{m\in[1,3]} m\cdot\delta_{\widetilde{\beta}}(p,i,m)=\widetilde{\beta}(p,i), 
 p\in [1,s^*\!+\! t^*],    i\in [2,n_\mathrm{tree}^{\mathrm{S}}],  \label{eq:AE4}
 \end{align}  

\begin{align}
 \sum_{m\in[0,3]} \delta_{\widetilde{\beta}}(t,1,q)= 1, ~
 \sum_{m\in[1,3]} m \cdot\delta_{\widetilde{\beta}}(t,1,m)= \widetilde{\beta}(t,1), &&
   t\in [1,t^*+1],     \label{eq:AE5} 
\end{align}  

\begin{align}
 \sum_{m\in[0,3]} \delta_{\widehat{\beta}}(s,t,m)= 1, ~
 \sum_{m\in[0,3]} m\delta_{\widehat{\beta}}(s,t,m)= \widehat{\beta}(s,t), &&
   s\in [1,s^*],   t\in [1,t^*], \label{eq:AE6}  
\end{align}   

\begin{align} 
\sum_{i\in E_B(s)} \widetilde{\beta}(i)  
+\sum_{t\in[1,t^*]}\widehat{\beta}(s,t)   
+\sum_{j\in\mathrm{Cld}_{\mathrm{S}}(1)}\widetilde{\beta}(s,j)
 \leq \sum_{\ta\in \Lambda}\val(\ta) 
  \cdot \delta_{\alpha}(s,1,\ta),  && s\in [1,s^*],  \label{eq:AE7} 
\end{align}  

\begin{align}
\sum_{s\in[1,s^*]}\widehat{\beta}(s,t)  
+\widetilde{\beta}(t,1) + \widetilde{\beta}(t\!+\!1,1)  
+\!\!\! \sum_{j\in\mathrm{Cld}_{\mathrm{T}}(1)}\!\!\!  \widetilde{\beta}(s^*\!+\! t,j)
 \leq \sum_{\ta\in \Lambda}\val(\ta) 
   \delta_{\alpha}(s^*\!+\! t,1,\ta),  &&  t\in [1,t^*],    \label{eq:AE8} 
\end{align}  

\begin{align}
\widetilde{\beta}(s,i) 
+\sum_{j\in\mathrm{Cld}_{\mathrm{S}}(i)}\widetilde{\beta}(s,j)  
 \leq \sum_{\ta\in \Lambda}\val(\ta)  \delta_{\alpha}(s,i,\ta), 
  &&  s\in [1,s^*], i\in [2, n_\mathrm{tree}^{\mathrm{S}}], \label{eq:AE9} 
\end{align}  

\begin{align}
\widetilde{\beta}(s^*\!+\!t,i) 
+\sum_{j\in\mathrm{Cld}_{\mathrm{T}}(i)}\widetilde{\beta}(s^*\!+\!t,j)  
 \leq \sum_{\ta\in \Lambda}\val(\ta) 
 \delta_{\alpha}(s^*\!+\!t,i,\ta), 
  &&  t\in [1,t^*], i\in [2, n_\mathrm{tree}^{\mathrm{T}}]. \label{eq:AE_last} 
\end{align}

\subsection{Descriptors on Mass, the Numbers of Elements and Bonds} 
\label{sec:NE}

We include constraints to compute descriptors $\overline{\mathrm{ms}}(G)$, 
$\ce_\ta(G)$ ($\ta\in \Lambda)$, 
$\bd_m(G)$ ($m\in [2,3]$) and $n_{\tt H}(G)$ 
according to the definitions in Section~\ref{sec:chemical_model}.

 \medskip
\noindent
{\bf constants: } \\ 
~~ A function $\mathrm{mass}^*:\Lambda\to \mathbb{Z}$ 
(we let $\mathrm{mass}(\ta)$ denote  the observed mass of a chemical element \\
~~~~ 
$\ta\in \Lambda$, and define 
   $\mathrm{mass}^*(\ta)=\lfloor 10\cdot \mathrm{mass}(\ta)\rfloor$); \\

\noindent
{\bf variables: } \\ 
~~ $\mathrm{Mass}\in \mathbb{Z}$: 
$\mathrm{Mass}$ represents $\sum_{v\in V} \mathrm{mass}^*(\alpha(v))$; \\
~~ $\bd(m)\in[0,2n^*]$, $m\in[1,3]$; \\ 
~~ $\mathrm{n}_{\tt H}\in [0,4n^*]$: the number $n_{\tt H}(G)$
   of hydrogen atoms  to be included to  $G$; \\

\noindent
{\bf constraints: } 
\begin{align}
 \sum_{p\in [1,s^*\!+\! t^*]} \delta_{\alpha}(p,1,\ta) =  \ce^\inn(\ta), ~~~  
 \sum_{p\in [1,s^*\!+\! t^*],  
                 i\in [2, n_\mathrm{tree}^{\mathrm{S}}]  }\!\!\!  \delta_{\alpha}(p,i,\ta)=
  \ce^\ex(\ta), &&    \ta\in\Lambda,   \label{eq:NE_first} 
\end{align}  

\begin{align}   
\sum_{\ta\in\Lambda}\mathrm{mass}^*(\ta)
  (\ce^\inn(\ta)+\ce^\ex(\ta)) =\mathrm{Mass}, &&    \label{eq:NE2}  
\end{align}  

\begin{align}   
 \sum_{i\in [1,c^*]} \delta_{\widetilde{\beta}}(i,q)
 +  \sum_{s\in [1,s^*], t\in [1,t^*]} \delta_{\widehat{\beta}}(s,t,q)  
 +    \sum_{ t\in [2,t^*]}\delta_{\widetilde{\beta}}(t,1,q)
    = \bd^\inn(m),  &&   m\in[1,3],  \label{eq:NE3} 
\end{align}  

\begin{align}   
   \sum_{p\in [1,s^*\!+\! t^*],  
                 i\in [2, n_\mathrm{tree}^{\mathrm{S}}]  } \delta_{\widetilde{\beta}}(p,i,m)
                 = \bd^\ex(m),  &&   m\in[1,3],  \label{eq:NE4}   
%
\end{align}  

\begin{align}   
\sum_{\ta\in \Lambda}\val(\ta) (\ce^\inn(\ta) + \ce^\ex(\ta) )  
   - 2(n^*-1+\bd^\inn(2)+\bd^\ex(2)
   + 2\bd^\inn(3) +  2\bd^\ex(3) ) =\mathrm{n}_{\tt H}.  
   \label{eq:NE_last} 
\end{align}  
\subsection{Descriptor for the  Number of Specified Degree} 
\label{sec:ND}

We include constraints to compute descriptors $\dg_i(G)$ ($i\in[1,4]$) 
according to the definitions in Section~\ref{sec:chemical_model}.
We also add constraints so that the maximum degree of a  vertex in $H$
is at most 3 (resp., equal to 4) when $\dmax=3$ (resp., $\dmax=4)$. 
 
\noindent
{\bf variables: } \\
~~    $\deg(p,i)\in [0,4]$, 
 $p\in [1,s^*\!+\! t^*]$, $i\in [1,n_\mathrm{tree}^{\mathrm{S}}]$: \\
~~~~~~  $\deg(p,i)$ represents $\deg_H(u_{p,i})$ for $p\leq s^*$
 or $\deg_H(v_{p-s^*,i})$ for $p> s^*$; \\   
~~   $\delta_{\deg}(p,i,d)\in\{0,1\}$, $p\in [1,s^*\!+\! t^*]$, 
$i\in [1,n_\mathrm{tree}^{\mathrm{S}}]$,
  $d\in[0,4]$:   \\
~~~~~~ 
       $\delta_{\deg}(p,i,d)=1$ $\Leftrightarrow$ 
       $\deg(p,i)=d$; \\

\noindent
{\bf constraints: } 
\begin{align} 
\sum_{ i\in E_B(s)}a(i)+ \sum_{t\in [1,t^*]  }(e(s,t)+e(t,s)) 
+ \sum_{j\in \mathrm{Cld}_{\mathrm{S}}(1)}u(s,j) 
=\deg(s,1),  &&   s\in [1,s^*],  \label{eq:ND_first} 
\end{align}  

\begin{align} 
u(s,i)+ \sum_{j\in \mathrm{Cld}_{\mathrm{S}}(i)}u(s,j)=\deg(s,i), 
&&  s\in[1,s^*], i\in [2,n_\mathrm{tree}^{\mathrm{S}}], \label{eq:ND2} \\ 
2v(t,1) + \sum_{j\in \mathrm{Cld}_{\mathrm{T}}(1)}v(t,j)=\deg(s^*\!+\! t,1), && t\in[1,t^*], \label{eq:A_0} \\
v(t,i)+ \sum_{j\in \mathrm{Cld}_{\mathrm{T}}(i)}v(t,j)=\deg(s^*\!+\! t,i), 
&&  t\in[1,t^*], i\in [2,n_\mathrm{tree}^{\mathrm{T}}], \label{eq:ND3} 
\end{align}  

\begin{align} 
  \sum_{d\in [0,4]} \delta_{\deg}(p,i,d)= 1, ~~
  \sum_{d\in [1,4]}d\cdot \delta_{\deg}(p,i,d)=  \deg(p,i),
  &&   p\in[1,s^*\!+\! t^*], i\in[1,\ntreeS],    \label{eq:ND4} 
\end{align}  

\begin{align} 
 \sum_{p\in[1,s^*\!+\! t^*]}\delta_{\deg}(p,1,d) = \dg^\inn(d), ~~
 \sum_{p\in[1,s^*\!+\! t^*], i\in[2,\ntreeS]}\delta_{\deg}(p,i,d) 
 = \dg^\ex(d), &&   d\in [1,4], \label{eq:ND5}
\end{align}  

\begin{align} 
\dg^\inn(4)+ \dg^\ex(4)\geq 1 \mbox{   (resp., $= 0$)}
    &&  \mbox{ when $\dmax=4$ (resp., $=3$). }    \label{eq:ND_last}  
\end{align}

\subsection{Descriptor for the Number of Adjacency-configurations} 
\label{sec:NAC}

We include constraints to compute descriptors 
$\ac_{\gamma}(G)$ ($\gamma=({\tt a,b},m)\in \Gamma$) 
according to the definitions in Section~\ref{sec:chemical_model}.

 \medskip
\noindent
{\bf constants: } \\ 
~~ A set $\Gamma=\Gamma_{<}\cup\Gamma_{=}\cup \Gamma_{>}$ 
of proper tuples $({\tt a,b},m)\in \Lambda\times  \Lambda\times [1,3]$; \\
~~ The set $\Gamma_0=\{({\tt a,b},0)\mid {\tt a,b}\in \Lambda\cup\{\epsilon\}\}$; \\

\noindent
{\bf variables: } \\
~~   $\delta_{\tau}(i,\gamma )\in\{0,1\}$,   $i\in [1,c^*]$, 
    $\gamma\in \Gamma\cup\Gamma_0 $:  \\
~~~~~~       $\delta_{\tau}(i, \gamma)=1$ $\Leftrightarrow$  
       edge $a_{i}$ is assigned tuple $\gamma$; i.e., 
     $\gamma=(\widetilde{\alpha}(\mathrm{tail}(i),1),
      \widetilde{\alpha}(\mathrm{head}(i),1),
      \widetilde{\beta}(i))$; \\
~~   $\delta_{\tau}(t,1,\gamma )\in\{0,1\}$,  $t\in [2,t^*]$,
         $\gamma\in \Gamma\cup\Gamma_0 $:  \\
~~~~~~
       $\delta_{\tau}(t,1, \gamma)=1$ $\Leftrightarrow$ 
      edge $e_{t,1}$  is assigned tuple   $\gamma$; i.e., 
     $\gamma=(\widetilde{\alpha}(s^*\!+\! t-1,1), \widetilde{\alpha}(s^*\!+\! t,1),
      \widetilde{\beta}(t,1))$; \\
~~   $\delta_{\tau}(p,i,\gamma )\in\{0,1\}$,  $p\in [1,s^*\!+\! t^*]$, 
$i\in [2,n_\mathrm{tree}^{\mathrm{S}}]$, 
   $\gamma\in \Gamma\cup\Gamma_0 $:   \\
~~~~~~ 
       $\delta_{\tau}(p,i, \gamma)=1$ $\Leftrightarrow$ 
       edge $e'_{p,i}$, $p\leq s^*$ (or  $e_{p-s^*,i}$, $p> s^*$) is assigned tuple 
     $\gamma$; i.e., \\ ~~~~~~~~
     $\gamma=(\widetilde{\alpha}(p,\mathrm{prt}(i)), \widetilde{\alpha}(p,i),
      \widetilde{\beta}(p,i))$; \\
~~   $\delta_{\widehat{\tau}}(s,t,\gamma)\in\{0,1\}$,  
   $s\in [1,s^*]$, $t\in [1,t^*]$,  $\gamma\in \Gamma\cup\Gamma_0 $:   \\
~~~~~~ 
       $\delta_{\widehat{\tau}}(s,t, \gamma)=1$ $\Leftrightarrow$ 
       edge $u_{s,1}v_{t,1}$ is assigned tuple 
     $\gamma$; i.e., 
     $\gamma=(\widetilde{\alpha}(s,1), \widetilde{\alpha}(s^*\!+\! t,1),
      \widehat{\beta}(s,t))$; \\ 

\noindent
{\bf constraints: } 
\begin{align} 
 \sum_{\gamma\in \Gamma \cup\Gamma_0 } 
 \delta_{\tau}(i,\gamma)= 1,   ~~~ 
 \sum_{({\tt a,b},m)\in \Gamma\cup\Gamma_0 }\!\!\!  
  [\ta]\delta_{\tau}(i,({\tt a,b},m))= \widetilde{\alpha}(\mathrm{tail}(i),1),    
 && \notag \\
 \sum_{({\tt a,b},m)\in \Gamma\cup\Gamma_0  } \!\!\! 
  [{\tt b}]\delta_{\tau}(i,({\tt a,b},m))= \widetilde{\alpha}(\mathrm{head}(i),1),  ~~~ 
 \sum_{({\tt a,b},m)\in \Gamma\cup\Gamma_0 }\!\!\!  
  m\cdot \delta_{\tau}(i,({\tt a,b},m))=\widetilde{\beta}(i), 
  && i\in [1,c^*],  \label{eq:NAC_first}  
  \end{align}  

\begin{align} 
 \sum_{\gamma\in \Gamma \cup\Gamma_0 }  \delta_{\tau}(t,1,\gamma)=1 ,  ~~~
  \sum_{({\tt a,b},m)\in \Gamma\cup\Gamma_0   }\!\!\! 
   [\ta]  \delta_\tau(t,1,({\tt a,b},m) )=  \widetilde{\alpha}(s^*\!+\! t-1,1), 
    &&   \notag \\
  \sum_{({\tt a,b},m)\in \Gamma\cup\Gamma_0   }\!\!\!\!\!\! 
   [{\tt b}]  \delta_\tau(t,1,({\tt a,b},m) )=  \widetilde{\alpha}(s^*\!+\! t,1), ~~ 
\sum_{({\tt a,b},m)\in \Gamma\cup\Gamma_0   }\!\!\!\!\!\! 
  m\cdot \delta_\tau(t,1,({\tt a,b},m) )=  \widetilde{\beta}(t,1), &&
    t\in [2,t^*],    \label{eq:NAC2}  
\end{align} 

\begin{align} 
 \sum_{\gamma\in \Gamma \cup\Gamma_0 }  \delta_{\tau}(p, i,\gamma)=1 , ~~ 
  \sum_{({\tt a,b},m)\in \Gamma\cup\Gamma_0   }\!\!\! 
  [\ta]  \delta_\tau(p,i, ({\tt a,b},m) )= \widetilde{\alpha}(p,\mathrm{prt}(i)), 
   \hspace{1cm}   \notag \\ 
   \sum_{({\tt a,b},m)\in \Gamma\cup\Gamma_0   }\!\!\! 
  [{\tt b}]  \delta_\tau(p,i, ({\tt a,b},m) )= \widetilde{\alpha}(p,i),  ~~ 
\sum_{({\tt a,b},m)\in \Gamma\cup\Gamma_0   }\!\!\! 
  m\cdot \delta_\tau(p,i,({\tt a,b},m) )=  \widetilde{\beta}(p,i),  \hspace{1cm}   \notag \\ 
   p\in [1,s^*\!+\! t^*], i\in [2, \ntreeS],  \label{eq:NAC3} 
\end{align}  

\begin{align}
 \sum_{\gamma \in \Gamma\cup\Gamma_0  }  \delta_{\widehat{\tau}}(s,t,\gamma)= 1, ~~
 \sum_{({\tt a,b},m)\in \Gamma\cup\Gamma_0  }\!\!\!
   [\ta]\delta_{\widehat{\tau}}(s,t,({\tt a,b},m))= \widetilde{\alpha}(s,1),  \hspace{1cm}   \notag \\ 
 \sum_{({\tt a,b},m)\in \Gamma\cup\Gamma_0  }\!\!\!
   [{\tt b}]\delta_{\widehat{\tau}}(s,t,({\tt a,b},m))= \widetilde{\alpha}(s^*\!+\! t,1), ~~
 \sum_{({\tt a,b},m)\in \Gamma\cup\Gamma_0  }\!\!\!
  m\cdot \delta_{\widehat{\tau}}(s,t,({\tt a,b},m))= \widehat{\beta}(s,t), \hspace{1cm}   \notag \\ 
    s\in [1,s^*], t\in [1,t^*],  \label{eq:NAC4}      
\end{align}

\begin{align}
\sum_{i\in [1,c^*]}
    (\delta_\tau(i, \gamma)+\delta_\tau(i, \overline{\gamma} ) )  
+\sum_{s\in [1,s^*], t\in [1,t^*]}
    (\delta_{\widehat{\tau}}(s,t, \gamma)
    +\delta_{\widehat{\tau}}(s,t, \overline{\gamma} ) )  && \notag \\
+\sum_{t\in [2,t^*]}
    (\delta_\tau(t,1, \gamma)+\delta_\tau(t,1, \overline{\gamma} ) )
        = \ac^\inn(\gamma),  &&  
   \gamma\in \Gamma_{<},   \label{eq:NAC5} 
\end{align}  

\begin{align}
\sum_{i\in [1,c^*]}
    \delta_\tau(i, \gamma)  
+\sum_{s\in [1,s^*], t\in [1,t^*]}\delta_{\widehat{\tau}}(s,t, \gamma) 
+ \sum_{t\in [2,t^*]}     \delta_\tau(t,1, \gamma) 
        = \ac^\inn(\gamma),   
    &&    \gamma\in \Gamma_{=},  \label{eq:NAC6} 
\end{align}  

\begin{align}
\sum_{p\in [1,s^*\!+\! t^*],  
                 i\in [2, n_\mathrm{tree}^{\mathrm{S}}] }
    (\delta_\tau(p,i, \gamma)+\delta_\tau(p,i, \overline{\gamma} ) )
        = \ac^\ex(\gamma), 
    &&    \gamma\in \Gamma_{<},  \label{eq:NAC7} 
\end{align}  

\begin{align}
\sum_{p\in [1,s^*\!+\! t^*],  
                 i\in [2, n_\mathrm{tree}^{\mathrm{S}}]  }
     \delta_\tau(p,i, \gamma)   = \ac^\ex(\gamma), 
    &&    \gamma\in \Gamma_{=}.  \label{eq:NAC_last}
%
\end{align}

\subsection{Descriptor for Bond-configuration}  
\label{sec:NBC}
 
We include constraints to compute descriptor for bond-configuration 
$\bd_{\mu}(G)$, $\mu\in \Bc$  according to the definition.

 \medskip
\noindent
{\bf variables: } \\
~~   $\bc(\mu)\in[0,n^*-1]$, $\mu\in \Bc$; \\   
~~   $\delta_{\mathrm{dc}}(i,d,d',m)\in\{0,1\}$,
 $i\in [1,c^*]$,   $d,d'\in[0, 4]$, $m\in [0,3]$:  \\
~~~~~~         $\delta_{\mathrm{dc}}(i,d,d',m)=1$ $\Leftrightarrow$  
   $\deg_H(u_{\tail(i)})=d$,   
    $\deg_H(u_{\head(i)})=d'$ and  $\beta(a_i)=m\in[1,3]$ in $G$;  \\
~~   $\delta_{\mathrm{dc}}(t,1,d,d',m)\in\{0,1\}$,
 $t\in [2,t^*]$,   $d,d'\in[0, 4]$, $m\in [0,3]$:   $\delta_{\mathrm{dc}}(t,1,d,d',m)=1$ $\Leftrightarrow$       \\
~~~~~~     
   $\deg_H(v_{t-1,1})=d$, $\deg_H(v_{t,1})=d'$ and 
     $\beta(e_{t,1})=m\in[1,3]$ in $G$; \\
~~   $\delta_{\mathrm{dc}}(p,i,d,d',m)\in\{0,1\}$,
 $p\in [1,s^*\!+\! t^*]$, $i\in [2,n_\mathrm{tree}^{\mathrm{S}}]$,   
  $d,d'\in[0, 4]$, $m\in [0,3]$:  \\
~~~~~~ 
      $\delta_{\mathrm{dc}}(p,i,d,d',m)=1$ $\Leftrightarrow$     
   $\deg_H(u_{p,\mathrm{prt}(i)})=d$
    $\deg_H(u_{p,i})=d'$ and $\beta(e'_{p,i})=m\in[1,3]$  for $p\leq s^*$   \\
~~~~~~  
   (or 
   $\deg_H(v_{p-s^*,\mathrm{prt}(i)})=d$, 
    $\deg_H(v_{p-s^*,i})=d'$ 
     and $\beta(e_{p-s^*,i})=m\in[1,3]$  for $p> s^*$) in $G$; \\
~~   $\delta_{\widehat{\mathrm{dc}}}(s,t,d,d',m)\in\{0,1\}$,
 $s\in[1,s^*]$, $t\in [1,t^*]$,  $d,d'\in[0, 4]$, $m\in [0,3]$:   \\
~~~~~~ 
       $\delta_{\widehat{\mathrm{dc}}}(s,t,d,d',1)=1$ $\Leftrightarrow$ 
  $\deg_H(u_{s,1})=d$,   $\deg_H(v_{t,1})=d'$ 
  and $\beta(u_{s,1}v_{t,1})=m\in[1,3]$ in $G$; \\

\noindent
{\bf constraints: }   
\begin{align}  
 \sum_{d,d'\in [0, 4],m\in[0,3]}  \delta_{\mathrm{dc}}(i,d,d',m)=1, ~~ 
 \sum_{d,d'\in [0, 4],m\in[0,3] }\!\!\!  m\cdot \delta_{\mathrm{dc}}(i,d,d',m)=\widetilde{\beta}(i), 
&& \notag \\
 \sum_{d\in[1,4],d'\in[0,4],m\in[0,3]}\!\!\! 
  d\cdot \delta_{\mathrm{dc}}(i,d,d',m)=\deg(\mathrm{tail}(i),1), && \notag \\
 \sum_{d\in[0,4],d'\in[1,4],m\in[0,3]}\!\!\! 
  d'\cdot \delta_{\mathrm{dc}}(i,d,d',m)=\deg(\mathrm{head}(i),1), 
  &&   i\in [1,c^*],    \label{eq:NBC_first} 
\end{align}

\begin{align}
 \sum_{d,d'\in [0, 4],m\in[0,3] }  \delta_{\mathrm{dc}}(t,1,d,d',m)= 1, ~~ 
 \sum_{d,d'\in [0, 4],m\in[0,3] }\!\!\!  m\cdot \delta_{\mathrm{dc}}(t,1,d,d',m)
   =\widetilde{\beta}(t,1),  && \notag \\
 \sum_{d\in[1,4],d'\in[0,4],m\in[0,3]}\!\!\! d\cdot \delta_{\mathrm{dc}}(t,1,d,d',m)
 =\deg(s^*\!+\! t-1,1),  && \notag \\
 \sum_{d\in[0,4],d'\in[1,4],m\in[0,3]}\!\!\!  d'\cdot \delta_{\mathrm{dc}}(t,1,d,d',m)
 =\deg(s^*\!+\! t,1), 
  &&   t\in [2,t^*], \label{eq:NBC2}  
\end{align}  
  
\begin{align} 
 \sum_{d,d'\in [0, 4],m\in[0,3] }  \delta_{\mathrm{dc}}(p,i,d,d',m)= 1, 
  &&   p\in [1,s^*\!+\! t^*],   i\in [2, \ntreeS],    \label{eq:NBC3} \\ 
 \sum_{d,d'\in [0, 4],m\in[0,3] }\!\!\!  m\cdot  \delta_{\mathrm{dc}}(s,i,d,d',m) 
    =\widetilde{\beta}(s,i), 
  &&   s\in [1,s^*],  i\in [2,n_\mathrm{tree}^{\mathrm{S}}],    \label{eq:NBC4} \\
 \sum_{d,d'\in [0, 4],m\in[0,3] }\!\!\!  m\cdot  \delta_{\mathrm{dc}}(s^*\!+\! t,i,d,d',m) 
   =\widetilde{\beta}(s^*\!+\! t,i), 
  &&   t\in [1,t^*],  i\in [2,n_\mathrm{tree}^{\mathrm{T}}],    \label{eq:NBC5}
\end{align}  
  
\begin{align} 
 \sum_{d\in[1,4],d'\in[0,4],m\in[0,3]}\!\!\! d \cdot \delta_{\mathrm{dc}}(p,i,d,d',m)
 =\deg(p,\mathrm{prt}(i)),  && \notag \\ 
 \sum_{d\in[0,4],d'\in[1,4],m\in[0,3]} d'\cdot \delta_{\mathrm{dc}}(t,i,d,d',m)
 =\deg(p, i), 
  && p\in [1,s^*\!+\! t^*],  i\in [2,n_\mathrm{tree}^{\mathrm{S}}],    \label{eq:NBC6} 
\end{align}  
  
\begin{align} 
 \sum_{d,d'\in [1, 4], m\in[0,3] }  \delta_{\widehat{\mathrm{dc}}}(s,t,d,d',m)=1,  && \notag \\
 \sum_{d,d'\in [1, 4], m\in[0,3] }
   m\cdot \delta_{\widehat{\mathrm{dc}}}(s,t,d,d',m)=\widehat{\beta}(s,t),  && \notag \\
 \sum_{d\in[1,4],d'\in[0,4], m\in[0,3]}d \cdot \delta_{\widehat{\mathrm{dc}}}(s,t,d,d',m)
 =\deg(s,1),   && \notag \\
 \sum_{d\in[0,4],d'\in[1,4], m\in[0,3]} d' \cdot \delta_{\widehat{\mathrm{dc}}}(s,t,d,d',m)
 =\deg(s^*\!+\! t,1),  
  &&    s\in [1,s^*],  t\in [1,t^*],    \label{eq:NBC7}  
\end{align}  

\begin{align}   
 \sum_{i\in [1,c^*]}\!(\delta_{\mathrm{dc}}(i,d,d',m) 
 \!+\!  \delta_{\mathrm{dc}}(i,d',d,m)) 
  +  \!\!\! 
 \sum_{t\in [2,t^*]}(\delta_{\mathrm{dc}}\!(t,1,d,d',m) 
  \!+\! \delta_{\mathrm{dc}}(t,1,d',d,m) )
\hspace{2cm}  \notag \\
 +  \sum_{  s\in [1,s^*],  t\in [1,t^*]  }     \!\!\!\!\!  
  ( \delta_{\widehat{\mathrm{dc}}}(s,t,d,d',m) +  \delta_{\widehat{\mathrm{dc}}}(s,t,d',d,m)) 
  =\bc^\inn(\mu),  
 \hspace{.7cm}  \notag \\
 \sum_{ p\in [1,s^*\!+\! t^*], i\in [2, \ntreeS] }
      \!\!\!\!\!  (   \delta_{\mathrm{dc}}(p,i,d,d',m) 
      + \delta_{\mathrm{dc}}(p,i,d',d,m) )
 =\bc^\ex(\mu),  
 \hspace{.7cm}  \notag \\
   \mu=(d,d',m)\in \Bc, d<d', \label{eq:NBC8} 
\end{align}  

\begin{align}   
 \sum_{i\in [1,c^*]}\delta_{\mathrm{dc}}(i,d,d,m) 
 +  \sum_{t\in [2,t^*]}\delta_{\mathrm{dc}}(t,1,d,d,m)  
  + \!\!\!\!\! \sum_{  s\in [1,s^*],  t\in [1,t^*]  }\!\!\!\!\!
           \delta_{\widehat{\mathrm{dc}}}(s,t,d,d,m) 
            =\bc^\inn(\mu),  
 \hspace{1cm}  \notag \\
  \sum_{ p\in [1,s^*\!+\! t^*], i\in [2, \ntreeS] }
     \!\!\!\!\! \delta_{\mathrm{dc}}(p,i,d,d,m)      
            =\bc^\ex(\mu),   \hspace{1cm}  \notag \\
             \mu=(d,d,m)\in \Bc. \label{eq:NBC_last}  
\end{align}

\newpage

\section{Descriptions of New Graph Search Algorithms}
\label{sec:graph_search_appendix}

\subsection{Frequency Vectors of Fictitious Trees}

Let $T$ be a chemical bi-rooted or tri-rooted tree,
where we regard a rooted tree $T$ as a bi-rooted tree
with $r_1(T)=r_2(T)$ for a notational convenience.
Recall that our algorithm generates 
a target graph $G\in \G(x^*)$ as a supergraph of $T$,
where one of terminals $r_1(T)$ and $r_2(T)$ can be a 2-branch  of $G$.
We assume that the second terminal
 $r_2(T)$ will be a 2-branch of $G$ in such a case in our algorithms. 
  
\begin{figure}[!ht] \begin{center}
\includegraphics[scale=0.53]{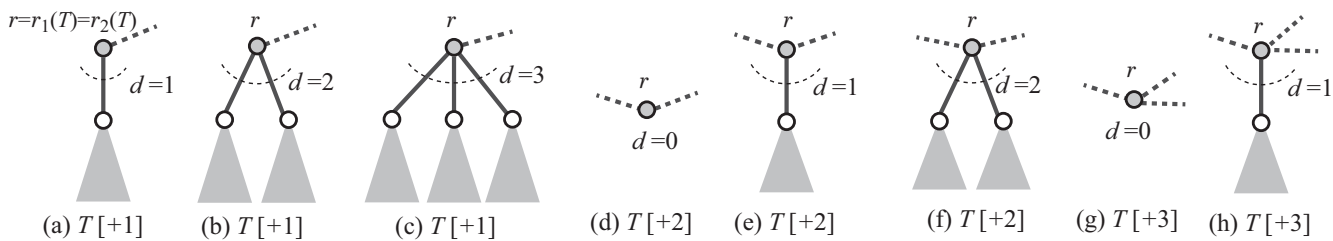}
\end{center}
\caption{An illustration of fictitious rooted trees $T[+p]$, $p\in [1,3]$
for rooted trees $T$ with $r=r_1(T)=r_2(T)$ and $d=\deg_T(r)$,
where a dashed line depicts a fictitious edge incident to the terminal $r_1(T)=r_2(T)$:   
	(a) $T[+1]$ and $d=1$; 
	(b) $T[+1]$ and $d=2$; 
	(c)  $T[+1]$ and $d=3$; 
	(d)  $T[+2]$ and $d=0$; 
	(e)  $T[+2]$ and $d=1$; 
	(f)  $T[+2]$ and $d=2$.; 
	(g)  $T[+3]$ and $d=0$; 
	(h)  $T[+3]$ and $d=1$. 
	}
\label{fig:generate_FT} \end{figure}

\begin{figure}[!ht] \begin{center}
\includegraphics[scale=0.55]{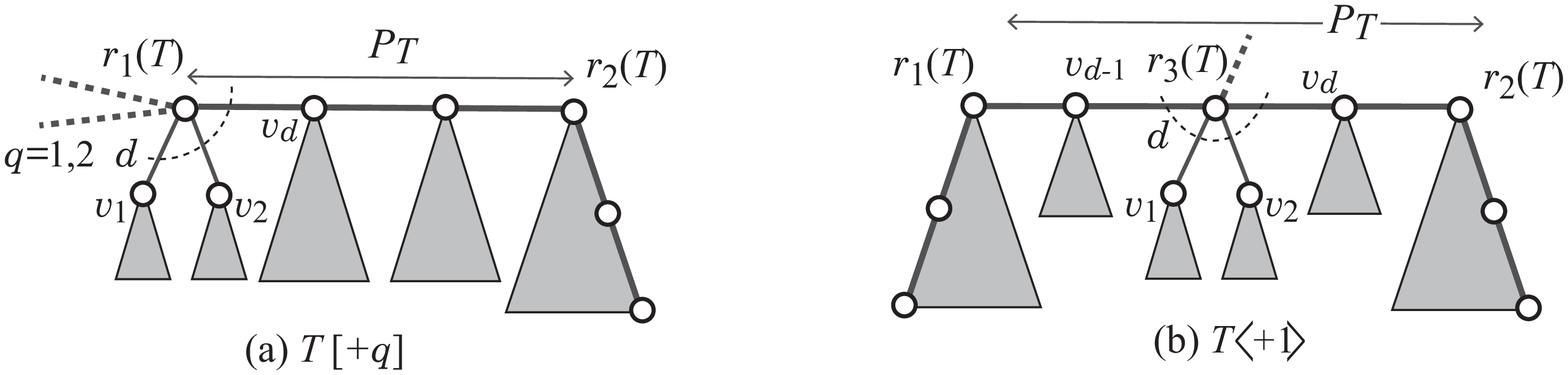}
\end{center}
\caption{An illustration of fictitious  trees
 $T[+q]$ and $T\langle +1\rangle$
for   bi-rooted tree and tri-rooted trees $T$:   
	(a) $T[+q]$ of a bi-rooted tree $T$; 	
	(b)  $T\langle +1\rangle$ of  a tri-rooted tree $T$. 	}
\label{fig:bi-rooted_trees} \end{figure}

For an integer $p\in[1,3]$, 
let   $T[+p]$  denote  a fictitious chemical graph obtained from $T$
by regarding the degree of terminal $r_1(T)$ as
$\deg_T(r_1(T))+p$.
Figure~\ref{fig:generate_FT}
 (resp.,  Figure~\ref{fig:bi-rooted_trees}(a)) illustrates fictitious trees $T[+p]$ 
in the case of  $r_1(T)=r_2(T)$ (resp., $r_1(T)\neq r_2(T)$).
The frequency  vectors $\f_{\inn}(T[+p])$ and 
$\f_{\ex}(T[+p])$ are obtained as follows: 
Let  $d=\deg_T(r_1(T))$, $v_i$, $i\in [1,d]$ denote the
neighbors of  $r_1(T)$, and $d_i=\deg_T(v_i)$, 
$m_i=\beta(r_1(T) v_i)$, 
and $\mu_i=(d, d_i, m_i)$, $\mu'_i=(d+p, d_i, m_i)$,  $i\in [1,d]$. \\
For  $r_1(T)=r_2(T)$ and $d'=d+p$, 
\[ \f_{\inn}(T[+p])
=  \f_{\inn}(T) +\1_{\dg  d'}-\1_{\dg d},  ~
 \f_{\ex}(T[+p])
=  \f_{\ex}(T) + \sum_{1\leq i\leq d}( \1_{\mu'_i}   -\1_{\mu_i}) . \]
For  $r_1(T)\neq r_2(T)$ and $d'=d+p$, where $v_d$ denotes the vertex in $P_T$, 
\[ \f_{\inn}(T[+1]) =  \f_{\inn}(T) 
        +\1_{\dg d'}-\1_{\dg d} +\1_{\mu'_d}-\1_{\mu_d},  ~
 \f_{\ex}(T[+1]) =  \f_{\ex}(T)  
      +\sum_{1\leq i\leq d-1}(\1_{\mu'_i}-\1_{\mu_i}). \]

Let $T$ be a chemical tri-rooted tree, where the third terminal
 $r_3(T)$ is in the backbone path
$P_T$ between vertices $r_1(T)$ and $r_2(T)$.
Let     $T\langle +1\rangle$   denote  a fictitious chemical graph obtained
 from $T$ by regarding the degree of  terminal $r_3(T)$  as
$\deg_T(r_3(T))+1$. 
Figure~\ref{fig:bi-rooted_trees}(b) illustrate a fictitious tri-rooted tree
  $T\langle +1\rangle$. 
The frequency vectors 
$\f_{\inn}(T\langle +1\rangle)$ and 
$\f_{\ex}(T\langle +1\rangle)$ are obtained as follows: 
Let  $d=\deg_T(r_3(T))$, $v_i$, $i\in [1,d]$ denote the
neighbors of  $r_3(T)$, where $v_d$ and $v_{d+1}$ are contained in
the path $P_T$.
For each index $i\in [1,d]$, let  $d_i=\deg_T(v_i)$, 
$m_i=\beta(r_3(T) v_i)$, 
  $\mu_i=(d, d_i, m_i)$ and $\mu'_i=(d+1, d_i, m_i)$. \\
Then 
\[ \f_{\inn}(T\langle +1\rangle)
=  \f_{\inn}(T) +\1_{\dg (d+1)}-\1_{\dg d}
 + \sum_{  i\in[ d-1,d]}( \1_{\mu'_i}   -\1_{\mu_i}) ,  ~
 \f_{\ex}(T\langle +1\rangle)
=  \f_{\ex}(T) + \sum_{ i\leq [1,d-2]}( \1_{\mu'_i}   -\1_{\mu_i}) . \]
  
\subsection{Sets of Frequency Vectors}

For an element $\ta \in \Lambda$ and integers $d \in [0, \dmax-2]$
and  $m\in[d,\val(\ta)-1]$,
 let   $\W_{\inl}^{(0)}(\ta, d, m)$
 (resp., $\W_{\inl+3}^{(0)}(\ta, d, m)$) denote the set of frequency vectors 
$(\f_{\inn}(T[+2]), \f_{\ex}(T[+2]))$ 
(resp.,    $(\f_{\inn}(T[+3]), \f_{\ex}(T[+3]))$) 
of a  chemical rooted tree $T$ such that
  \[\mbox{ $r_1(T)=r_2(T)$, the height of $T$ is at most 2, 
  $\alpha(r_1(T))=\ta$, $\deg_T(r_1(T))=d$,  and $\beta(r_1(T))=m$. }\]
   Recall that $\beta(u)=\sum_{uv\in E}\beta(uv)$ 
   defined in Section~\ref{sec:preliminary}. 
   
For an element $\ta \in \Lambda$ and integers $d \in [1, \dmax-1]$,
  $m\in[d,\val(\ta)-1]$  and $h\geq 0$,
 let   $\W_{\en}^{(h)}(\ta, d, m)$
 (resp.,   $\W_{\en+2}^{(h)}(\ta, d, m)$) denote the set of frequency vectors 
$(\f_{\inn}(T[+1]), \f_{\ex}(T[+1]))$ 
(resp.,   $(\f_{\inn}(T[+2]), \f_{\ex}(T[+2]))$)
of   chemical bi-rooted trees $T$ such that 
\[\mbox{  $\alpha(r_1(T))=\ta$, $\deg_T(r_1(T))=d$, 
   $\beta(r_1(T))=m$, $\ell(P_T)=h$ and}\]
\[\mbox{if $h=0$ then the height of the tree $T'$ rooted at 
$r_2(T)$ is  2. }\]

\subsection{Case of Two Leaf 2-branches}

\subsubsection{Step~1: Enumeration of 2-fringe-trees}
 \label{sec:fringe-tree}
 
The main task of Step~1 is to compute 
for each tuple $(\ta,d,m)$
 of an element $\ta \in \Lambda$ and 
 integers $d \in [1,\dmax-1]$ (resp., $d\in [0,\dmax-2]$) and
  $m\in[d,\val(\ta)-1]$  (resp., $m\in[d,\val(\ta)-2]$),
  the set $\W_{\en}^{(0)}(\ta, d, m)$ 
  (resp.,    $\W_{\inl}^{(0)}(\ta, d, m)$) of all frequency vectors 
  $\f(T[+1])$   (resp.,  $\f(T[+2])$) 
  of  chemical rooted trees $T$ such that  
$r_1(T)=r_2(T)$, 
  $\alpha(r_1(T))=\ta$, $\deg_T(r_1(T))=d$ and $\beta(r_1(T))=m$.
  
Step~1 first computes the set $\mathcal{FT}$ 
of all possible chemical rooted trees $T\in \T(\x^*)$
 (where $r_1(T)=r_2(T)$)
that can be a 2-fringe-tree of  a target graph $G\in \G(x^*)$.
For this, we design a branch-and-bound procedure 
where we append a new vertex one by one
to construct a rooted tree with only one child.
To design a bounding procedure,
we derive a property of the structure
of chemical rooted trees that can be a 2-fringe-tree of  a target graph

Let $G_0$ be a  chemical rooted tree with a terminal $r_0=r_1(G_0)=r_2(G_0)$,
where $\f_{\inn}(\alpha(r_0);G_0)=1$ 
and $\f_{\inn}(\ta;G_0)=0$, 
$\ta\in \Lambda\setminus\{\alpha(r_0)\}$
and $\f_{\inn}(\gamma;G_0)=0$,
 $\gamma\in \Gamma$.
For a vector $\x = (\x_\inn, \x_\ex)$ with  
$\x_\inn, \x_\ex\in \mathbb{Z}_+^{\LtoD}$, 
we call  $G_0$  {\em $\x$-extensible} if 
some chemical acyclic graph $G\in \G(\x)$
  contains $G_0$ as a subgraph of  a 2-fringe-tree $T$ rooted at $r_0$ in $G$.

We use the next condition as a bounding procedure when we 
generate chemical rooted trees in Step~1.  

\begin{lemma}\label{le:subgraph} 
For a branch-parameter $k$, let $\x^* = (\x_\inn^*, \x_\ex^*)$ 
be a vector with $\x_\inn^*, \x_\ex^*\in \mathbb{Z}_+^{\LtoD}$, 
and $G_0$ be a chemical rooted tree rooted 
at a vertex $r_0$ such that
$\f(G_0)\leq \x^*$.  
\begin{enumerate}
\item[{\rm (i)}] 
Graph $G_0$ is $\x^*$-extensible   only when the next holds for any   subset 
$\Lambda'\subseteq \Lambda$: 
\begin{equation}\label{eq:nec}
\displaystyle{  \sum_{\ta\in \Lambda'} 
            (\x_\ex^*(\ta)   - \f_\ex(\ta; G_0))  
  \leq  
 \sum_{\substack{\gamma=({\tt a,b},k)\in \Gamma: \\ \ta\in \Lambda',
           {\tt b}\in \Lambda\setminus \Lambda'}}  
      \!\!\!   (\x_\ex^*(\gamma)  - \f_\ex(\gamma; G_0))  
   +   2 \!\!\!\sum_{\substack{\gamma=({\tt a,b},k)\in \Gamma: \\ {\tt a,b}\in \Lambda'}}
          \!\!\!(\x_\ex^*(\gamma)  - \f_\ex(\gamma; G_0)). } 
\end{equation}
\item[{\rm (ii)}] 
Let $G_1$ denote the chemical rooted tree obtained from $G_0$
by appending a new atom with an element ${\tt b}\in \Lambda$
to an atom with an element $\ta\in \Lambda$ in $G_0$ 
with a multiplicity $q$; i.e., we join an atom $\ta$ in $G_0$ and a new atom ${\tt b}$  
with an adjacency-configuration $({\tt a,b},q)$.
Then $G_1$  is $\x^*$-extensible only when the next holds: 

\[  \x_\ex^*(\ta) -  \f_\ex(\ta; G_0)\leq \pmb{nb}(\ta) -1 \]  
for 
\[
\pmb{nb}(\ta)=  \displaystyle{ \sum_{\gamma=({\tt a,b},k)\in \Gamma: 
{\tt b\neq a}\in \Lambda }
\!\!\!\!\!\!
(\x_\ex^*(\gamma)  - \f_\ex(\gamma; G_0) )   
  + 2 \!\!\! \sum_{\gamma=({\tt a,a},k)\in \Gamma}
       (\x_\ex^*(\gamma) - \f_\ex(\gamma; G_0) ) }.
\]
\end{enumerate}
\end{lemma}
\noindent {\bf Proof.} 
(i) 
Assume that $G_0$ is a subgraph of a 2-fringe-tree $T$
 in some chemical graph $G\in \G(\x^*)$ 
so that   $T$ is rooted at $r_0$. 
The left-hand side means the number of the remaining $k$-external vertices with elements in
$\Lambda'$ in the 2-fringe-trees in $G$. 
Each of such atoms has a neighbor in the connected graph $G$.
The right-hand side indicates 
an upper bound on the number of $k$-external edges joining elements in $\Lambda'$ 
in the 2-fringe-trees in $G$. 

(ii) Note that 
$\f_{\ex [\Lambda\cup\Gamma]}(G_1)
= \f_{\ex [\Lambda\cup\Gamma]}(G_0)+\1_{\tt b}+\1_{\gamma}$.
For $\Lambda'=\{\ta\}$, 
the left-hand side in Eq.~(\ref{eq:nec}) is 
$\x^*_\ex(\ta)- \f_\ex(\ta; G_0)$, which 
remains unchanged if ${\tt a\neq b}$
(resp., reduces by 1 if ${\tt a= b}$); and  
the right-hand side in (\ref{eq:nec}) is $\pmb{nb}(\ta)$, which 
 reduces by 1 if ${\tt a\neq b}$ (resp., reduces by 2 if ${\tt a= b}$).
That is, the left-hand side minus the right-hand side in (\ref{eq:nec})  
always reduces by 1.
This gives the required necessary condition for $G_1$ to be $\x^*$-extensible.
\qed\bigskip 
   
\begin{figure}[!ht] \begin{center}
\includegraphics[width=.75\columnwidth]{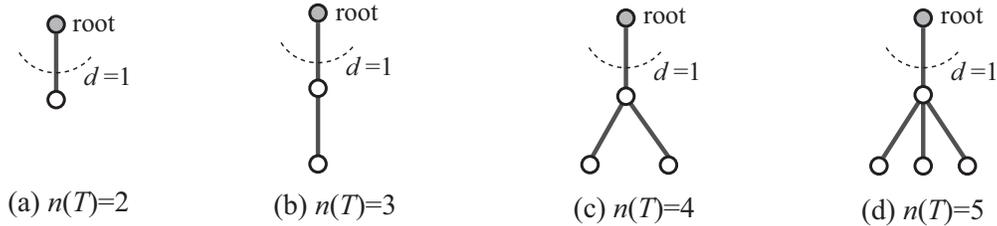}
\end{center} 
\caption{An illustration of rooted trees $T$ with height at most 2
and only one child satisfying the size constraint: 
(a) case of $n(T)=2$;
(b) case of $n(T)=3$;
(c) case of $n(T)=4$;
(d) case of $n(T)=5$. }
\label{fig:FT_one_child} \end{figure}

Figure~\ref{fig:FT_one_child} illustrates all graph structures of 
rooted trees $T$ with height at most 2
and only one child satisfying the size constraint (\ref{eq:fringe-size}).
For each element $\ta\in \Lambda$,
we enumerate chemical  trees $T\in \T(x^*)$ rooted a vertex $r$ with 
$\alpha(r)=\ta$ that has only one child 
by a branch-and-bound algorithm. 
Let $\T_\ta$ denote the set of resulting rooted trees
for each root element $\ta\in \Lambda$.

We next  enumerate chemical  trees $T\in \T(x^*)$ rooted a vertex $r$ with 
$\alpha(r)=\ta$ that has two or three children
by generating a combination of two or three graphs in $\T_\ta$.
During generating graphs, our bounding procedure tests whether the current graph 
satisfies the necessary condition in Lemma~\ref{le:subgraph}(ii).

Finally we compute the following sets: \\
for each element   $\ta\in \Lambda$, 
integers  $d\in [1,\dmax-1]$, $m\in[d,\val(\ta)-1]$,  
the set $\W_{\en}^{(0)}(\ta, d, m)$ 
of frequency vectors $\f(T[+1])$
for rooted trees $T\in \T_\ta$ with $\deg_T(r)=d$ and height 2; \\
for each element   $\ta\in \Lambda$, 
integers  $d\in [0,\dmax-2]$, $m\in[d,\val(\ta)-2]$,  
the set  $\W_{\inl}^{(0)}(\ta, d, m)$
of frequency vectors $\f(T[+2])$
for rooted trees $T\in \T_\ta$ with $\deg_T(r)=d$ and height  at most 2.  
         
For each vector $\w\in \W_{\en}^{(0)}(\ta, d, m)$ 
(resp.,   $\w\in\W_{\inl}^{(0)}(\ta, d, m)$),
we store a sample tree $T_{\w}$.  

\subsubsection{Step~2: Generation of Frequency Vectors of End-subtrees}
 \label{sec:internal-subtree}

The main task of Step~2 is to compute the following sets 
in the ascending order of $h=1,2,\ldots, \delta_2 $:\\
for elements $\ta \in \Lambda$, 
integers $d \in [1,\dmax-1]$,
$m \in [d ,\val(\ta )-1]$ 
 and $h\in[ 1,  \delta_2 ]$,
  the sets  $\W_{\en}^{(h)}(\ta,  d , m )$ 
of all frequency vectors  $\f(T[+1])$   
of chemical bi-rooted trees  $T\in \T(x^*)$ such that 
 $\alpha(r_1(T))=\ta$, 
 $\deg_T(r_1(T))=d $, $\beta(r_1(T))=m$
 and
 $\ell(P_T)=h$.   
  
\begin{figure}[!ht] \begin{center}
\includegraphics[width=.6\columnwidth]{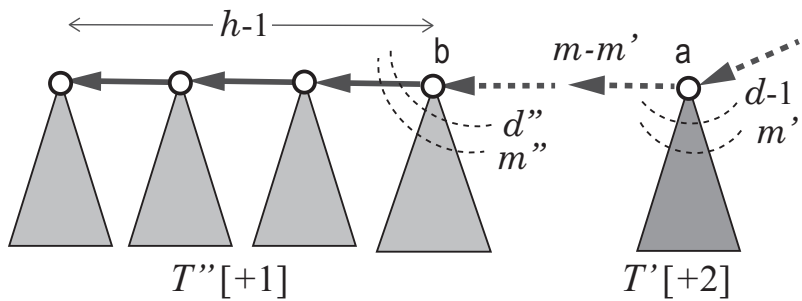}
\end{center} 
\caption{An illustration of appending a rooted tree $T'$ to a bi-rooted tree
$T''$ to compute a vector $\w\in \W_{\en}^{(h)}(\ta,  d , m )$
from the frequency vectors 
$\w'=\f(T'[+2])\in \W_{\inl}^{(0)}(\ta , d -1, m')$
  of a rooted tree $T'$   and 
$\w'' =\f(T''[+1])  \in \W_{\en}^{(h-1)}(\tb, d'', m'')$ 
  of a bi-rooted tree $T''$. }
\label{fig:compute_subtree} \end{figure}

Observe that 
 each vector $\w=(\w_{\inn},\w_{\ex})\in \W_{\en}^{(h)}(\ta,  d , m )$
 is obtained from a combination of vectors  
$\w'=(\w'_{\inn},\w'_{\ex})\in \W_{\inl}^{(0)}(\ta , d -1, m')$ 
 and 
$\w''=(\w''_{\inn},\w''_{\ex}) \in \W_{\en}^{(h-1)}(\tb, d'', m'')$ 
such that 
\[\begin{array}{l}
  m'  \leq \val(\ta ) - 2,  ~~~ 1\leq  m - m'  \leq \val(b) -m'' ,   \\
 \w_{\inn} = \w'_{\inn} + \w''_{\inn} + \1_\gamma + \1_\mu \le \x_{\inn}^*, 
 ~~~
 \w_{\ex} = \w'_{\ex} + \w''_{\ex}  \le \x_{\ex}^* \\
 \mbox{ for  } \gamma = (\ta , \tb, m - m' ) \in \Gamma \mbox{ and  } 
 \mu = (d+2,  d''+1, m - m') \in \Bc. 
\end{array}\] 
Figure~\ref{fig:compute_subtree} illustrates this process of computing
a vector $\w \in \W_{\en}^{(h)}(\ta,  d , m )$.

For each vector $\w\in \W_{\en}^{(h)}(\ta,  d , m )$
obtained from a combination  $\w'\in \W_{\inl}^{(0)}(\ta , d -1, m')$ 
 and  
$\w'' \in \W_{\en}^{(h-1)}(\tb, d'', m'')$, 
we construct  a sample tree $T_{\w}$ from their sample trees
$T_{\w'}$ and $T_{\w''}$.

\subsubsection{Step~3: Enumeration of Feasible  Vector Pairs}
\label{sec:feasible_pair}

A {\em feasible pair} of vectors is defined to be a pair of vectors
$\w^i= (\w^i_{\inn}, \w^i_{\ex})  \in \W_{\en}^{(\delta_i)} (\ta_i,d_i,m_i)$,
 $\ta_i\in \Lambda$, $d_i\in [1,\dmax-1]$, $m_i\in[d_i,\val(\ta_i)-1]$,  $i = 1,2$
that admits an adjacency-configuration $\gamma = (\ta_1, \ta_2, m) \in \Gamma$
and a bond-configuration $\mu = (d_1 + 1, d_2 + 1, m) \in \Bc$  
with an integer $m \in [1, \min\{3, \val(\ta_1) - m_1, \val(\ta_2) - m_2\} ]$ 
such that
\[\mbox{
$\x^*_{\inn} = \w^1_{\inn} + \w^2_{\inn} + \1_{\gamma} + \1_{\mu}$
and
$\x^*_{\ex} = \w^1_{\ex} + \w^2_{\ex}$, }\]
or equivalently $\w^1$ is equal to 
the vector $(\x^*_{\inn} - \w^2_{\inn}-\1_{\gamma} -\1_{\mu}, 
\x^*_{\ex} - \w^1_{\ex})$,
which we call  the {\em $(\gamma, \mu)$-complement}
of $\w^2$, and denote it by  $\overline{\w^2}$.

The main task of Step~3 is to enumerate all feasible vector pairs $(\w^1,\w^2)$, 
$\w^i \in \W_{\en}^{(\delta_i)}(\ta_i,d_i,m_i)$ 
 with $\ta_i\in \Lambda$, $d_i\in [1,\dmax-1]$, $m_i\in[d_i,\val(\ta_i)-1]$,  $i = 1,2$.
  
To efficiently search for a feasible pair of vectors
in two sets $\W_{\en}^{(\delta_i)}(\ta_i,d_i,m_i)$,   $i = 1,2$, 
we first compute the 
$(\gamma, \mu)$-complement vector $\overline{\w}$  
of each vector $\w\in \W_{\en}^{(\delta_2)}(\ta_2,d_2,m_2)$ for each pair of $ \gamma = (\ta_1, \ta_2, m) \in \Gamma$
and $\mu = (d_1 + 1, d_2 + 1, m) \in \Bc$  with 
$m \in [1, \min\{3, \val(\ta_1) - m_1, \val(\ta_2) - m_2\} ]$,
and denote by  $\overline{\W_{\en}^{(\delta_2)}}$ the set of the resulting 
$(\gamma, \mu)$-complement vectors.
Observe that $(\w^1,\w^2)$ is a feasible vector pair if and only if
 $\w_1=\overline{\w_2}$.
To find such pairs,  we merge the sets $\W_{\en}^{(\delta_1)}(\ta_1,d_1,m_1)$
 and $\overline{\W_{\en}^{(\delta_2)}}$
into a sorted list $L_{\gamma, \mu}$.
Then each consecutive pair of vectors $\z_1, \z_2 \in L_{\gamma, \mu}$
gives a feasible pair of vectors $\z_1$ and $\overline{\z_2}$.

\subsubsection{Step~4: Construction of Chemical Graphs}
 \label{sec:construct}
 
The task of Step~4 is to  construct for each feasible vector pair  
 $\w^i\in \W_{\en}^{(\delta_i)}(\ta_i,d_i,m_i)$,   $i = 1,2$ such that
  $\w^1$ is  equal to the $(\gamma= (\ta_1, \ta_2, m), \mu)$-complement 
vector $\overline{\w^2}$  of $\w^2$,
construct  a target graph  $T_{(\w_1, \w_2)}\in \G(\x^*)$ by combining
the sample trees $T_i=T_{\w^i}$ of vectors $\w^i$ with
an edge $e=r_1(T_1)r_1(T_2)$ such that $\beta(e)=m$.
Figure~\ref{fig:combine_two_2-branches} illustrates
two sample trees $T_i$, $i=1,2$ to be combined 
with a new edge $e=r_1(T_1)r_1(T_2)$.

\subsection{Case of Three Leaf 2-branches}

\subsubsection{Step~1: Enumeration of 2-fringe-trees}
 \label{sec:fringe-tree_3}
  
The main task of Step~1 is to compute the following sets: \\
for each tuple $(\ta,d,m)$
 of an element $\ta \in \Lambda$ and 
 integers $d \in [1,\dmax-1]$ (resp., $d\in [0,\dmax-2]$ and $d\in [0,\dmax-3]$) and
  $m\in[d,\val(\ta)-1]$ (resp., $m\in[d,\val(\ta)-2]$ and $m\in[d,\val(\ta)-3]$),
  the set $\W_{\en}^{(0)}(\ta, d, m)$ 
  (resp.,    $\W_{\inl}^{(0)}(\ta, d, m)$ and $\W_{\inl+3}^{(0)}(\ta, d, m)$)
   of all frequency vectors    $\f(T[+1])$ 
    (resp.,  $\f(T[+2])$ and $\f(T[+3])$) 
  of  chemical rooted trees $T$ such that 
$r_1(T)=r_2(T)$, 
  $\alpha(r_1(T))=\ta$, $\deg_T(r_1(T))=d$ and $\beta(r_1(T))=m$.
For each vector $\w\in \W_{\en}^{(0)}(\ta, d, m)$ 
  (resp., $\w\in \W_{\inl}^{(0)}(\ta, d, m)$ 
  and $\w\in \W_{\inl+3}^{(0)}(\ta, d, m)$),
we store a sample tree $T_{\w}$.  
 This step can be designed in a similar way of Step~1 for the case of $\bl_2(G)=2$.

\subsubsection{Step~2: Generation of Frequency Vectors of End-subtrees}
 \label{sec:internal-subtree_3}

Analogously with  Step~2 for the case of $\bl_2(G)=2$,
 Step~2  computes the following sets 
 in the ascending order of $h=1,2,\ldots,  \dia^* -6 -\delta_3$:\\
for elements $\ta \in \Lambda$, 
integers $d  \in [1,\dmax-1]$,
$m \in [d ,\val(\ta )-1]$, $i=1,2$
 and $h\in[ 1,  \dia^* -6 -\delta_3]$,
  the sets  $\W_{\en}^{(h)}(\ta,d,m)$ 
of all frequency vectors  $\f(T[+1])$   
of chemical bi-rooted trees  $T\in \T(x^*)$ such that 
 $\alpha(r_1(T))=\ta $, 
 $\deg_T(r_1(T))=d$, $\beta(r_1(T))=m$ 
 and  $\ell(P_T)=h$.    
  
For each vector $\w\in \W_{\en}^{(h)}(\ta, d, m)$, 
we construct  a sample tree $T_{\w}$ from their sample trees
$T_{\w'}$ and $T_{\w''}$.

\subsubsection{Step~3:   Generation of Frequency Vectors of End-subtrees
with Two Fictitious Edges}

The main task of Step~3 is to compute the following sets: \\
for elements $\ta \in \Lambda$,
integers $d  \in [1,\dmax-2]$, $m\in [d,\val(\ta)-2]$  
 and $h\in  [\lceil \dia^*/2\rceil -2,  \dia^* - 5 -\delta_3] $,
 the  sets $\W_{\en+2}^{(h)}(\ta,d,m)$ 
of all frequency vectors of bi-rooted trees $T[+2]$ such that
  $\alpha(r_1(T))=\ta$, $\deg_T(r_1(T))=d$, 
   $\beta(r_1(T))=m$ and  $\ell(P_T)=h$. 
For each vector $\w\in \W_{\en+2}^{(h)}(\ta,d,m)$,
we store  a sample tree $T_{\w}$.   
 This step can be designed
  in a similar way of Step~3 for the case of $\bl_2(G)=2$.

\subsubsection{Step~4: Enumeration of Frequency Vectors of Main-subtrees}

For an element $\ta \in \Lambda$, and integers $d \in   [2,\dmax-1]$, 
$m \in [d, \val(\ta) -1]$,  and 
$\delta_1 \in[\lceil \dia^*/2\rceil -3,  \dia^* - 6 -\delta_3]$,
define   
$\W_{\mathrm{main}}^{(\delta_1+1)}(\ta, d, m)$ to be the 
set of  the frequency vectors $\f(T\langle+1\rangle)$ 
of chemical tri-rooted trees $T$ such that 
 \[\mbox{
 $\alpha(r_1(T))=\ta$, 
 $\deg_T(r_1(T))=d$, $\beta(r_1(T))=m$, $\ell(P_T)=\dia^*-4$ and }\]
\[\mbox{
 the length of the path $P_{r_2(T),r_3(T)}$ between
 vertices $r_2(T)$ and $r_3(T)$ is $\delta_1+1$.
 }\]
 See Figure~\ref{fig:combine_three_2-branches} for the structure of
 a main-tree. 
 Such a chemical tri-rooted graph $T$ corresponds to the main-subtree
 of a target graph $G\in \G(x^*)$. 

\begin{figure}[!ht] \begin{center}
\includegraphics[width=.75\columnwidth]{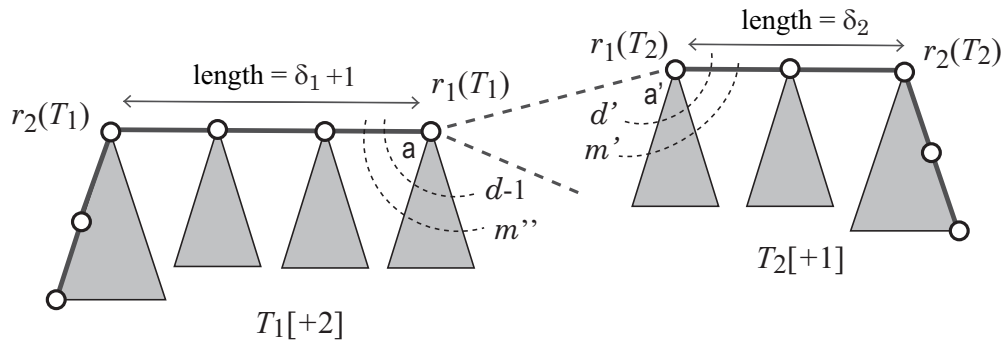}
\end{center}
\caption{An illustration of computing
the frequency vector $\w=\f(T\langle+1\rangle)
\in \W_{\mathrm{main}}^{(\delta_1+1)}(\ta, d, m)$ 
of a tri-rooted tree $T$ from the  frequency vectors
$\w^1=\f(T_1[+2])\in \W_{\en+2}^{(\delta_1+1)}(\ta, d-1, m'')$ and 
$\w^2=\f(T_2[+1])\in \W_{\en}^{(\delta_2)}(\ta', d', m')$
for  bi-rooted trees $T_1$ and $T_2$.
}
\label{fig:main-subtree} 
\end{figure}

The main task of Step~4 is to compute the sets 
 $\W_{\mathrm{main}}^{(\delta_1+1)}(\ta, d, m)$,
 $\ta \in \Lambda$, $d \in  [2,\dmax-1]$, 
$m \in [d, \val(\ta) -1]$,   
$\delta_1 \in[\lceil \dia^*/2\rceil -3,  \dia^* - 6 -\delta_3]$.
Each vector $\w\in \W_{\mathrm{main}}^{(\delta_1+1)}(\ta, d, m)$
can be obtained from a combination of 
vectors $\w^1\in  \W_{\en+2}^{(\delta_1+1)}(\ta, d-1, m'')$ and 
$\w^2\in \W_{\en}^{(\delta_2)}(\ta', d', m')$ such that 
$\delta_1 + \delta_2 = \dia^* - 4$ and $\delta_1 \geq \delta_2 $, 
as illustrated in Figure~\ref{fig:main-subtree}.
For each vector $\w\in \W_{\mathrm{main}}^{(\delta_1+1)}(\ta, d, m)$,
we store a sample tree $T_{\w}$. 
 This step can be designed
  in a similar way of Step~3 for the case of $\bl_2(G)=2$.

\subsubsection{Step~5: Enumeration of Feasible Vector Pairs}
\label{sec:feasible_pair_3}

Analogously with the case of  $\bl_2(G)=2$,
a {\em feasible pair} of vectors is defined to be a pair of vectors
$\w^1= (\w^1_{\inn}, \w^1_{\ex})
  \in \W_{\mathrm{main}}^{(\delta_1+1)}(\ta_1,d_1, m_1)$,   
and 
$\w^2= (\w^2_{\inn}, \w^2_{\ex})  \in \W_{\en}^{(\delta_3)}(\ta_2,d_2, m_2)$,
$\delta_1 \in [\lceil \dia^*/2\rceil -3,  \dia^* - 6 -\delta_3]$,
 $\ta_i\in \Lambda$, $d_i\in [1,\dmax-1]$, $m_i\in[d_i,\val(\ta_i)-1]$,  $i = 1,2$
that admits an adjacency-configuration $\gamma = (\ta_1, \ta_2, m) \in \Gamma$
and a bond-configuration $\mu = (d_1 + 1, d_2 + 1, m) \in \Bc$  
with an integer $m \in [1, \min\{3, \val(\ta_1) - m_1, \val(\ta_2) - m_2\} ]$ 
such that
\[\mbox{
$\x^*_{\inn} = \w^1_{\inn} + \w^2_{\inn} + \1_{\gamma} + \1_{\mu}$
and
$\x^*_{\ex} = \w^1_{\ex} + \w^2_{\ex}$. }\]
Step~5 computes the set all feasible vector pairs $(\w^1,\w^2)$
by using a sorting algorithm as in the Step~4 for the case of  $\bl_2(G)=2$.

\subsubsection{Step~6: Construction of Chemical Graphs}
 \label{sec:construct_3}
 
 Analogously with Step~4 for the case of  $\bl_2(G)=2$, 
  Step~6  constructs   a target graph  $T_{(\w_1, \w_2)}\in \G(\x^*)$ 
  for each feasible vector pair  $(\w^1,\w^2)$    by combining
the sample trees $T_i=T_{\w^i}$ of vectors $\w^i$ with
a new edge $e=r_1(T_1)r_1(T_2)$. 
 
\end{document}